\documentclass[aps,pra,reprint,amsmath,amssymb,superscriptaddress,onecolumn,longbibliography,nofootinbib,notitlepage]{revtex4-2}
\usepackage{mathtools}
\usepackage{siunitx}
\usepackage[hidelinks]{hyperref}
\usepackage{svg}
\usepackage{float}
\usepackage[braket, qm]{qcircuit}
\usepackage{bbm}
\usepackage{bm}
\usepackage{lipsum}
\usepackage{tocloft}
\usepackage{titletoc}
\usepackage{multirow}
\usepackage{adjustbox}
\usepackage{xspace}
\usepackage[normalem]{ulem}
\usepackage[commentmarkup=uwave]{changes}
\usepackage[dvipsnames]{xcolor}

\newcommand{\nocontentsline}[3]{}
\let\origcontentsline\addcontentsline
\newcommand\stoptoc{\let\addcontentsline\nocontentsline}
\newcommand\resumetoc{\let\addcontentsline\origcontentsline}

\newcommand{\Udec}{\mathcal{U}_{\rm meas}}
\newcommand{\avg}[1]{\langle #1 \rangle}

\usepackage{xcolor}

\begin{document}
\title{Quantum computational displacement sensing}

\author{Sridhar~Prabhu}
\thanks{Contact: svp36@cornell.edu, pmcmahon@cornell.edu}
\affiliation{School of Applied and Engineering Physics, Cornell University, Ithaca, NY 14853, USA}
\affiliation{Department of Physics, Cornell University, Ithaca, NY 14853, USA}

\author{Saeed~A.~Khan}
\affiliation{School of Applied and Engineering Physics, Cornell University, Ithaca, NY 14853, USA}

\author{Xingrui~Song}
\affiliation{School of Applied and Engineering Physics, Cornell University, Ithaca, NY 14853, USA}

\author{Mathieu~Ouellet}
\affiliation{School of Applied and Engineering Physics, Cornell University, Ithaca, NY 14853, USA}

\author{Ryotatsu~Yanagimoto}
\affiliation{School of Applied and Engineering Physics, Cornell University, Ithaca, NY 14853, USA}
\affiliation{Physics \& Informatics Laboratories, NTT Research, Inc., Sunnyvale, CA 94085, USA}

\author{Saswata~Roy}
\affiliation{School of Applied and Engineering Physics, Cornell University, Ithaca, NY 14853, USA}
\affiliation{Department of Physics, Cornell University, Ithaca, NY 14853, USA}

\author{Alen~Senanian}
\thanks{Present address: HRL Laboratories, 3011 Malibu Canyon Road, Malibu, CA 90265, USA}
\affiliation{School of Applied and Engineering Physics, Cornell University, Ithaca, NY 14853, USA}
\affiliation{Department of Physics, Cornell University, Ithaca, NY 14853, USA}

\author{Logan~G.~Wright}
\thanks{Present address: Department of Applied Physics, Yale University, New Haven, CT 06520, USA}
\affiliation{School of Applied and Engineering Physics, Cornell University, Ithaca, NY 14853, USA}
\affiliation{Physics \& Informatics Laboratories, NTT Research, Inc., Sunnyvale, CA 94085, USA}

\author{Valla~Fatemi}
\affiliation{School of Applied and Engineering Physics, Cornell University, Ithaca, NY 14853, USA}

\author{Peter~L.~McMahon}
\thanks{Contact: svp36@cornell.edu, pmcmahon@cornell.edu}
\affiliation{School of Applied and Engineering Physics, Cornell University, Ithaca, NY 14853, USA}
\affiliation{Kavli Institute at Cornell for Nanoscale Science, Cornell University, Ithaca, NY 14853, USA}

\begin{abstract}

Quantum computational sensing combines quantum sensing with quantum computing to extract task-relevant information from the physical world. Quantum computational sensing can in principle achieve an accuracy advantage for specific tasks versus the alternative of raw-signal estimation using conventional quantum sensing followed by task-specific classical postprocessing. Here we report the experimental demonstration of quantum computational displacement sensing (QCDS) with a superconducting circuit comprising a qubit coupled to an oscillator. We consider binary classification sensing tasks, where the goal is to predict the class label of a single complex-valued displacement sensed once by the oscillator. Rather than estimating the displacement, our computational-sensing protocol -- using parameterized quantum circuits before and after sensing -- attempts to determine the binary class label using quantum processing and map it onto the ground or excited state of the qubit. A single measurement of the qubit therefore directly outputs the prediction. We implemented circuits with up to 24 entangling gates and 38 free parameters, which were trained in silico. We show that increasing the circuit depth systematically improves expressivity and classification accuracy. We experimentally obtained an accuracy advantage over a suite of protocols that first use conventional quantum sensing to estimate the displacement before using classical postprocessing to perform prediction. For certain tasks, our protocol achieves a $15$-percentage-points higher classification accuracy than most conventional approaches. Our results establish the feasibility of quantum computational sensing with noisy superconducting hardware and illustrate how integrating quantum computation with quantum sensing can enhance performance when the goal is to estimate a property or function of a signal rather than to estimate the signal.

\end{abstract}
\maketitle

\section{Introduction}
\label{sec:intro}

Quantum sensing beyond the standard quantum limit (SQL)~\cite{degen2017quantum, giovannetti2004quantum, giovannetti2011advances, giovannetti2006quantum} has been demonstrated across a wide range of platforms and applications, including gravitational-wave detection~\cite{jia2024squeezing}, biomedical imaging~\cite{taylor2013biological}, precision timekeeping~\cite{ludlow2015optical}, and dark-matter searches~\cite{backes2021quantum}. In many sensing applications, however, the goal is not ultimately to reconstruct the full signal but rather to perform some downstream task based on the signal. Mathematically, this corresponds to computing a function $\mathcal{F}(\boldsymbol{\alpha})$ of the sensed signal $\boldsymbol{\alpha}$, rather than estimating $\boldsymbol{\alpha}$ itself~\cite{khan2025quantum}. The performance of a protocol in such tasks is therefore determined by how accurately $\mathcal{F}(\boldsymbol{\alpha})$ can be inferred from the sensor measurement outcomes. In this work, we consider functions $\mathcal{F}(\boldsymbol{\alpha})$ whose outputs take one of two possible values (corresponding to binary classification tasks). The signal we sense, $\boldsymbol{\alpha} = (\alpha_x,\alpha_p)$, is represented by a 2D vector denoting the in-phase (position) and quadrature (momentum) components of a displacement of a superconducting microwave bosonic mode. We experimentally study this task in the single-shot regime~\cite{sinanan2024single, liao2024quantum, khan2025quantum_novel, meyer2025quantum, chin2025quantum, marciniak2022optimal, ilias2025end}, where the quantum system is allowed to sense the displacement only once.

In conventional quantum sensing, the sensor is designed to efficiently estimate the displacement $\boldsymbol{\alpha}$. For example, a sensor that performs a heterodyne measurement simultaneously outputs an estimate of both $\alpha_x$ and $\alpha_p$. However, since the operators associated with the position and momentum components do not commute, their measurement outcomes must have at least an additional half a photon of noise~\cite{caves1982quantum, clerk2010introduction}. This yields a noisy estimate $(\widetilde{\alpha}_x, \widetilde{\alpha}_p)$, which would then be processed by a classical computer -- for instance, by a neural network -- to predict the class from which the displacement originated~\cite{khan2025quantum_novel}. For small displacements, where quantum noise dominates, this indirect approach cannot achieve perfect classification accuracy, regardless of the expressivity (the complexity of the space of functions that can be represented)~\cite{raghu2017expressive} of the classical postprocessing. The limitation arises from the finite information that is extracted by the quantum sensor in a single shot -- a constraint that applies to all quantum sensing protocols.

\begin{figure}[h!]
    \centering
    \includegraphics[width=\textwidth]{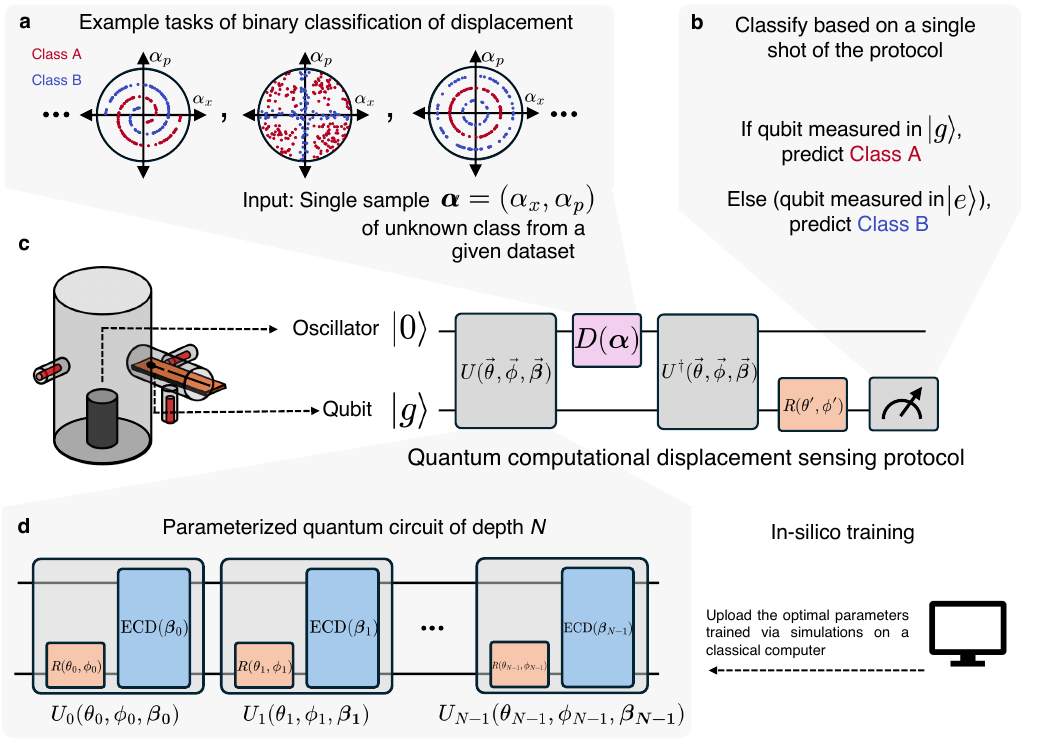}
    \caption{\textbf{Overview of the quantum computational displacement sensing (QCDS) protocol.} We consider sensing tasks where an oscillator experiences a displacement $\boldsymbol{\alpha}$ originating from either class A or B. The goal is to predict which class the displacement originated from, after sensing it once. \textbf{a.} Examples of tasks we consider. A randomly selected subset of the dataset used for training is plotted to aid visualization. Each class consists of a distribution of displacements, with position $(\alpha_x)$ and momentum $(\alpha_p)$ quadratures. A single randomly selected a priori unknown displacement data point $\boldsymbol{\alpha}$ is sensed. \textbf{b.} The output of the protocol is a single bit corresponding to the outcome of the qubit measurement at the end. The value of the bit directly corresponds to the predicted class. \textbf{c.} Left: Our quantum computational sensor is a superconducting circuit comprising a transmon qubit coupled to the bosonic mode of a 3D cavity (oscillator) (see Appendix~\ref{app:experiment_setup} for more details). Right: Circuit diagram of the protocol, which consists of a parameterized quantum circuit $U(\vec{\theta}, \vec{\phi}, \vec{\boldsymbol{\beta}})$ before the displacement sensing, and $U^\dagger(\vec{\theta}, \vec{\phi}, \vec{\boldsymbol{\beta}})$ afterwards. The qubit is measured in its computational basis after a final single qubit rotation. \textbf{d.} $U(\vec{\theta}, \vec{\phi}, \vec{\boldsymbol{\beta}})$ is composed of a sequence of gates $U_i(\theta_i, \phi_i, \boldsymbol{\beta_i})$ of length $N$, each of which is in turn composed of a single qubit rotation $R(\theta_i, \phi_i)$ and echoed conditional displacement $\mathrm{ECD}(\boldsymbol{\beta}_i)$ gate with trainable parameters. The optimal task-dependent parameters are found via a simulation of the protocol on a classical computer. Increasing the depth of the protocol increases the number of trainable parameters, allowing the protocol to better discriminate displacements from the two datasets and achieve a higher classification accuracy.}
    \label{fig:main1}
\end{figure}

Instead of reconstructing the signal $\boldsymbol{\alpha}$, quantum computational sensors \cite{khan2025quantum} perform task-dependent computations directly within the quantum domain, prior to measurement, such that measurement outcomes yield an estimate of the desired function. This direct approach can produce a more accurate estimate of $\mathcal{F}(\boldsymbol{\alpha})$ over conventional approaches, achieving a quantum computational-sensing advantage (QCSA) in classification accuracy. Here, we experimentally realize a quantum computational sensor by coupling a transmon qubit to a single microwave bosonic mode. We implement a quantum-computational-displacement-sensing (QCDS) protocol, based on theoretical proposals in Refs.~\cite{sinanan2024single, liao2024quantum, khan2025quantum_novel}, which employs parameterized quantum circuits applied before and after the sensing interaction, followed by one measurement of the single qubit. The circuit parameters are trained so that the final measurement outcome produces an estimate of the function directly: the class is inferred from whether the qubit collapses to the ground ($\ket{g}$) or excited ($\ket{e}$) state. The circuit depth $N$ determines the number of trainable parameters and the expressivity of the protocol. Increasing $N$ generally allows the protocol to better approximate $\mathcal{F}(\boldsymbol{\alpha})$ and achieve higher classification accuracy. $\mathcal{F}(\boldsymbol{\alpha})$ can be output without error, in a single shot, when all displacements of one class are mapped to $\ket{g}$, and all displacements of the other class are mapped to $\ket{e}$. 

The structure of this paper is as follows. We begin by describing the QCDS protocol and its training procedure. We then present three example tasks to illustrate how increasing the circuit depth improves the protocol’s ability to approximate the target function. Finally, we compare the experimental performance of our protocol against several conventional quantum sensing baselines, for a family of tasks. We chose baseline sensing protocols that are designed to estimate either one component of the displacement $\boldsymbol{\alpha}$ (such as those involving squeezed states or cat-states) or both components (such as those involving heterodyne measurements, two-mode squeezed vacuum (TMSV) interferometry or Gottesman-Kitaev-Preskill (GKP) states). For each baseline protocol considered, we evaluated its performance in one of two ways. For baselines compatible with our experimental setup, we obtained their performance in experiment. For those that are not, we simulated their performance assuming state-of-the-art parameters reported in the literature (e.g., an amplifier with $44\%$ quantum efficiency~\cite{zhong2018results}). The QCDS protocol in experiment outperformed all baselines except one: TMSV interferometry, which can provide an arbitrarily accurate estimate of $\boldsymbol{\alpha}$ with infinite squeezing, and at practically realistic values of squeezing performs well versus our experiment. However, the TMSV interferometry protocol requires higher average photon number in the probe state than QCDS to achieve the same accuracy. We show in simulation that QCDS with a more coherent system could outperform TMSV interferometry at relatively high squeezing values.

\section{Quantum computational displacement sensing (QCDS) protocol}

The QCDS protocol is summarized in Fig.~\ref{fig:main1}. The device (see Fig.~\ref{fig:main1}c) consists of a transmon qubit capacitively coupled to a high-purity aluminum cavity supporting a microwave bosonic mode, with an additional on-chip resonator used for qubit measurement (see Appendix~\ref{app:experiment_setup}). Similar superconducting circuit architectures have been widely used to demonstrate conventional displacement and phase sensing beyond the SQL~\cite{hua2026quantum, deng2024quantum, vlastakis2013deterministically, krisnanda2025direct, wang2019heisenberg, pan2025realization}. The system initially starts with the qubit in the ground state, and the oscillator in the vacuum state: $\ket{0,g}$. The protocol begins by applying a parameterized unitary $U(\vec{\theta}, \vec{\phi}, \vec{\boldsymbol{\beta}})$, which prepares an entangled probe state. The oscillator then senses a displacement $D(\boldsymbol{\alpha}) = \exp{\left( 
\left( \alpha_x + i\alpha_p \right)\hat{a}^\dagger -  \left( \alpha_x - i\alpha_p \right)\hat{a}\right)}$, with $\hat{a}$ defined as the annihilation operator. The displacement $\boldsymbol{\alpha}$, implemented using a calibrated microwave pulse, is known to originate from one of two distributions (see Fig.~\ref{fig:main1}a), and the task is to infer from which distribution the sensed displacement originated. After sensing, we apply a second parameterized unitary $U^\dagger(\vec{\theta}, \vec{\phi}, \vec{\boldsymbol{\beta}})$ which -- together with the first unitary -- computes a function of $\boldsymbol{\alpha}$. A final qubit rotation $R(\theta',\phi')$ aligns this information along the Pauli Z axis of the Bloch sphere of the qubit. The state of the sensor at the end of the protocol is therefore:
\begin{equation}
    \ket{\psi(\boldsymbol{\alpha})} = R(\theta',\phi')U^\dagger(\vec{\theta}, \vec{\phi}, \vec{\boldsymbol{\beta}}) D(\boldsymbol{\alpha}) U(\vec{\theta}, \vec{\phi}, \vec{\boldsymbol{\beta}})\ket{0,g},
\end{equation}
making explicit the dependence of the state on $\boldsymbol{\alpha}$. A projective measurement of the qubit collapses it to either the ground state $\ket{g}$ or the excited state $\ket{e}$, which we interpret as the binary prediction of the class label (see Fig.~\ref{fig:main1}b).

We parameterize the unitary as a sequence of $N$ layers, $U(\vec{\theta}, \vec{\phi}, \vec{\boldsymbol{\beta}}) = \prod_{i=0}^{N-1} U_i(\theta_i, \phi_i, \boldsymbol{\beta_i})$. Each layer is made up of an arbitrary single-qubit rotation $R(\theta_i, \phi_i) = \exp [i(\theta_i/2)( \sigma_x \cos \phi_i + \sigma_y \sin \phi_i)]$ and an entangling echoed conditional displacement $\mathrm{ECD}(\boldsymbol{\beta}_i) = D(\boldsymbol{\beta}_i/2)\ket{g}\bra{e} + D(-\boldsymbol{\beta}_i/2)\ket{e}\bra{g}$, which imparts a displacement of a particular amplitude and phase on the oscillator dependent on the qubit state~\cite{eickbusch2022fast}. Therefore, $U_i(\theta_i, \phi_i, \boldsymbol{\beta_i}) = R(\theta_i, \phi_i)\mathrm{ECD}(\boldsymbol{\beta}_i)$. Such a gate set is universal in its ability to realize any unitary on the Hilbert space of the system, provided sufficient depth~\cite{eickbusch2022fast}. For each layer, the trainable parameters are the qubit rotation amplitude $\theta_i$ and phase $\phi_i$, and the phase of the conditional displacement $\boldsymbol{\beta}_i$. We fix the magnitude of the conditional displacements $|\boldsymbol{\beta}_i| = 0.24$ for all layers $i$ (see Appendix~\ref{app:quantum_computational_sensing_protocol} for additional details). In total, for a protocol of depth $N$, the number of trainable parameters is $3N + 2$ (the additional $2$ comes from the final qubit rotation). 

Our choice of circuit architecture is inspired by the theoretical bosonic quantum signal processing (BQSP) protocol introduced in Ref.~\cite{sinanan2024single}. BQSP shows that a sequence of interleaved single-qubit rotations and fixed echoed conditional displacements can distinguish between two distributions of single-quadrature displacements $\alpha_x$. For a protocol of depth $N$, the resulting qubit-excitation probability is a degree-$N$ Laurent polynomial. Increasing $N$ improves the ability of the circuit to approximate the desired function $\mathcal{F}(\alpha_x)$. Our experimental protocol extends this idea to the two-dimensional displacement plane. By allowing echoed conditional displacements along an arbitrary axis for each layer (determined by the trainable phases of $\boldsymbol{\beta_i}$), our protocol combines information from both the position and momentum components. This yields an expressive family of quantum circuits capable of approximating nonlinear decision boundaries in $(\alpha_x,\alpha_p)$ space, thereby enabling classification tasks that cannot be addressed by the original single-component BQSP construction.

\begin{figure}[h!]
    \centering
    \includegraphics[width=\textwidth]{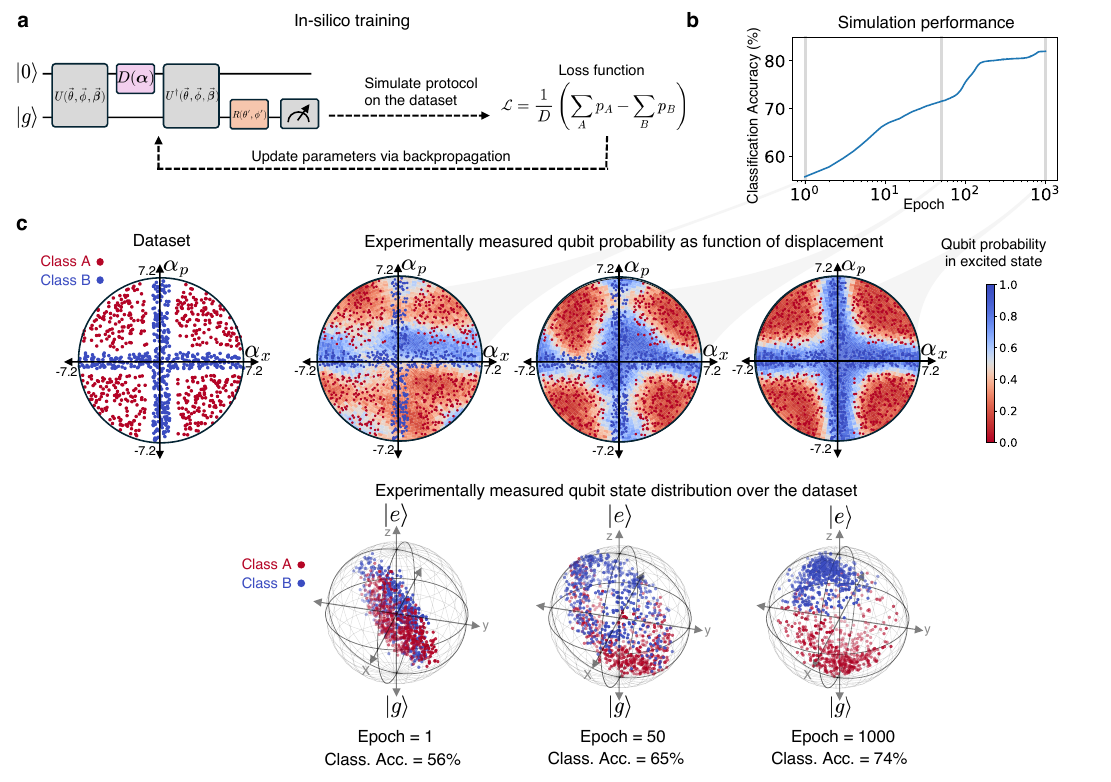}
    \caption{\textbf{Training procedure and simulated performance of the quantum computational displacement sensing (QCDS) protocol.} \textbf{a.} We train the parameters of the protocol in-silico. This allows us to compute the exact probability of the qubit for each displacement in the dataset. The loss function is the difference in the averaged qubit probability in the excited state $p_A$ and $p_B$ for class A and B respectively. Minimizing the function pushes $p_A \to 0$ and $p_B \to 1$. Therefore, during inference, if the qubit is measured in the ground state, we predict class A. Else, we predict class B. The parameters are updated by backpropagating through the simulation. \textbf{b.} Simulated classification accuracy of the protocol on the test dataset of depth $N$ over the training epochs. The classification accuracy is equal to $\frac{1}{2} + \frac{1}{2}(p_B - p_A)$. As the training proceeds, the loss is minimized and the classification accuracy increases. \textbf{c.} Experimentally measured qubit probability as a function of displacement of the bosonic mode at the $1^{\rm st}$, $50^{\rm th}$ and $1000^{\rm th}$ (final) epochs of training (overlaid is the dataset of the two classes). As the protocol trains, the qubit probability better approximates the functional form of the dataset. Below are plots of the qubit states represented on the Bloch sphere over the dataset for each class, measured in experiment. Minimizing the loss separates the two distributions towards opposite poles of the Bloch sphere, increasing the classification accuracy.}
    \label{fig:main2}
\end{figure}

We train the parameters of the QCDS protocol using gradient descent via a simulation on a classical computer of the pulse sequence (Fig.~\ref{fig:main2}a). The experiment-calibrated simulation captures the analog oscillator-qubit dynamics under the large displacements that occur during the ECD gate (see Appendix~\ref{app:system_hamiltonian_and_parameters} for details of the experiment calibration and Appendix~\ref{app:trainable_digital_model} for details of the simulation). Each training epoch consists of a forward and backward pass. In the forward pass, we compute the qubit-excitation probability for every displacement in the dataset. To quantify the class separation, we define the loss $\mathcal{L}$ as the difference in the simulated qubit-excitation probability of the two classes, averaged over the dataset, $\mathcal{L} = \left( \sum_{A} p_A - \sum_{B} p_B \right)/D$. Here $p_A$ and $p_B$ are the qubit excitation probabilities for a displacement of class A and B respectively, and $D$ is the (equal) size of the dataset of each class. Minimizing the loss drives $p_A$ towards zero and $p_B$ towards one. This provides the basis of our prediction -- if we measure the qubit in the excited state at the end of the protocol, we predict the displacement to have originated from class B. Consequently, if we measure the qubit in the ground state at the end of the protocol, we predict class A. In the backward pass, gradients of the loss with respect to all variational parameters are evaluated through automatic differentiation and used to update the protocol. For each task, we train multiple independent instances of the protocol with random parameter initialization, and select the instance that achieves the higher training classification accuracy. Additional details of the training procedure are provided in Appendix~\ref{app:training_procedure}.

As an example, we consider the task of distinguishing the two displacement distributions shown in Fig.~\ref{fig:main2}c. In Fig.~\ref{fig:main2}b, we plot the classification accuracy during the training phase of the simulation, evaluated on the test dataset, for a protocol of depth $N = 10$. The performance steadily improves and converges within 1000 epochs. We then evaluate the performance of the trained protocol in experiment. Fig.~\ref{fig:main2}c illustrates the experimentally-measured qubit-excitation probability as a function of oscillator displacements at the $1^{\rm st}$, $50^{\rm th}$ and $1000^{\rm th}$ (final) epoch, along with the experiment qubit states right before measurement across the dataset. As optimization proceeds, the state of the qubit in the quantum computational sensor under displacements from class A moves towards the south pole of the Bloch sphere $(\ket{g})$. On the other hand, for displacements from class B the qubit state moves towards the north pole $(\ket{e})$. Consequently, the measured qubit-excitation probability in the 2D space of sensed displacement evolves to mimic the labeled dataset. 

\section{Results}

\begin{figure}[h!]
    \centering
    \includegraphics[width=0.82\textwidth]{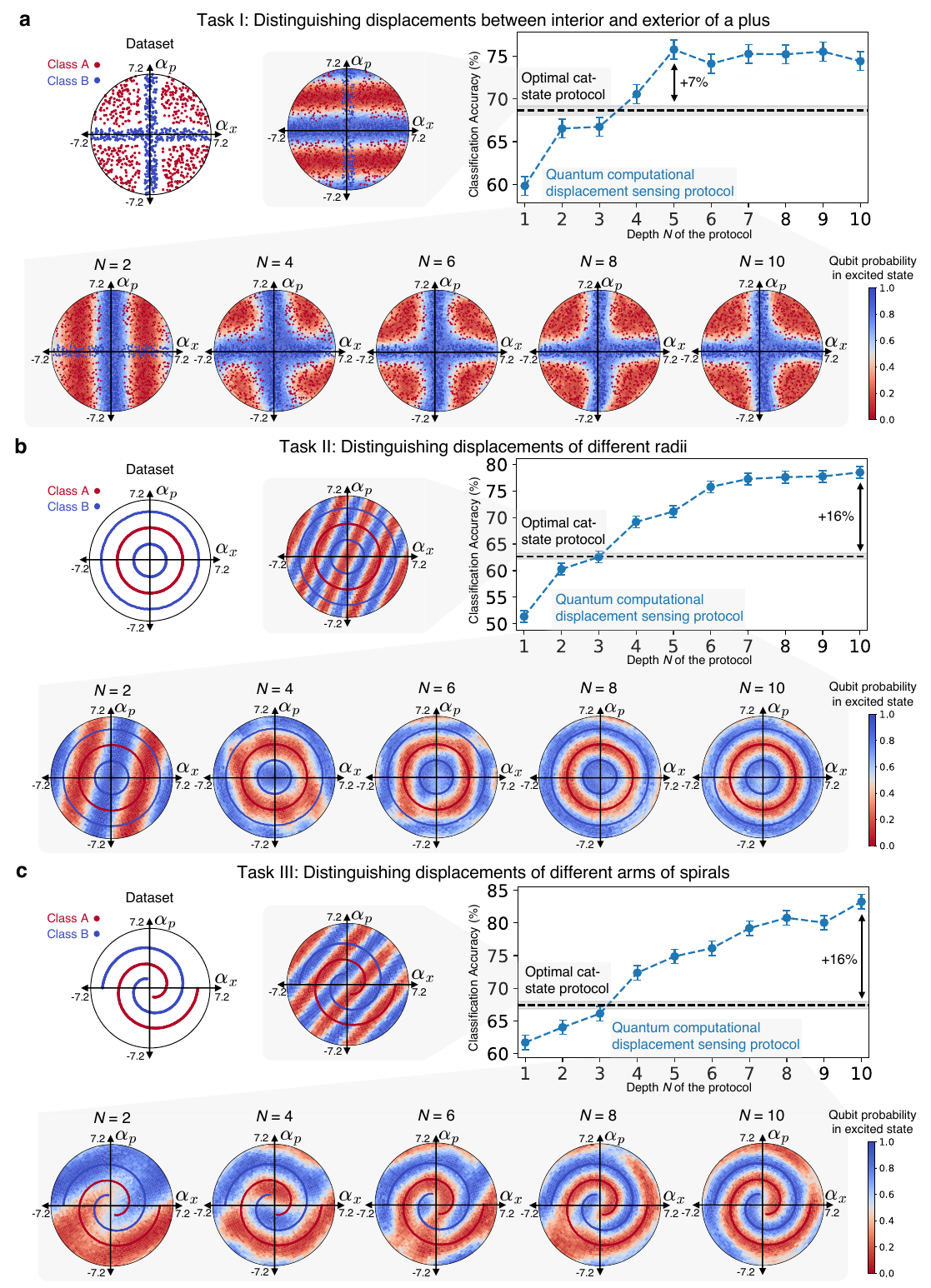}
    \caption{\textbf{Experimental performance of the quantum computational displacement sensing (QCDS) protocol on three different tasks.} For each task, we train the parameters of the protocol for depths $N$ ranging from $1$ to $10$. The classification accuracy tends to increase with $N$, as the protocol is better able to distinguish the two classes, as visualized in the plots of the measured qubit probability as a function of the displacement for select values of $N$. As a benchmark, we compare against the performance of the optimal cat-state sensing protocol, which is an example of a conventional displacement sensing protocol. While this protocol is highly sensitive for estimating small displacements, it fails to classify effectively. In contrast, the quantum computational sensor achieves a higher classification accuracy given sufficient depth.}
    \label{fig:main3}
\end{figure}

In Fig.~\ref{fig:main3}, we show the experimental performance of the QCDS protocol on three representative classification tasks. For each task, we plot the experimental classification accuracy as a function of protocol depth $N$. In the simulation of the protocol -- which doesn't account for decoherence -- the accuracy improves monotonically with depth, consistent with the growing expressive power of the circuit. The experimental performance follows the same qualitative trend but plateaus at lower accuracies. This is primarily due to two reasons. First, each additional layer increases the total protocol duration by $880~\rm ns$, leading to increased decoherence of both the qubit and oscillator and a corresponding reduction in sensitivity. This results in an optimal depth of highest accuracy: for the task in Fig.~\ref{fig:main3}a, performance peaks at $N=5$, whereas for the tasks in Fig.~\ref{fig:main3}b and c, the accuracy continues to improve up to $N=10$. Second, the imperfect qubit-readout fidelity (an error which doesn't scale with $N$) limits the achievable classification accuracy. Our single-shot measurement discriminates the excited state with $95\%$ fidelity and the ground state with $96\%$. This sets an upper bound on the maximum classification accuracy the protocol can achieve. In Fig.~\ref{fig:main3}, we also plot the measured qubit-excitation probability as function of the oscillator's displacement, overlaid on the corresponding dataset, for each task and select circuit depths $N$. As $N$ increases, the protocol is better at matching the distribution of the underlying dataset. To highlight the versatility of our protocol, we demonstrate its performance on other tasks in Appendix~\ref{app:quantum_computational_sensing_protocol:examples}.

To understand how these functions of displacements emerge, it is instructive to begin with the simplest case of a depth $N=1$ circuit, which reduces to the standard cat-state displacement sensing protocol~\cite{sinanan2024single}. The unitary $U$ prepares an entangled cat state, $U\ket{0,g} = c_g\ket{\boldsymbol{\beta}/2,g} + c_e\ket{-\boldsymbol{\beta}/2,e}$, so that the oscillator is displaced along $\pm \boldsymbol{\beta}/2$ conditional on the qubit state. Such states are only sensitive to the component of the unknown displacement orthogonal to $\boldsymbol{\beta}$, which we denote as $\alpha_\perp$. After $U^\dagger$, this displacement imparts a qubit phase proportional to $|\boldsymbol{\beta}| \alpha_{\perp}$, which is estimated through the final qubit measurement. The sensitivity increases with the size of the cat $|\boldsymbol{\beta}|$. However, due to the modular nature of the qubit phase, this increased sensitivity comes at the expense of a reduction in the dynamic range of displacements which can be uniquely resolved. The cat-state sensing protocol is therefore a conventional sensing strategy optimized for estimating small displacements known a priori to lie along a particular quadrature.

In Fig.~\ref{fig:main3}, we include the experimentally optimized cat-state performance for each task. The optimal $\boldsymbol{\beta}$ depends on the task and reflects a compromise between sensitivity and dynamic range, as well as the fact that the protocol senses only a single displacement axis. In Appendix~\ref{app:benchmark:cat_state_sensor}, we describe how the optimal cat-state protocol is determined and implemented experimentally. Due to the ability of the output of the QCDS protocol to conform to the requirements of the task, it surpasses the performance of the optimal cat-state protocol by $7$ percentage points for the task in Fig.~\ref{fig:main3}a and by $16$ percentage points for the tasks in Figs.~\ref{fig:main3}b,~c.

\section{Quantum computational-sensing advantage (QCSA)}

\begin{figure}[h!]
    \centering
    \includegraphics[width=0.9\textwidth]{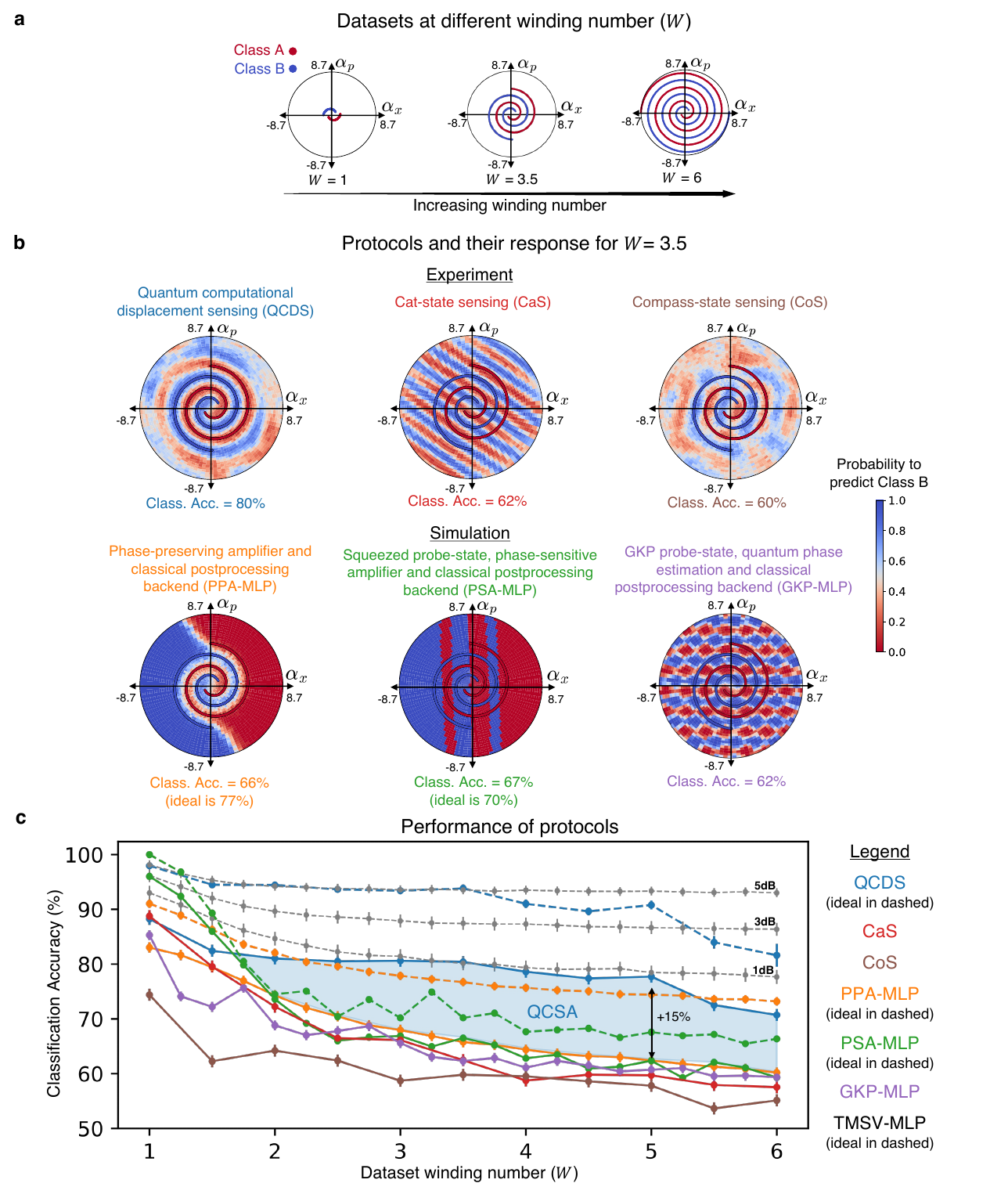}
    \caption{\textbf{Performance of the quantum computational displacement (QCDS) sensing protocol versus conventional quantum displacement sensing protocols.} \textbf{a.} We consider the task of distinguishing displacements from one of two arms of a spiral distribution with varying winding number $W$, which serves as a proxy for the complexity of the task. \textbf{b.} Response of the various protocols we compare, for the task with $W=3.5$, illustrated by the probability of protocol to predict class B as a function of the sensed displacement. For the protocols compatible with our device, we plot the experimental results. For others, we simulate their implementation. The plots for the protocols involving either the phase-preserving or phase-sensitive amplifiers are for the idealized version. \textbf{c.} Classification accuracy versus the winding number for the various protocols. The shaded region shows the quantum computational-sensing advantage (QCSA) over the next-best protocol either simulated with state-of-the-art parameters from the experimental literature (see main text for details) or performed in experiment by us. We achieve up to $15$-percentage-points improvement in accuracy.}
    \label{fig:main4}
\end{figure}

A QCSA is achieved when the quantum computational displacement sensing protocol classifies more accurately than a conventional displacement sensing protocol. We benchmark our protocol against several sensing protocols that are optimized for estimating one or both components of a displacement of an electromagnetic microwave bosonic mode, and highlight regimes where QCSA is obtained over those protocols (see Appendix~\ref{app:benchmark} for more details). We consider a family of tasks defined by distinguishing displacements drawn from one of two arms of a spiral distribution (similar to the task in Fig.~\ref{fig:main3}c). The datasets are parameterized by a continuous winding number $W$ that sets the number of spiral turns, as shown in Fig.~\ref{fig:main4}a, and is a proxy for task complexity. Increasing the winding number increases the nonlinearity of the decision boundary and consequently the intrinsic difficulty of the task.\footnote{While the classification accuracy of all the protocols generally decreases with winding number $W$, it is non-monotonic for some. The dependence of the classification accuracy on $W$ depends on the expressivity of the protocol.} The displacements span magnitudes up to $8.7$, which is large compared with typical displacement (such as phase-shift-keying) discrimination tasks~\cite{tsujino2011quantum, sasaki1996optimum, han2020helstrom}. Nevertheless, quantum sampling noise remains the dominant factor limiting the achievable accuracy for both our protocol and the conventional sensing protocols we consider, due to the added complexity in computing $\mathcal{F}(\alpha_x, \alpha_p)$. 

We calculate the performance of conventional displacement sensing strategies in one of two ways. For protocols that can be implemented using the same gate primitives as the QCDS protocol, we evaluate their performance directly in experiment. The cat-state and compass-state sensing protocols fall within this category. For protocols that are incompatible with our protocol, we estimate their performance through simulation using state-of-the-art experimental numbers. The GKP-state sensing protocol and protocols involving an amplifier (either a phase-preserving or phase-sensitive) fall within this category. The resulting accuracies are summarized in Fig.~\ref{fig:main4}c. Across a wide range of winding numbers explored, our quantum computational sensor achieves a QCSA of up to $15$-percentage-points over most sensing protocols. Furthermore, for some winding numbers, our protocol outperforms an ideal phase-preserving amplifier by up to $4$-percentage-points. In Appendix~\ref{app:benchmark:additional_examples}, we consider several additional sensing tasks. For all of these tasks, the QCDS protocol approaches perfect classification accuracy with increasing circuit depth in simulations of the protocol running on ideal hardware. However, whether an experimental setup can achieve a QCSA for a given task, and how large the advantage will be, depends on the task and on the nature and magnitude of the experimental errors. \\

\underline{\emph{Cat-state sensing (CaS)}} -- Since this protocol (introduced in the previous section) is sensitive only to the displacement component orthogonal to its cat axis, its performance is limited on high-winding-number tasks, which require simultaneous sensitivity to both quadratures $\alpha_x$ and $\alpha_p$ (see Appendix~\ref{app:benchmark:cat_state_sensor} for more details). \\

\underline{\emph{Compass-state sensing (CoS)}} -- The compass-state sensing protocol is a natural extension of the cat-state sensing protocol that uses four-component superpositions of coherent states to gain sensitivity to both quadratures~\cite{toscano_sub-planck_2006}. Despite this ability, the space of realizable functions remains limited, preventing it from matching the decision boundary required for these spiral-classification tasks. As shown in Fig.~\ref{fig:main4}c, the cat-state and compass-state protocols (red and brown curves, respectively) underperform relative to the quantum computational sensing protocol (blue curve) across all winding numbers investigated. Their corresponding experimentally measured qubit-excitation probability landscapes are shown in Fig.~\ref{fig:main4}b, where the excitation probability directly encodes the predicted likelihood of class B. We discuss more details in Appendix~\ref{app:benchmark:compass_state_sensor}. \\

\underline{\emph{Phase-preserving amplifier and classical postprocessing backend (PPA-MLP)}} -- Another widely-used strategy for sensing displacements is heterodyne detection, which provides single-shot estimates of both components of the displacement $\boldsymbol{\alpha} = (\alpha_x, \alpha_p)$. In microwave experiments, this can be realized by coupling the bosonic mode to a transmission line and routing the signal through a phase-preserving, quantum-limited amplifier such as a traveling-wave parametric amplifier (TWPA)~\cite{macklin2015near}. According to Caves' theorem, this approach adds at least a single photon worth of added noise (including the half-photon noise of the vacuum state) on top of the signal $\boldsymbol{\alpha}$. This architecture was employed in the first phase of the HAYSTAC dark matter axion-search experiment to detect extremely small displacements of a microwave mode across a broad frequency range~\cite{zhong2018results, al2017design}. In that setting, the figure of merit is the scan rate, which is determined largely by the total system noise photon number of the measurement chain. HAYSTAC achieved a value of $2.3$ photons on average, which is the value we use in the simulation~\cite{zhong2018results}. In Fig.~\ref{fig:main4}c, we show (in orange) the simulated classification performance of heterodyne sensing using both an ideal amplifier (a total system noise photon number of $1$) and an amplifier with the noise photon number of the HAYSTAC measurement chain. Heterodyne detection yields a noisy estimate $(\widetilde{\alpha}_x, \widetilde{\alpha}_p)$, which must then be processed classically. To map these estimates to class labels, we train a multi-layer perceptron (MLP). The performance of this heterodyne-MLP pipeline is limited not by the expressivity of the MLP, but by the quantum noise added by the amplifier. Indeed, with access to the noise-free displacement $(\alpha_x,\alpha_p)$, a trained MLP would achieve perfect classification over the range of spiral tasks considered. This limitation is evident in Fig.~\ref{fig:main4}b, which shows the MLP-predicted probability of class B as a function of the displacement for an ideal amplifier: near class boundaries the output approaches $0.5$, indicating the region where most misclassifications occur. Across the full range of winding numbers, our quantum computational sensor outperforms the heterodyne baseline evaluated at the HAYSTAC parameters. Furthermore, it also surpasses the simulated performance of the heterodyne protocol even when assuming an ideal phase-preserving amplifier for winding numbers considered between $2.5$ and $5$. We provide more details of the simulation of this protocol in Appendix~\ref{app:benchmark:phase_preserving_amplifier}. \\

\underline{\emph{Squeezed probe-state, phase-sensitive amplifier and classical postprocessing backend (PSA-MLP)}} -- The second phase of the HAYSTAC experiment implemented squeezed-state sensing, where a squeezed oscillator state is measured along its low-noise quadrature using a phase-sensitive amplifier to enhance sensitivity after displacement~\cite{backes2021quantum, bai2025dark}. Similar to the cat-state sensor, this protocol is sensitive to only a single quadrature of the displacement. Unlike the cat-state protocol, however, it does not suffer from the trade-off between sensitivity and dynamic range. The measurement outcomes are classically postprocessed using an MLP to perform classification. In Fig.~\ref{fig:main4}c, we show the performance of this protocol (in green) under two scenarios: an idealized case with an infinitely squeezed state and a noiseless phase-sensitive amplifier, and a realistic case corresponding to the parameters of the HAYSTAC experiment (which achieves a squeezing of up to $4~\rm dB$ below the vacuum noise floor, after taking into account measurement inefficiency)~\cite{bai2025dark}. For winding numbers below $1$, where the two classes are linearly separable, the idealized case achieves perfect classification. For higher winding numbers, the spirals become nonlinearly separable, requiring information from both components for accurate classification. In this regime, even the idealized protocol fails, as illustrated in Fig.~\ref{fig:main4}b, demonstrating the limitations of single-quadrature sensing (see Appendix~\ref{app:benchmark:squeezed_phase_sensitive_amplifier} for more details). \\

\underline{\emph{GKP probe-state, quantum phase estimation and classical postprocessing backend (GKP-MLP)}} -- We consider sensing with a Gottesman-Kitaev-Preskill (GKP) probe-state. By applying the quantum phase estimation (QPE) algorithm to an ideal GKP state after sensing, both quadratures of the displacement can be estimated simultaneously without error~\cite{duivenvoorden2017single}. This is possible because QPE measures modular observables of $\alpha_x$ and $\alpha_p$ that commute, allowing their values to be extracted without noise. Therefore, in general, a GKP probe-state with sufficiently large photon number followed by sufficiently many noiseless rounds of QPE will produce accurate enough estimates of the displacement quadratures $\widetilde{\alpha}_x$ and $\widetilde{\alpha}_p$ to allow accurate classification with postprocessing. Practically, however, the sensitivity of this approach is fundamentally limited by the photon number of experimentally prepared GKP states. Ref.~\cite{sivak2023real} reports an experimental demonstration of quantum error correction using a GKP state (with a mean photon number of approximately $4$) in superconducting circuits. As shown theoretically in Ref.~\cite{labarca_quantum_2025}, this protocol can be naturally extended to perform quantum phase estimation (QPE) on a displaced GKP state. We adopt the simulation framework of Ref.~\cite{labarca_quantum_2025} to estimate the sensitivity of such a protocol for displacement sensing, using GKP states with parameters matched to those achieved in experiment. In these simulations, we consider $10$ rounds of QPE (each round corresponding to one measurement per quadrature), yielding a bitstring of length $20$. The simulation neglects additional experimental imperfections and therefore provides an optimistic upper bound on experimentally achievable performance. We then train an MLP to map the measurement outcomes to class labels. As shown in Fig.~\ref{fig:main4}b, the resulting probability of predicting class B exhibits strongly oscillatory behavior, with periodicity $\sqrt{\pi}$ (for the convention adopted) in both quadratures, reflecting the modular structure of the measured observables. This periodicity severely limits performance when the displacement range exceeds this range, as in the spiral distributions considered here, illustrated in Fig~\ref{fig:main4}c in purple. Additional details are provided in Appendix~\ref{app:benchmark:gkp_state}, where we also analyze the expected performance of a full GKP displacement-sensing protocol implemented in trapped-ion systems~\cite{valahu2025quantum}, which achieves the best sensitivity for a GKP state with a mean photon number of approximately $3$. \\

\underline{\emph{Two-mode squeezed vacuum state, phase-sensitive amplifier and classical postprocessing backend (TMSV-MLP)}} -- Finally, we consider the scenario where the probe-state is a two-mode squeezed vacuum state, where the sensing oscillator is entangled with an ancilla oscillator. TMSV state followed by ideal homodyne permit simultaneous estimation of both quadratures of a displacement with an error that decreases with increasing squeezing~\cite{braunstein_dense_2000, steinlechner_quantum-dense_2013}. In principle, a TMSV state with sufficiently large squeezing would yield estimates $\widetilde{\alpha}_x$ and $\widetilde{\alpha}_p$ with small enough errors that a classical postprocessor would be able to achieve accurate classifications. In Fig.~\ref{fig:main4}c (in black), we plot the classification accuracy under ideal conditions for three different levels of squeezing. $1$~dB of squeezing achieves comparable performance with our QCDS experiment. However, generating such highly squeezed joint states of bosonic modes of electromagnetic microwave oscillators~\cite{qiu2023broadband, liu2022noise} and implementing the displacement sensing protocol is experimentally demanding. Experimental inefficiencies of real amplifiers limit the reduction in noise obtainable, even in the limit of infinite squeezing, setting an upper bound on the maximum classification accuracy obtainable. To the best of our knowledge, there has been no demonstration of this protocol in displacement sensing of an electromagnetic microwave oscillator. In Appendix~\ref{app:benchmark:two_mode_squeezed_phase_sensitive_amplifier}, we show with simulations that the probe-state photon-number required to achieve the same classification accuracy as achievable with the QCDS protocol can be substantially larger.\\

In summary, the comparison with these sensing protocols highlights the central principle behind the observed quantum computational-sensing advantage: the ability of the QCDS protocol to quantum-compute the target function $\mathcal{F}(\alpha_x, \alpha_p)$ before measurement. This results in the QCDS protocol outperforming most protocols considered for a wide range of $W$.

\stoptoc

\section{Discussion}
\label{sec:discussion}

\subsection{Summary of the work}

We have experimentally realized a quantum computational displacement sensor based on superconducting circuits. We considered classification tasks in which the goal was to identify which of two classes a displacement -- sensed only once -- originated. Our approach was to design a sensor to directly output the label of the class, rather than an estimate of the displacement. Our protocol, which consists of a qubit entangled with an oscillator and employs parameterized quantum circuits before and after sensing, was trained such that a single qubit measurement at the end of the sequence directly predicted the class label, eliminating the need for any classical postprocessing. This approach enabled a quantum computational–sensing advantage (QCSA) – higher classification accuracy – over most protocols that rely on a conventional sensor designed to estimate displacements. We provided examples of tasks where we experimentally realized an advantage of up to 15 percentage points. The only protocol considered that can outperform the QCDS experiment--two-mode squeezed-vacuum interferometry--has, to the best of our knowledge, not yet been demonstrated for sensing displacements of microwave electromagnetic modes.

Via numerical simulations, we show that improvements in the experiment parameters (presented in Appendices~\ref{app:quantum_computational_sensing_protocol:examples} and ~\ref{app:benchmark:two_mode_squeezed_phase_sensitive_amplifier}) or different choices of gates (presented in Appendix~\ref{app:future:snap}) can further increase the achievable QCSA. An increase in the coherence times of the qubit and the oscillator modes can support larger echoed-conditional-displacement gates and/or deeper circuits. Such improvements can enable representing more expressive functions $\mathcal{F}(\boldsymbol{\alpha})$, which vary on a smaller displacement scale. However, our experimental results already demonstrate the promise of quantum computational sensors, even when realized with current quantum hardware.

The classification tasks we considered in this work are computationally easy: for each, the input is only two-dimensional ($\boldsymbol{\alpha} = (\alpha_x, \alpha_p)$), the output is just a binary value, and the learned classification function ($\mathcal{F}(\boldsymbol{\alpha}) \in \{0,1\}$) could be realized using a very simple neural network provided that the input $\boldsymbol{\alpha}$ were given as noiseless classical data instead of as a displacement to a quantum system. Despite this task simplicity, our experimental results demonstrate that a sizable quantum computational-sensing advantage can already be achieved. We now describe three possible extensions of our work to more sophisticated tasks. 

\subsection{Future direction: Multi-valued-output quantum computational displacement sensors}

A natural extension of our work is to consider tasks where the output of $\mathcal{F}$ can take on more than two values. In many applications, the goal is to classify signals drawn from one of several distributions -- a multiclass classification problem. In this setting, the present quantum computational sensor cannot achieve high accuracy because it is fundamentally limited to outputting a single bit of information. One approach to overcome this bottleneck is to entangle multiple qubits with the bosonic mode and measure all of them at the end of the protocol, thereby producing a sufficient number of bits to support single-shot multiclass classification. Another approach is to perform a small number of sensing rounds; for example, an adaptive strategy could use our quantum computational sensor to narrow down the correct class.

Quantum computational displacement sensing need not be limited to classifying the displacement $\boldsymbol{\alpha}$: one could also design protocols to compute functions $\mathcal{F}(\boldsymbol{\alpha})$ that take on continuous values. Quantum-computational-sensing protocols based on quantum phase estimation (QPE)~\cite{nielsen_quantum_2010} could potentially achieve the Heisenberg limit (HL), where the error in the estimate of $\mathcal{F}(\boldsymbol{\alpha})$ scales as $1/S$ in the number of samples $S$ (beyond the SQL of $1/\sqrt{S}$). Such a protocol could have a QCSA (producing more accurate estimations with the same $S$) over conventional protocols that perform quantum sensing to first estimate $\boldsymbol{\alpha}$ (e.g., with HL sensitivity by using QPE) and then compute $\mathcal{F}(\boldsymbol{\alpha})$ in classical post-processing, since the error in $\boldsymbol{\alpha}$ can, for certain functions $\mathcal{F}$ and values of $\boldsymbol{\alpha}$, propagate poorly to error in $\mathcal{F}(\boldsymbol{\alpha})$~\cite{khan2025quantum_novel, tellinghuisen2001statistical}.

Finally, another approach is to replace the ancilla qubit(s) with a single ancilla bosonic mode, with programmable non-Gaussian interactions between the ancilla and sensing bosonic modes. A homodyne (or heterodyne) measurement outputs a continuous value. Such a quantum computational sensor realizes a nonlinear bosonic amplifier, which was theoretically analyzed in Ref.~\cite{khan2025quantum_novel}.

\subsection{Future direction: Distributed quantum computational displacement sensors}

In distributed quantum sensing, the goal is to infer a global property of signals sensed across several quantum modes~\cite{zhang2021distributed}. Ref.~\cite{zhuang2019physical} theoretically introduced supervised learning assisted by an entangled sensor network (SLAEN) -- a quantum computational sensor -- that directly computes the sign of a linear combination of single-mode displacements $\mathcal{F} = {\rm sign}(\sum_i \mathcal{W}_i \alpha_i + b)$. SLAEN estimates the sum $\sum_i \mathcal{W}_i \alpha_i + b$ at the HL in the number of modes by entangling the bosonic modes via trained Gaussian quantum operations. This reduction in the sum translates to an exponential reduction in the error in predicting $\mathcal{F}$ over unentangled sensing protocols that estimate the displacement of each mode individually, which was demonstrated experimentally by Ref.~\cite{xia2021quantum}. However, SLAEN relies solely on Gaussian quantum operations, which restricts its expressive power to linear functions of the signal parameters. By introducing a single ancilla qubit, Ref.~\cite{liao2024quantum} theoretically demonstrates how non-Gaussian operations can be implemented that -- together with the displacement sensing modes -- can realize nonlinear and more sophisticated decision functions, substantially broadening the expressivity. Our work can be understood as an experimental demonstration of the limiting case of a single sensing bosonic mode. A natural extension would be to experimentally demonstrate the case with multiple modes.

Finally, distributed quantum computational displacement sensors can also be composed of multiple ancilla qubits entangled with multiple sensing bosonic modes, combining the features of Ref.~\cite{liao2024quantum} and multi-valued-output sensors. Such sensors, comprising a few bosonic modes and ancilla qubits, could potentially be realized with current hardware~\cite{zhou2023realizing, wang2020efficient, chou2018deterministic}. We discuss this further in Appendix~\ref{app:future:multi_mode_multi_qubit}.

\subsection{Future direction: Quantum computational microwave receivers}

While our experiment is a proof-of-principle demonstration of QCSA for displacement sensing, it can be naturally extended to the computational sensing of traveling microwave signals. In this setting, the transmission line carrying the incoming signal (for example, from an antenna) is strongly coupled to the drive port of the oscillator~\cite{krantz2019quantum}, enabling efficient transfer of the signal into the quantum system.\footnote{A strongly coupled port is required to maximize the efficiency of signal transfer into the quantum system, and can be equivalently understood as impedance matching between the transmission line and the oscillator.} The intracavity field of the oscillator is then continuously driven by the incoming signal, such that computing functions of the oscillator displacement directly corresponds to computing functions of the incoming signal. This architecture could be applied, for example, to the discrimination of classical communication symbols encoded in the complex displacements~\cite{tsujino2011quantum, sasaki1996optimum, han2020helstrom}.

Ref.~\cite{senanian2024microwave} took a step in this direction by experimentally realizing a quantum reservoir computer to classify time-dependent radio-frequency signals associated with different digital modulation schemes~\cite{wong2020rfml}. However, in both that work and our experiment reported in the present paper, the drive port was only weakly coupled to the transmission line, such that a large fraction of the incident signal was reflected (approximately $90\%$), limiting the effective signal capture. This conventional design minimizes decoherence induced by the transmission line but is suboptimal for receiver applications. Realizing a quantum computational microwave receiver will therefore require carefully balancing signal capture efficiency against decoherence of the oscillator.

\section*{Data and code availability}

All experiment and simulation data and code used in this work are available at \url{https://doi.org/10.5281/zenodo.19520012}

\section*{Acknowledgements}

The authors would like to thank Fernanda H\"uller, Ed Kluender, Vladimir Kremenetski, J\'er\'emie Laydevant, Pengcheng Liao, Tatsuhiro Onodera, Shyam Shankar, Kent Shirer, Jasmine Sinanan-Singh, Mandar Sohoni, Hakan T\"ureci, Wayne Wang and Taekwan Yoon for helpful discussions and comments. The authors would also like to thank Bradley Cole, Clayton Larson, Britton Plourde, Eric Yelton, and Luojia Zhang for the fabrication of the transmon and on-chip resonator, Chris Wang for the design of the transmon, the on-chip resonator and the 3D superconducting cavity, and Nord Quantique for the fabrication of the 3D superconducting cavity. We gratefully acknowledge MIT Lincoln Laboratory for supplying the Josephson traveling-wave parametric amplifier (TWPA) used in our experiments. This paper is based upon work supported by the Air Force Office of Scientific Research under award number FA9550-22-1-0203 and Army Research Office under award numbers W911NF-25-1-0261. We gratefully acknowledge a DURIP award with AFOSR award number FA9550-22-1-0080 for equipment used in this work. M.O. acknowledges support from the \href{https://ror.org/00w3qhf76}{Fonds de recherche du Québec (FRQ)} Postdoctoral Research Scholarship, \href{https://doi.org/10.69777/366727} {Grant No.~366727}. The authors wish to thank NTT Research for their financial and technical support. P.L.M. acknowledges membership in the CIFAR Quantum Information Science Program as an Azrieli Global Scholar.

\section*{Author contributions}
S.P. carried out the experiments. S.P. and S.A.K. designed the protocol, with contributions from X.S. and R.Y.. S.P. and S.A.K. coded the simulation, with contributions from M.O.. S.A.K. and S.P. analyzed the conventional sensing protocols benchmarks. V.F. oversaw the design and fabrication of the superconducting devices by S.R. and others, which were characterized by S.R., A.S. and S.P.. S.P., S.A.K. and P.L.M. conceived the project, with initial contributions from L.G.W.. S.P., S.A.K. and P.L.M. wrote the manuscript with input from all authors. P.L.M. supervised the project.

\bibliographystyle{mcmahonlab}
\bibliography{references}

\resumetoc

\appendix

\startcontents[Appendices]
\addtocontents{toc}{\protect\setcounter{tocdepth}{0}} 
\section*{Appendices} 
\addtocontents{toc}{\protect\setcounter{tocdepth}{2}} 
\printcontents[Appendices]{l}{1}{}

\makeatletter
\let\toc@pre\relax
\let\toc@post\relax
\makeatother

\newpage

\section{Experiment setup}
\label{app:experiment_setup}

\begin{figure}[h!]
    \centering
    \includegraphics[width=\textwidth]{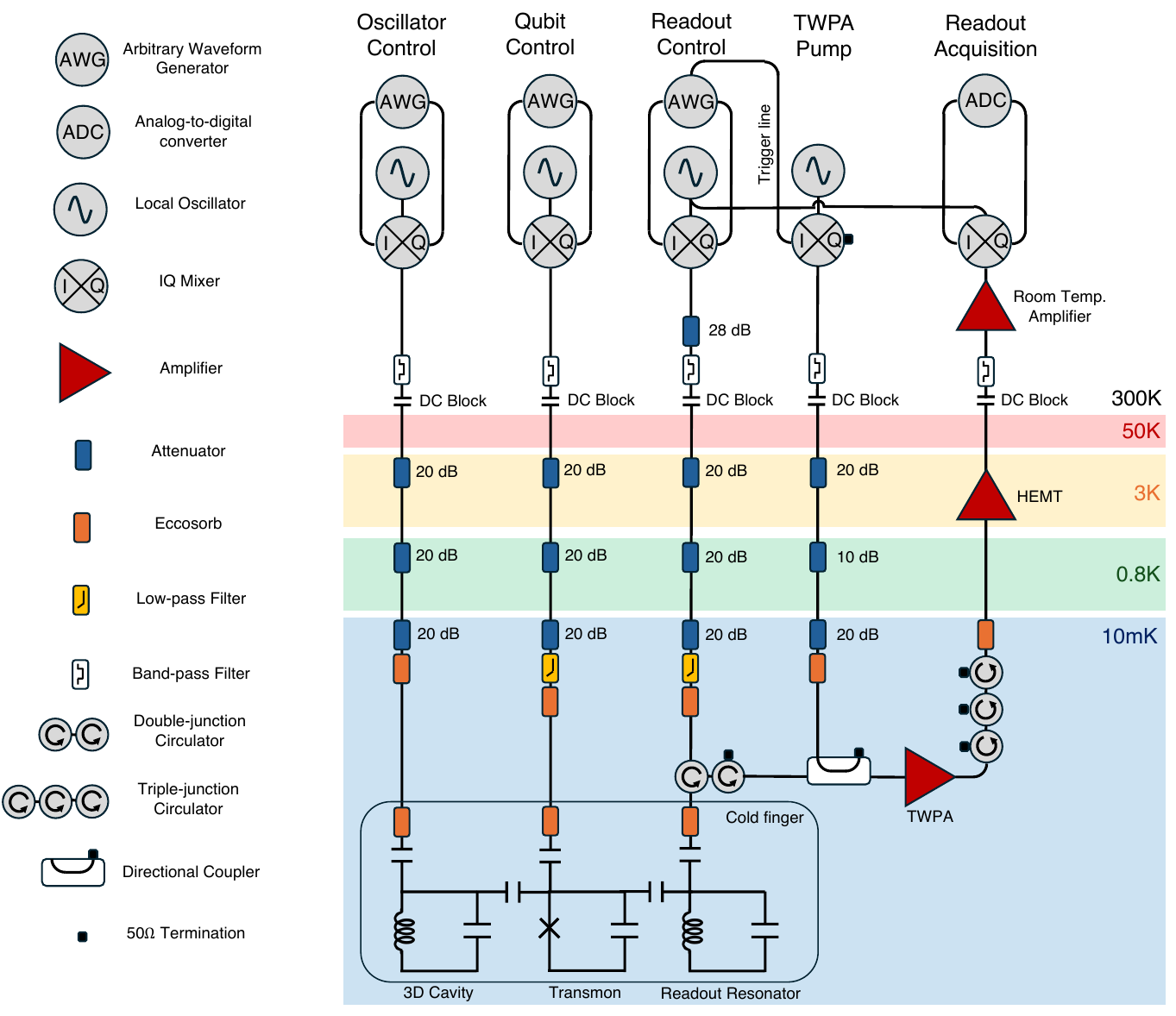}
    \caption{\textbf{Wiring Diagram of the experiment.} Experiment setup for microwave hardware and cabling for the device used in this work.}
    \label{fig:wiring}
\end{figure}

\begin{figure}[h!]
    \centering
    \includegraphics[width=\textwidth]{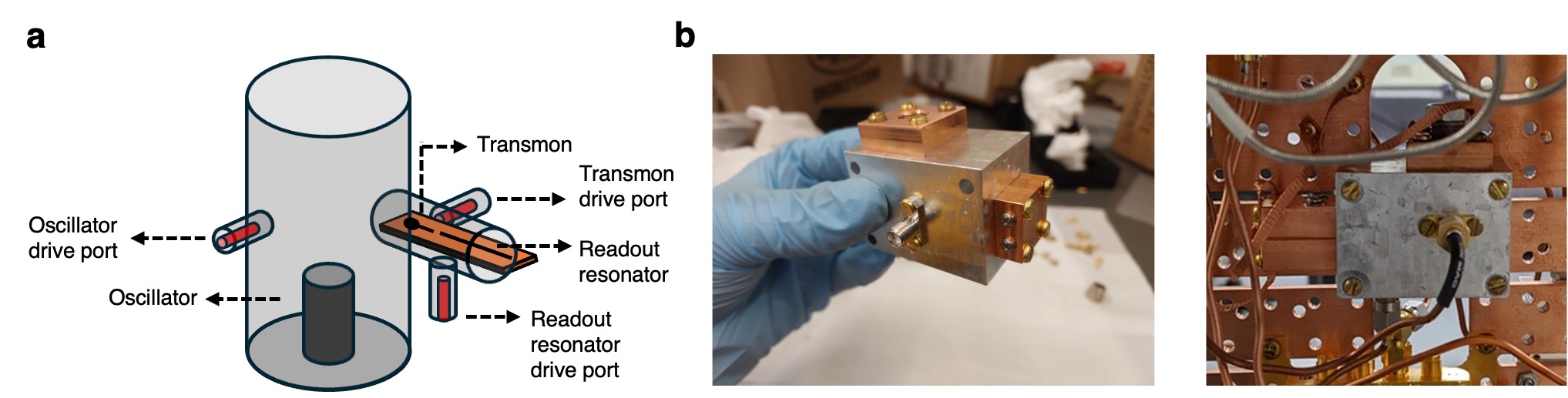}
    \caption{\textbf{Schematic illustration and photographs.} \textbf{a.} Schematic illustration of the superconducting device, showing the 3D oscillator, the transmon qubit, and the transmon's readout resonator (along with their corresponding drive ports). \textbf{b.} Photographs of the device and the cold finger it is mounted on.}
    \label{fig:image}
\end{figure}

In this work, we use a transmon qubit coupled to the fundamental mode of a 3D stub-post cavity made of high-purity 4N Aluminum. The transmon qubit (the same device used in Ref.~\cite{senanian2024microwave}) is fabricated on resistive silicon chip, and is made of Niobium, with a Aluminum - Aluminum Oxide Josephson junction. The chip also contains a readout resonator also made out of Niobium. This chip is inserted in the Aluminum cavity and is held by a copper clamp thermalized to the cold finger made of Oxygen-free high conductivity (OFHC) Copper (see Fig.~\ref{fig:image}). The cold finger is enclosed in a Berkeley black coated Copper shield and an Aluminum shield. The outermost can of the dilution fridge is lined with a cryoperm shield on the inside. The device is mounted to the cold finger at the base plate of a dilution fridge and is cooled to $10~\rm mK$. Details of the cryogenic setup are shown in Fig.~\ref{fig:wiring}.

Microwave pulses are generated using Zurich Instruments (ZI) HDAWG at baseband. Acquisition of microwave pulses from the devices are received by ZI UHFQA. All up and down conversions are performed using Rohde \& Schwarz SGS100A, with in-built IQ mixers. The readout pulse generation and acquisition use the same SGS100A local oscillator (using a splitter). This is done with external Marki IQ mixers (MMIQ-0416LSM-2). The pump for the traveling-wave Parametric Amplifier (TWPA) is gated with a marker line from the ZI HDAWG which is sent to the corresponding SGS100A which produces the pump signal. The readout signal, after amplification by the TWPA, is further amplified by a high-electron mobility transistor (HEMT) at the $4\rm K$ stage (LNF-LNC$0.3\_14$B from Low Noise Factory), and finally at room temperature outside the fridge with a high gain amplifier (ZVA-1W-10 3+ from Mini Circuits). The microwave lines into and out of the fridge are filtered with a series of DC blocks and Mini Circuits bandpass filters centered around the frequency of the tones. For qubit readout, a constant pulse of $0.5~\rm \mu\text{s}$ duration is sent, and the signal is measured for $1~\rm \mu\text{s}$. The TWPA is gated for $2~\rm \mu\text{s}$. We wait for $4~\rm ms$ between runs of the protocol to provide plenty of time for the qubit to thermalize to the ground state, and the oscillator to the vacuum state.

\clearpage
\section{System Hamiltonian and parameters}
\label{app:system_hamiltonian_and_parameters}

In this section, we describe how to characterize the system Hamiltonian and parameters when the interaction strength of the transmon qubit with the oscillator is much less than the linewidth of the transmon mode. Our characterization mostly follows the protocol introduced in Ref.~\cite{eickbusch2022fast}, which operates in a similar regime. In this regime, the Hamiltonian of our device is well-described by:
\begin{equation}
    \hat{H}/\hbar = \omega_q \hat{q}^\dagger \hat{q} + \omega_a \hat{a}^\dagger \hat{a} - K_q \hat{q}^{\dagger 2} \hat{q}^2 - K_a \hat{a}^{\dagger 2} \hat{a}^2 -  \chi \hat{q}^\dagger \hat{q} \hat{a}^\dagger \hat{a} - \chi' \hat{q}^\dagger \hat{q} \hat{a}^{\dagger 2} \hat{a}^2 + \Omega(t)\hat{q}^\dagger + \varepsilon(t)\hat{a}^\dagger + h.c.,
    \label{appeq:hsysq}
\end{equation}
where $\hat{q}$ is the annihilation operator in the transmon Hilbert space, while $\hat{a}$ is the annihilation operator in the oscillator Hilbert space, describing a transmon and an oscillator mode with frequencies $\omega_q$ and $\omega_a$ respectively. The transmon anharmonicity $K_q$ is large relative to its drive bandwidth, allowing the transmon mode to be approximated as a two-level qubit system, with spin-$Z$ operator $\hat{\sigma}_z = 1 - 2\hat{q}^\dagger \hat{q}$. The interaction between the transmon and the oscillator is described by the cross-Kerr interaction $\chi$, which has strength of $O(10)~$kHz; the precise value is measured via a specific calibration procedure detailed below. In this situation, the higher-order terms of oscillator anharmonicity $K_a$ and second-order cross-Kerr interaction strength $\chi'$ are even smaller (of $O(10)~$Hz). While these higher-order terms do affect the exact dynamics of the system under large displacements, their effect can be taken into account by operating with fixed displacement strengths (see Appendix~\ref{app:trainable_digital_model}). The term $\Omega(t)$ describes a generally time-dependent pulse on the transmon, while $\varepsilon(t)$ describes a generally time-dependent pulse on the oscillator; their precise forms will be introduced in due course. In Table~\ref{table:system_parameters}, we list the measured values of the relevant terms of the Hamiltonian, alongside other parameters.

\begin{table}[h!]
    \centering  
    \setlength{\tabcolsep}{8pt}            
    \renewcommand{\arraystretch}{1.2}      

    \begin{tabular}{|l|l|}
    \hline
    \textbf{Parameter} & \textbf{Value} \\ 
    \hline\hline 

    Transmon g--e transition frequency & $\omega_q = 2\pi \times 7.309~\rm GHz$ \\ 
    \hline
    
    Transmon self-Kerr & $K = 2K_q = 2\pi \times 340~\rm MHz$ \\ 
    \hline

    Transmon relaxation time & $T_1 = 30~\rm \mu\text{s}$ \\
    \hline

    Transmon Ramsey coherence time & $T_{2R} = 30~\rm \mu\text{s}$ \\
    \hline

    Transmon Echo coherence time & $T_{2E} = 40~\rm \mu\text{s}$ \\
    \hline

    Transmon readout fidelity for $\ket{g}$ & $96\%$ \\
    \hline

    Transmon readout fidelity for $\ket{e}$ & $95\%$ \\
    \hline

    Readout frequency & $\omega_r = 2\pi \times 8.920~\rm GHz$ \\
    \hline

    Readout-Transmon dispersive shift & $\chi_{rq} \sim 2\pi \times 1 \rm~MHz$ \\
    \hline

    TWPA pump frequency & $\omega_{\rm pump} = 2\pi \times 8.550~\rm GHz$ \\
    \hline

    Oscillator frequency & $\omega_a = 2\pi \times 6.072~\rm GHz$ \\
    \hline

    Oscillator-Transmon dispersive shift  & $\chi = 2\pi \times (13.8 \pm 0.3)~\rm kHz$\\
    \hline

    Ratio of displacement to the AWG amplitude & $s = 24.2 \pm 0.3$ \\
    \hline

    Oscillator relaxation time & $T_{1c} = 250~\rm \mu\text{s}$ \\
    \hline
    
    \end{tabular}
    
    \caption{\textbf{Measured experiment parameters and readout fidelity.} These parameters are obtained using standard spectroscopic and time-resolved measurements following~\cite{chou2018teleported, eickbusch2022fast} and Appendix~\ref{app:system_hamiltonian_and_parameters:characterization}.}
    \label{table:system_parameters}
\end{table}

\subsection{Characterization of displacement and cross-Kerr interaction}
\label{app:system_hamiltonian_and_parameters:characterization}

\begin{figure}[h!]
    \centering
    \includegraphics[width=\textwidth]{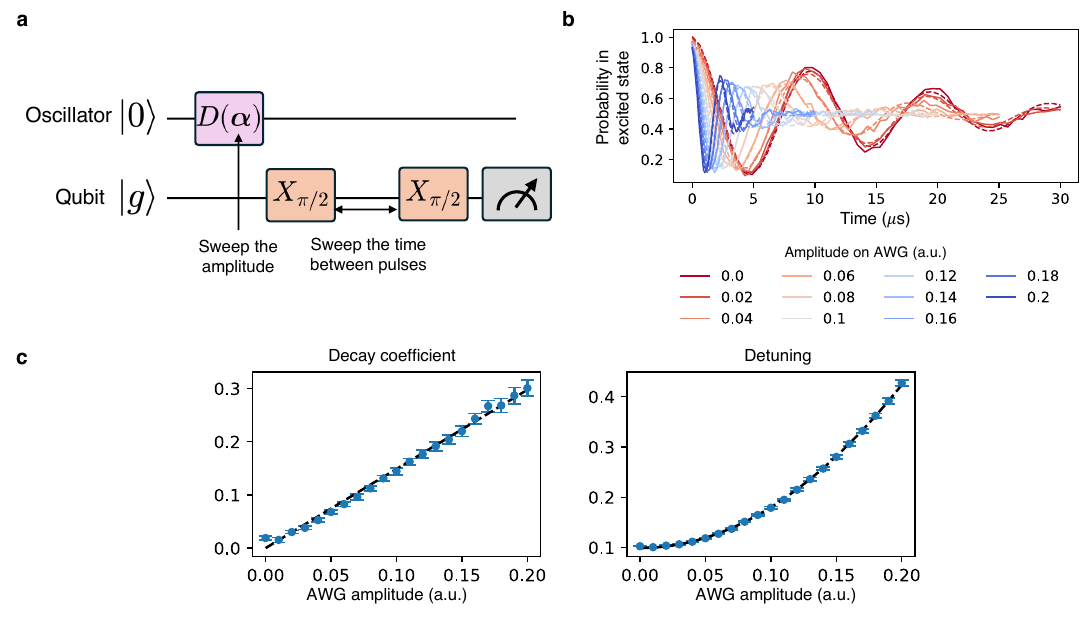}
    \caption{\textbf{Characterizing the parameters of the experiment.} \textbf{a.} Quantum circuit to measure the cross-Kerr interaction strength and the displacement on the oscillator. The amplitude of the pulse on the oscillator producing the displacement, and the time between the two $\pi/2$ pulses on the qubit are swept. The experiment is repeated multiple times to obtain the probability of the qubit in the excited state at the end of the protocol. \textbf{b.} Plot of the qubit probability as a function of the time between the two $\pi/2$ pulses for a select range of amplitude level of the AWG generating the pulse to displace the oscillator. \textbf{c.} Extracted decay coefficient and detuning from fits to data in \textbf{b}. From these parameters, the cross-Kerr and displacement strength can be estimated.}
    \label{fig:characterization}
\end{figure}

Measuring the value of the displacement we perform on the oscillator for a given pulse shape $\varepsilon(t)$ is an important quantity. This value is important for benchmarking the sensitivity of our protocols for displacement sensing with respect to other protocols. It is also important in enabling accurate numerical simulations of the dynamics of the system, which we use to train the parameters of the protocol (which is discussed in Appendix~\ref{app:trainable_digital_model}). In this section, we discuss a novel characterization technique to determine this quantity, along with the cross-Kerr interaction strength. The experimental parameters we obtain from this characterization are consistent with the values obtained from the characterization procedure introduced in Ref.~\cite{eickbusch2022fast}. A benefit of our protocol is that, unlike the procedure of Ref.~\cite{eickbusch2022fast}, it does not require fine-tuned classical simulation of the dynamics of the system.

Since the Ramsey protocol involves short $\pi/2$ pulses (which are high-fidelity), and not a continuous pulse, it is a convenient method to measure the frequency of the qubit mode. In our context, we measure the changes in the qubit frequency due to the cross-Kerr interaction with the oscillator, when the oscillator is displaced. However, since the cross-Kerr interaction is also unknown, this will not allow us to obtain the displacement value. Therefore, to obtain both values, we make use of the entanglement of the qubit with the oscillator. The protocol begins with the system in the ground state, and at the end of the first $\pi/2$ pulse, the state of the system is:
\begin{equation}
    \ket{\psi} = \frac{1}{\sqrt{2}}(\ket{\alpha,g} + \ket{\alpha,e}).
\end{equation}
Under the evolution of the interaction Hamiltonian, the two qubit-conditioned coherent states will evolve differently in Wigner space. After a time $t$, the general state is:
\begin{equation}
    \ket{\psi(t)} = \frac{1}{\sqrt{2}}\left(\ket{\alpha e^{i\chi t/2},g} + \ket{\alpha e^{-i\chi t/2},e}\right),
\end{equation}
which corresponds to the oscillator frequency shifted by the qubit state. After the second $\pi/2$ pulse, the final system state is:
\begin{equation}
    \ket{\psi(t)} = \frac{1}{2}\left(\ket{\alpha e^{i\chi t/2},g} + \ket{\alpha e^{-i\chi t/2},g}\right) + \frac{1}{2}\left(\ket{\alpha e^{i\chi t/2},e} + \ket{\alpha e^{-i\chi t/2},e}\right),
\end{equation}
following which the probability of measuring the qubit in the excited state can be obtained as $P_e(t) = \left|\langle \psi(t)|e\rangle \right|^2$,
which can be simplified to:
\begin{equation}
    P_e(t) = \frac{1}{4} \left(2 + e^{-\alpha^2}\left(e^{\alpha^2 e^{i\chi t}} + e^{\alpha^2 e^{-i\chi t}} \right) \right),
\end{equation}
to obtain:
\begin{equation}
    P_e(t) = \frac{1}{2} + \frac{1}{2}e^{-\alpha^2(1 - \cos{\chi t})}\cos{\alpha^2 \sin{\chi t}}.
    \label{appeq:characterization}
\end{equation}
This equation is exact and we can use this to directly fit to the experimental data of Fig.~\ref{fig:characterization}b. However, since $\chi t \ll 1$, we can approximate as:
\begin{align}
    P_e(t) \approx \frac{1}{2} + \frac{1}{2} e^{-(\alpha \chi t)^2/2}\cos{\alpha^2 \chi t}.
    \label{appeq:characterizationTaylor}
\end{align}

To take into account experimental imperfections, such as decoherence and readout fidelity, we fit the data at each displacement to: 
\begin{equation}
    P_{\rm fit}(t) = A + \left(A_0 e^{-(Ct)^2} \cos{2\pi ft}\right)e^{-t/T_2}.    
\end{equation}
Here $C$ is the decay coefficient while $f$ represents the detuning, as illustrated in Fig.~\ref{fig:characterization}c. Let $z$ be the value of the amplitude on the AWG used to generate the displacement. Then the displacement $\alpha$ caused is linearly related: $\alpha = sz$. Here $s$ is a unitless number which can be thought as a `scaling factor' between the amplitude set on the AWG and the corresponding displacement caused. This value is pulse-shape dependent (and we use the same pulse shape to enact the sensing displacement). Our goal is to estimate this value, along with the interaction strength $\chi$. From Eq.~(\ref{appeq:characterizationTaylor}), we can see the linear dependence of the decay coefficient with $z$: $C = m_1 z =\frac{1}{\sqrt{2}}sz\chi$, and the quadratic dependence of the detuning with $z$: $f = f_0 + m_2z^2 =f_0 + \frac{1}{2\pi}s^2z^2\chi$. Here $f_0$ is the detuning with no displacement (which is the protocol for measuring the Ramsey $T_2$). A finite detuning $f_0$ is conventionally used for obtaining a better fit. We then fit the data in Fig.~\ref{fig:characterization}c to obtain estimates of $m_1$ and $m_2$. From this, and using the prior expressions for $C$ and $f$, we can solve for the cross-Kerr strength and the scaling factor:
\begin{equation}
\chi = \frac{m_1 ^ 2}{m_2 \pi}
   \quad\mathrm{and}\quad 
s = \frac{\sqrt{2} \pi m_2}{m_1}.
\end{equation}
Based on these relations, we obtain the values the listed in Table~\ref{table:system_parameters}. We also verify the expression in Eq.~(\ref{appeq:characterizationTaylor}) by simulating this protocol using QuTiP~\cite{qutip5}, included in the linked repository. The scaling factor for the displacement $(s = 24.2 \pm 0.3)$ is calibrated for the sensing displacement, which is a $4\sigma$ Gaussian pulse of $100~\rm ns$. The AWG amplitude $z$ has a range between $0$ and $1$. Hence, the experiment can displace the oscillator at most $|\boldsymbol{\alpha}| = s$ from one pulse. For other oscillator displacement pulses -- such as in the ECD gate -- we choose different pulse shapes. However, the calibration of the scaling factor for the above pulse shape also allows us to accurately simulate arbitrary pulse shapes in the digital model of the experiment. We discuss this further in Appendix~\ref{app:trainable_digital_model}.

\clearpage
\section{Quantum computational displacement sensing (QCDS) protocol}
\label{app:quantum_computational_sensing_protocol}

In this section, we discuss the experimental realization of the QCDS protocol. We also present the performance of the protocol on various tasks, in simulation and experiment.

\subsection{Overview}
\label{quantum_computational_sensing_protocol:overview}

\subsubsection{Creating a superposition of coherent states}
We first discuss the motivation for the design of the quantum computational displacement protocol. We build intuition by considering the probe state prepared in the oscillator-qubit space just before the sensing. This is best done by representing the state of the state in the coherent state basis, conditioned on the state of the qubit. Each echoed conditional displacement operator imparts a displacement conditioned on the qubit state. Therefore, in general, the number of basis states doubles. On the other hand, single qubit rotations mix the coefficients of the wavefunction on these basis states. At the end of a sequence of depth $N$, the general form of the state can be written as:
\begin{equation}
    \ket{\psi} = \sum_{i=0}^{2^N/2 - 1} c_{ig}\ket{\boldsymbol{\beta}_i,g} + c_{ie}\ket{-\boldsymbol{\beta}_i,e}.
\end{equation}
Though there are $2^N$ coefficients, the maximum extent of the values of $|\boldsymbol{\beta}_i|$ are constrained to be within $N\beta$, where $\beta$ is the (fixed) magnitude of each conditional displacement. The various values of $\boldsymbol{\beta}_i$ arise from linear combinations of the conditional displacements in the $2$D space of vector additions with positive and negative signs based on the state of the qubit. Importantly, for every coherent state centered at $\boldsymbol{\beta}_i$ conditioned on qubit ground state, there is a corresponding coherent state (with a generally different coefficient) centered at $-\boldsymbol{\beta}_i$ conditioned on qubit excited state. The action of the displacement to be sensed $D(\boldsymbol{\alpha})$ can be understood by noting the commutation of displacement operators: $D(\boldsymbol{\alpha})D(\boldsymbol{\beta}) = e^{(\boldsymbol{\alpha} \boldsymbol{\beta}^* - \boldsymbol{\alpha}^*\boldsymbol{\beta})/2} D(\boldsymbol{\alpha + \beta})$. Therefore, acting on the probe state, each coherent state picks up a phase proportional to the sensed displacement, the position of the coherent state and their relative angle,
\begin{equation}
    \ket{\psi(\boldsymbol{\alpha})} = \sum_{i=0}^{2^N/2 - 1} c_{ig} e^{(\boldsymbol{\alpha} \boldsymbol{\beta}^* - \boldsymbol{\alpha}^*\boldsymbol{\beta})/2} \ket{\boldsymbol{\beta}_i + \boldsymbol{\alpha},g} + c_{ie} e^{-(\boldsymbol{\alpha} \boldsymbol{\beta}^* - \boldsymbol{\alpha}^*\boldsymbol{\beta})/2} \ket{-\boldsymbol{\beta}_i + \boldsymbol{\alpha},e}.
\end{equation}
Note the sign difference in the phase, due to the fact that the coherent states are positioned at $\boldsymbol{\beta}_i$ and $-\boldsymbol{\beta}_i$. Therefore, each pair picks up a different weighting of the displacement, each of which can be thought of as being sensitive to a single Fourier component in the $2$D displacement space. As the number of pairs increases with depth, the protocol has more Fourier components it can use to distinguish the two distributions. The application of the inverse unitary $U^\dagger$ completes the protocol. This is relevant since right after sensing, the qubit is generally entangled with the oscillator. A qubit measurement at this stage will reveal little information, since tracing out the oscillator will leave the qubit in (potentially highly) mixed state. The goal of $U^\dagger$ can be thought of as imparting the information of the geometric phase due to the commutation of the displacement operators onto the qubit phase (weighted optimally), such that the qubit is near the poles of the Bloch sphere (that is, the ground and excited states) depending on the class of the sensed displacement. This protocol can therefore be thought of as a generalized interferometer (as highlighted by Ref.~\cite{sinanan2024single}). 

Ref.~\cite{sinanan2024single} formalizes this rigorously using the framework of quantum signal processing, in the scenario of the conditional displacement acting on a single quadrature. The space of functions which can be realized via Ref.~\cite{sinanan2024single}'s protocol form a complete basis. Therefore, in the ideal scenario, the protocol can classify well-behaved distributions along a single quadrature without error, in the limit of sufficiently large depth. Extending their framework to the sensing displacement along both quadratures is an interesting and open question, but is beyond the scope of this work.

\subsubsection{Estimating an observable nonlinear in position and momentum}

The choice of our protocol can also be understood by noting that the interleaved sequence of echoed conditional displacements and single-qubit rotations can construct any arbitrary unitary provided sufficient depth. This can be shown by the analysis of Ref.~\cite{eickbusch2022fast}, which illustrates how the full Lie algebra for polynomial operators on the qubit-oscillator Hilbert space is generated. The generator associated with the primitive gates are $\{\sigma_i, x\sigma_i p\sigma_i\}$, where $i \in \{x,y,z\}$. By considering the commutation of their operators, any generator of the form $x^j p^k \sigma_i$ and $x^j p^k$ for integer $j,k$ can be constructed. The depth $N$ sets a limit on the largest degree of generator which can be formed. To motivate why this framework is useful, consider the requirements of classifying a given dataset. In classical machine learning, this can be thought of constructing a hyperplane in a generally nonlinear manifold. Given the components of the displacement along the position $x$ and momentum $p$ quadratures, this can be mathematically expressed as computing some function $\mathcal{F}(x,p)$, which is a polynomial of $x$ and $p$. Specifically, if this function evaluates to $\mathcal{F} > 0$, then we predict class A. Else, we predict class B.

In the quantum setting (since the displacement is sensed by a quantum system), the function now results as computing the expectation of an operator associated with the oscillator Hilbert space, which can be expressed as a polynomial of the position and momentum operators. Importantly, since the task is a binary classification task, estimating the expectation of a single operator is sufficient. Hence there is no fundamental noise limit in estimating this quantity due to Heisenberg's uncertainty principle. Our protocol can be thought of as a measurement of this operator, by mapping the expectation of the operator onto the phase of the qubit. For instance, in the case of $N = 1$ protocols, the space of operators that can be measured is spanned by linear combinations of the position and momentum operator. As $N$ increases, the space of accessible basis functions increases, enabling more complex operators to be estimated. 

The ability of achieving a quantum computational-sensing advantage can be understood by considering how a conventional displacement sensor would perform the same task. Such sensors are designed to only estimate the position and momentum operators. A classical postprocessor is then required to estimate the relevant function $\mathcal{F}$ based on these estimates. Crucially, since an estimate of both quadratures is required, this protocol usually suffers from a fundamental noise floor since these two operators do not commute. An example of this is using a phase preserving amplifiers. On the other hand, a protocol involving GKP states and phase estimation can circumvent this issue but at the expense of finite dynamic range. Two-mode squeezed states can circumvent the noise issue by utilizing an additional bosonic mode, such that the displacement is mapped onto a larger Hilbert space where the associated operators commute with each other. We provide additional details of conventional schemes in Appendix~\ref{app:benchmark}.

\subsubsection{Results of training}

\begin{figure}[h!]
    \centering
    \includegraphics[width=0.8\textwidth]{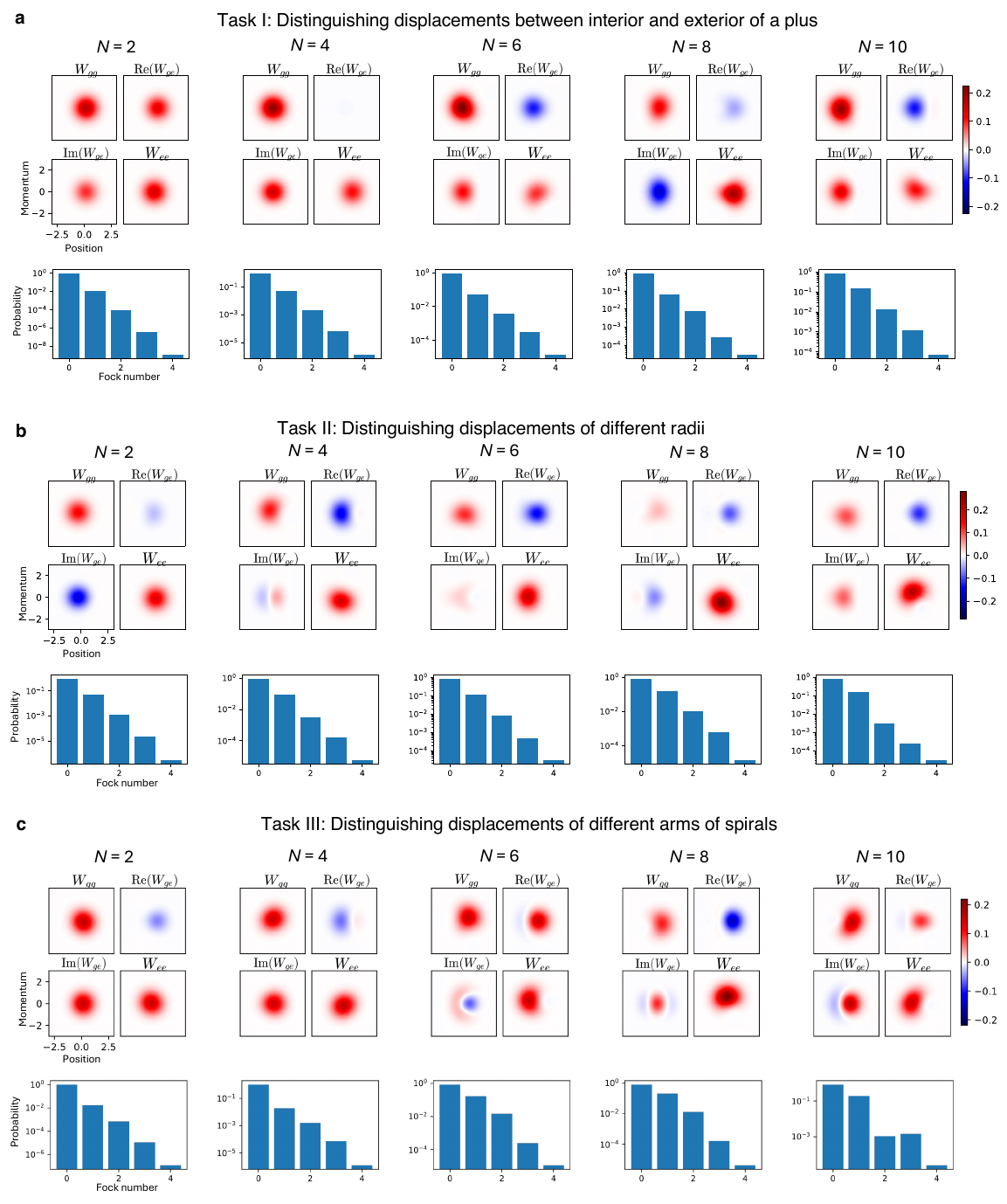}
    \caption{\textbf{Simulation probe states of the entangled qubit-oscillator system before sensing. The data are for different depths $N$ for the three tasks defined in Fig.~\ref{fig:main3}.} For each task, the top panel plots $W_{gg}$ (which is the Wigner function of $\bra{g}\rho \ket{g}$), $W_{ee}$ (which is the Wigner function of $\bra{e}\rho \ket{e}$) and real and imaginary components of the Wigner function of $W_{ge}$ (which is the Wigner function of $\bra{e}\rho \ket{g}$). $\rho$ is the density matrix of the system just before sensing. All plots share the same position and momentum range, which is labeled in the bottom-left plot. Underneath each are plots of the corresponding probability of occupation in Fock basis. Even with support on a few Fock states, the protocol can perform classification.}
    \label{fig:wigner}
\end{figure}

In Fig.~\ref{fig:wigner}, we plot the simulated probe states of the combined qubit-oscillator system for three different tasks (those introduced in Fig.~\ref{fig:main3}. Unlike experiments such as Ref.~\cite{eickbusch2022fast} that focus on state preparation, our protocol generates in general an entangled state between the qubit and oscillator. Therefore, to define the entire state, we require $4$ different Wigner functions, each which describes the density matrix of the oscillator mode in different qubit subspace (illustrated in the top panel of each subfigure of Fig.~\ref{fig:wigner}). These are the Wigner function of the density matrix $\rho$: conditioned on the ground state ($\bra{g} \rho \ket{g}$), conditioned on the excited state ($\bra{e} \rho \ket{e}$, and real and imaginary components of the off-diagonal contribution ($\bra{e} \rho \ket{g}$). Interestingly, from the simulation results, we notice that the Wigner negativity is predominantly in the off-diagonal component, for all the tasks presented. The off-diagonal component is relevant for the response of the protocol since it relies on interferometry of oscillator states conditioned on the ground versus excited states. Furthermore, similar to the conclusions of Ref.~\cite{sinanan2024single}, there is no obvious phase space distribution of the Wigner functions and the classification task. In the bottom panel of each subfigure of Fig.~\ref{fig:wigner}, we instead plot the photon probability. Interestingly, there is not a significant photon occupation beyond $5$ photons, even for the largest protocols. This highlights the native efficiency of the QCDS protocol in terms of photon occupation, which is a metric of quantum resource. We notice qualitatively similar forms of probe state for other example tasks considered. The average photon number of the probe state (and corresponding support on Fock states) will tend to increase as the range of displacement of the dataset considered are smaller and/or finer features are required to be resolved to distinguish the two datasets (consistent with the intuition developed in the previous subsections).

\subsection{Pulse sequence}
\label{app:quantum_computational_sensing_protocol:pulse_sequence}

\begin{figure}[h!]
    \centering
    \includegraphics[width=0.75\textwidth]{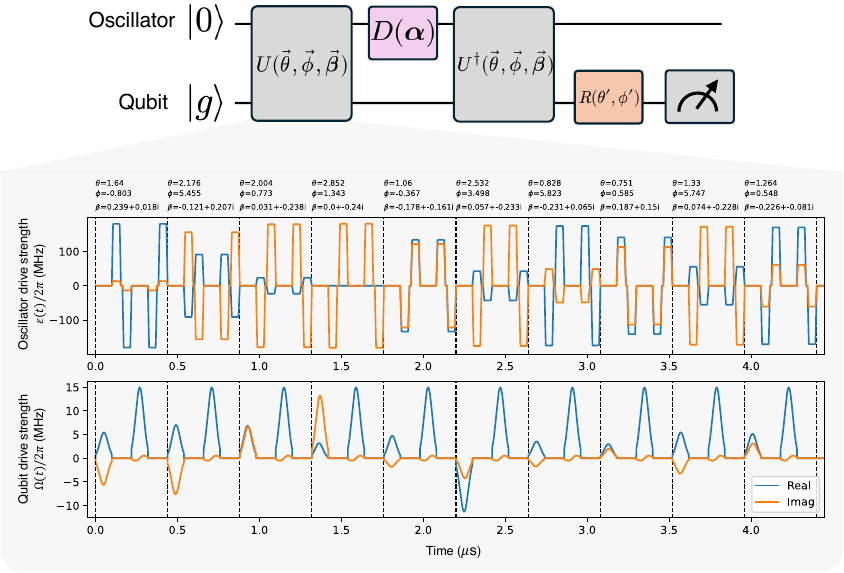}
    \caption{\textbf{Pulse sequence of the oscillator and qubit drives for an $N=10$ quantum computational displacement sensing protocol.} The pulse sequence corresponds to the best-performing $N=10$ protocol for the task defined in Fig.~\ref{fig:main3}c. Vertical dashed lines indicate the pulses for each unit. Each unit consists of a qubit rotation and an echoed conditional displacement gate. By reversing the order of pulses, along with a $\pi$ shift in the phase of the pulses, we realize the subsequent unitary $U^\dagger$. The sensing displacement and final qubit rotations are also realized by similar microwave pulses.}
    \label{fig:pulse}
\end{figure}

Fig.~\ref{fig:pulse} illustrates a typical pulse sequence for realizing the unitary $U(\vec{\theta}, \vec{\phi}, \vec{\boldsymbol{\beta}})$ from Fig.~\ref{fig:main1} of the main text. The pulse sequence consists of single qubit rotations and displacements on the oscillator. Single qubit rotations are implemented using $4\sigma$ Gaussian pulses of $100~\rm ns$. The amount of qubit rotation is controlled by tuning the amplitude of the pulse. We notice systematic phase errors which play a non-negligible role for the qubit rotations. To mitigate this, we use the DRAG scheme where an additional component in the orthogonal phase of the original pulse is added~\cite{krantz2019quantum}. This additional component is proportional to the derivative of the original pulse. We sweep the coefficient of the scaling of this correction $\beta$. We find the optimal value to be $\beta = 0.015$ in minimizing the phase error. We use this pulse shape for all qubit rotations. Usually, such a correction isn't required for high-fidelity single-qubit rotations of durations $100~\rm ns$. We believe our device experiences this effect more strongly than usual due to the additional spurious peaks we notice in the vicinity of the qubit frequency in the spectroscopy results. We hypothesize some of the nearby peaks to be two-photon transitions due to higher-order coupling terms of the transmon-oscillator-readout system; however a systematic study of this mode is beyond the scope of this work. 

The echoed conditional gate is composed of large displacements, with a qubit pi-pulse in between. We use a similar implementation to Ref.~\cite{eickbusch2022fast}. We optimize the pulse shape and durations of the protocol. Each pulse is a flattop Gaussian pulse shape, where the flattop region in the middle of the pulse is $70\%$ of the entire pulse duration. This is flanked by the half-profile of a Gaussian pulse, to smoothen the pulse shape. Each half-profile Gaussian pulse extends $2\sigma$ (so that an entire Gaussian pulse stitched together would be $4\sigma$). The entire pulse duration is $55~\rm ns$. We fix the amplitude of this pulse, and only set the phase as a trainable parameter. This forms the building block for the out-and-back implementation of the ECD gate, as illustrated in Fig.~\ref{fig:pulse}. The pulses are separated by $10~\rm ns$ of idle time. To implement the Hermitian-conjugate of $U$ (i.e. $U^\dagger$), we implement the time-reversed pulse sequence, along with a $\pi$ phase flip for all pulses. This, to a high fidelity, implements $U^\dagger$. We test the fidelity of this implementation using the cat-state sensing protocol (see Appendix~\ref{app:quantum_computational_sensing_protocol:cat_state}). This is possible since we operate in the weak-interaction regime, and therefore the contribution from the cross-Kerr interaction to the unitary is small.

We optimized the parameters of these pulses to optimize for fast gates such that large depths $N$ can be executed with high fidelity. As seen in Fig.~\ref{fig:pulse}, implementing $U$ for an $N=10$ protocol takes $4.4~\rm \mu\text{s}$. In total, the entire protocol takes about $10~\rm \mu\text{s}$ (including the sensing displacement and final single qubit rotation). The entire protocol duration is therefore a significant fraction of the coherence time of the qubit. Since this plays a detrimental role in the performance of the protocol, the gate speed plays a significant role in determining the maximum depth of protocols which can be experimentally implemented which still results in a net improvement in the accuracy.

We briefly discuss the differences between the protocol implemented in this work versus previous theoretical works. In Ref.~\cite{khan2025quantum_novel}, the protocol involves an additional qubit rotation in the structure of $U$, while not including a final qubit rotation. This choice of convention doesn't yield a dramatic difference in the expressivity. We empirically find that including a final qubit rotation made the protocol slightly more stable during training. Next, while the protocol introduced in Ref.~\cite{khan2025quantum_novel} uses conditional displacement gates, our protocol implements the echoed conditional displacement. This choice was motivated by experiment, since we are able to implement the echoed conditional displacement with higher fidelity. Finally, the set of trainable parameters are different between the two protocols. The protocol in Ref.~\cite{khan2025quantum_novel} involves training $3$ parameters of each qubit rotation, along with the amplitudes and phases of the conditional displacements. In our protocol, we only train $2$ parameters of each qubit rotation, since the third parameter -- which is a Pauli-Z rotation, can be implemented virtually~\cite{krantz2019quantum}. We only train the phases of the echoed conditional displacements due to the experimental ability to implement fixed-amplitude high-fidelity echoed conditional displacements.

As discussed in the main text, our protocol is a natural extension of that introduced in Ref.~\cite{sinanan2024single}. The main difference between the two protocols is that in our protocol, we allow the phases of the qubit rotations and the echoed conditional displacements to be trainable. Finally, the main difference between our protocol and that introduced in Ref.~\cite{liao2024quantum} is while our protocol implements $U^\dagger$ after sensing, the protocol in Ref.~\cite{liao2024quantum} considers the more general version of implementing a parameterized unitary $V$ which is not constrained to be equivalent to $U^\dagger$.

\subsection{Classification accuracy}
\label{app:quantum_computational_sensing_protocol:classification_accuracy}

\begin{figure}[h!]
    \centering
    \includegraphics[width=\textwidth]{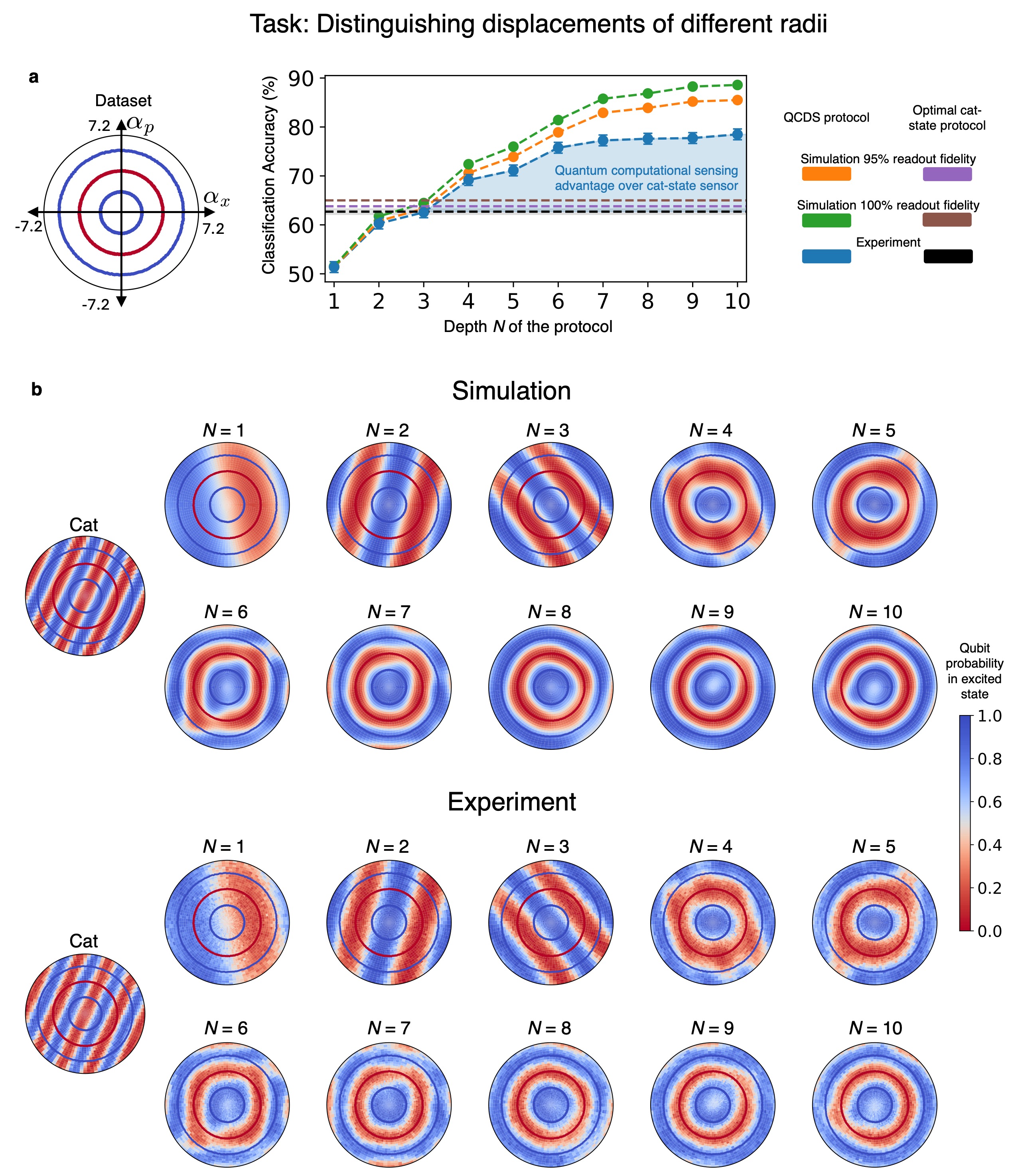}
    \caption{\textbf{Performance of the QCDS and cat-state protocols as a function of depth $N$ in distinguishing displacements of different radii (same as Fig.~\ref{fig:main3}b).} \textbf{a.} Dataset and classification accuracy of the simulation and experiment as a function of $N$. \textbf{b.} Simulated and experimental plots of the qubit excitation probability as a function of the sensed displacement.}
    \label{fig:circles2_detail}
\end{figure}

\begin{figure}[h!]
    \centering
    \includegraphics[width=\textwidth]{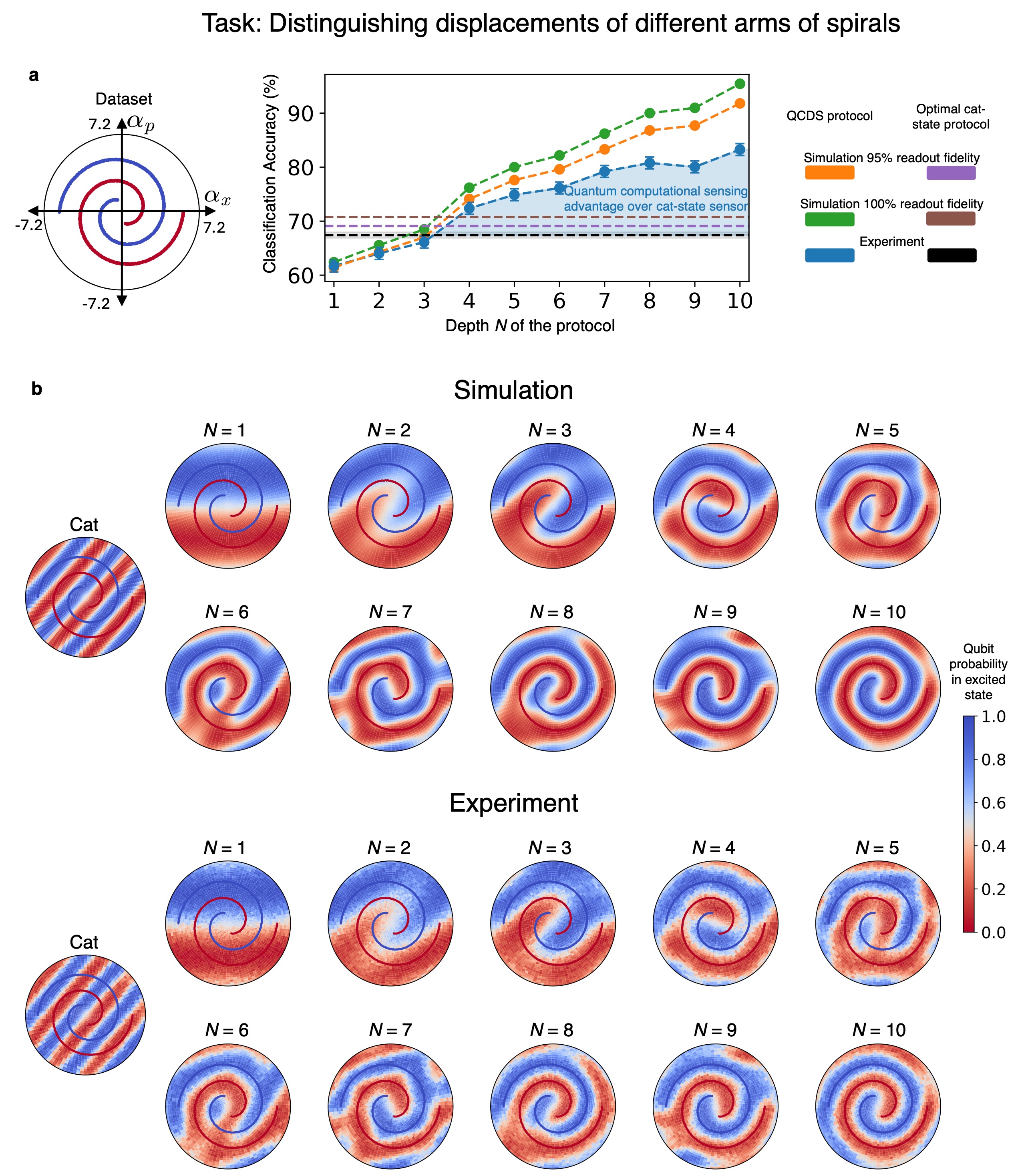}
    \caption{\textbf{Performance of the QCDS and cat-state protocols as a function of depth $N$ in distinguishing displacements of different arms of spirals (same as Fig.~\ref{fig:main3}c).} \textbf{a.} Dataset and classification accuracy of the simulation and experiment as a function of $N$. \textbf{b.} Simulated and experimental plots of the qubit excitation probability as a function of the sensed displacement.}
    \label{fig:spirals2_detail}
\end{figure}

In this subsection, we discuss how we obtain the classification accuracy based on experimental measurement data. Table~\ref{table:classification_accuracy} lists how the classification accuracy can be obtained from the average qubit-excitation probability for each class A and B ($p_A$ and $p_B$ respectively). The performance only depends on the difference between the two quantities. When the two probabilities are equal, then the two classes cannot be distinguished. On the other hand, a perfect classification accuracy is achieved when $p_A = 0$ and $p_B = 1$, as can be seen by the formula:
\begin{equation}
    P_{\rm CA} = \frac{1}{2} + \frac{1}{2}(p_B - p_A),
    \label{appeq:classification_accuracy}
\end{equation}
where $P_{\rm CA}$ is the classification accuracy. The dataset distribution of the two classes are defined by either closed-form functions, or a sampling algorithm. In both cases, we include a random number generator to define a separate training and testing displacement dataset. The random number generator is seeded with different numbers, one for training and one for testing (we choose a seed to ensure reproducibility). The training and testing dataset each consists of $512$ data points, for tasks defined on a small manifold, such as those considered in Figs.~\ref{fig:main4},~\ref{fig:circles2_detail} and~\ref{fig:spirals2_detail}. All others consist of $2048$ data points.

Once this is done, the experiment is evaluated using displacements from the testing dataset. For each displacement, we collect $2^5$ samples. Each sample consists of the measurement record, which can either be a $0$ (corresponding to a measurement of $\ket{g}$) or a $1$ (corresponding to a measurement of $\ket{e}$). These are the outcomes of the qubit measurement, measured using the standard dispersive readout method, described in Appendix~\ref{app:experiment_setup}. The prediction is as follows: for each measurement of $1$, we predict class B, and for each measurement of $0$, we predict class A. Based on this, we can compute the classification accuracy. This is mathematically equivalent to computing the classification accuracy based on Eq.~(\ref{appeq:classification_accuracy}), where we define $p_B$ as the estimated probability of the measuring a $1$ for displacements of class B, from the entire testing dataset (and similarly for $p_A$). Furthermore, the error bars for the estimate of the classification plotted represents the standard deviation in the error. This is obtained from the error in the estimate of $p_A$ and $p_B$, and then propagating this error via the definition of the classification accuracy. Finally, the plots of the qubit-excitation probability sweeps as a function of displacement are produced by a 2D grid sweep. We choose a resolution of $30$ points in the radial direction and $100$ in the azimuthal direction. For each grid point, we estimate the probability using $2^7$ shots. 

\begin{table}[h!]
    \centering  
    \setlength{\tabcolsep}{2pt}
    \renewcommand{\arraystretch}{1.2}

    \begin{tabular}{|l|c|}
    \hline
    \textbf{Probabilities} & \textbf{Value} \\ 
    \hline\hline

    P(class A) $=$ P(class B) & $\frac{1}{2}$ \\ 
    \hline
    
    P(Measuring $\ket{g}$ $|$ $D(\boldsymbol{\alpha})$ from class A) & $1 - p_A$ \\ 
    \hline
    
    P(Measuring $\ket{e}$  $|$ $D(\boldsymbol{\alpha})$ from class B) & $p_B$\\ 
    \hline

    $\rm P_{gA} = $ P(Measuring $\ket{g}$ $|$ $D(\boldsymbol{\alpha})$ from class A) x P(class A) & $\frac{1}{2}(1 - p_A)$\\ 
    \hline

    $\rm P_{eB} = $ P(Measuring $\ket{e}$ $|$ $D(\boldsymbol{\alpha})$ from class B) x P(class B) & $\frac{1}{2}p_B$\\ 
    \hline

    P(Correctly identifying class) $=$ $\rm P_{gA} + \rm P_{eB}$ & $\frac{1}{2} + \frac{1}{2}(p_B - p_A)$\\
    \hline

    \end{tabular}
    
    \caption{\textbf{Estimating classification accuracy.} Bayesian analysis of the classification accuracy based on the measurement records from experiment.}
    \label{table:classification_accuracy}
\end{table}

\subsection{Cat-state sensing and simple binary classification tasks}
\label{app:quantum_computational_sensing_protocol:cat_state}

\begin{figure}[h!]
    \centering
    \includegraphics[width=\textwidth]{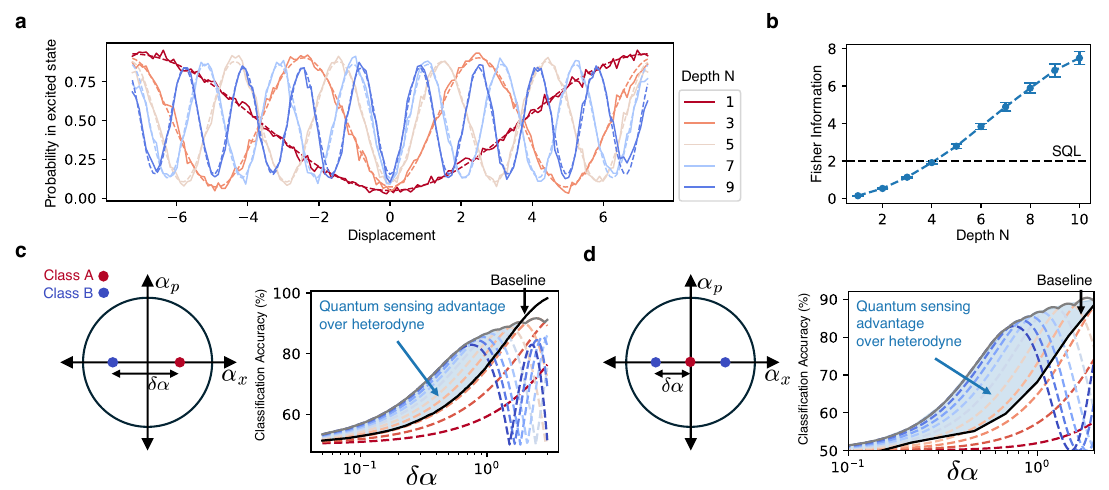}
    \caption{\textbf{Implementing the cat-state sensing protocol using the QCDS protocol.} \textbf{a.} Response of the cat-state sensing protocol to displacements orthogonal to the axis of the cat for different circuit depth $N$. Increasing the depth increases the size of the cat, and consequently, the frequency of the sinusoidal response. However, this comes at the expense of a reduction in the contrast. We use this protocol to calibrate and optimize the pulse sequence of the protocol. \textbf{b.} Fisher information as a function of the depth $N$ for the cat-state sensing protocol in sensing small single-component displacements. The Fisher information exceeds SQL (defined for displacement sensing using heterodyne on a vacuum state) for depths beyond $4$. \textbf{c.} and \textbf{d.} Pedagogical example tasks and the corresponding performance, demonstrating a quantum sensing advantage of the cat-state sensing protocol over an ideal phase-preserving amplifier.}
    \label{fig:cat}
\end{figure}

We use the fidelity of the depth-$N$ cat-state sensing protocol as a measure for the optimization of the pulse parameters. This protocol can be implemented by restricting to only a single qubit rotation in the beginning, and none during the intermediate layers. We choose a phase of $0$ for all odd layers and $\pi$ for all even layers of the ECD sequence. The resulting entangled cat-qubit state: $\ket{\psi} = \frac{1}{\sqrt{2}}(\ket{\boldsymbol{\beta}/2,g} + \ket{-\boldsymbol{\beta}/2,e})$ has $\beta$ which linearly increases with $N$. This is illustrated in Fig.~\ref{fig:cat}a, which plots the qubit probability in the excited state as a function of the component of the sensed displacement that is orthogonal to the axis of the cat. Dashed lines are best fit sinusoidal function to the data of the form
\begin{equation}
    P_e(x) = A_0 + A_1 \cos{(fx + \phi)},
\end{equation}
where $P_e(x)$ is the probability of the qubit in the excited state at the end of the protocol after sensing a single-axis displacement of amount $x$. $A_0$, $A_1$, $f$ and $\phi$ are parameters which we fit for. As $N$ increases, the frequency of the response increases. However, the contrast reduces. This is due to the greater susceptibility of larger cat states to decoherence. Finally, the qubit excitation probability at no displacement is close to $0$ (within the range of the contrast). This verifies the ability of the protocol to effectively implement $U^\dagger$, since the total action on the state should be the identity unitary in this case.

In Fig.~\ref{fig:cat}b, we illustrate the Fisher information of this protocol for single-axis displacement sensing. The Fisher information can be computed from the fit: $F = 4A_1^2f^2$. As a baseline, we include the Fisher information for displacement sensing a vacuum state with heterodyne (see Appendix~\ref{app:benchmark:phase_preserving_amplifier}), which has a theoretical value of $2$. Our protocol is able to surpass this limit for $N > 4$, demonstrating the ability of our protocol to achieve a quantum sensing advantage.

The cat-state sensing protocol can be used to solve simple binary classification tasks. We consider families of two simple tasks. In Fig.~\ref{fig:cat}c, we consider the task of distinguishing between two displacements (which is theoretically considered in Ref.~\cite{liao2024quantum}). From the functional fit, we can estimate the classification accuracy of the cat-state protocol as a function of the displacement distance between the two points $\delta \alpha$, for different values of $N$. Due to the modular nature of the qubit-excitation probability response, the classification accuracy is not monotonic. On the other hand, for the baseline method of a heterodyne measuring (using a phase-preserving amplifier), the accuracy increases monotonically. For a range of $\delta \alpha$, the cat-state sensing protocol achieves a higher classification accuracy (and hence a sensing advantage) than the baseline.

We also consider an extension of this task, illustrated in Fig.~\ref{fig:cat}d. Here one class consists of no displacement, while the other class consists of a displacement of $\delta \alpha$, but whose phase could either be $0$ or $\pi$. In this situation, the optimal protocol with a cat-state sensor is when the parameters are set such that the situation of no displacement is centered at the minimum (or alternatively at the maximum) of the response (similar to variance sensing~\cite{degen2017quantum}). In this case, the improvement in classification accuracy over the heterodyne baseline is larger. This illustrates one reason how the amount of advantage depends on the task. Since the baseline is a linear measurement of the displacement, a non-linear postprocessing step is required to distinguish the two classes (for e.g. by computing the square of the displacement). On the other hand, the response of the cat-state protocol, centered at the stationary points, is naturally nonlinear and suitable for performing the required computation in the quantum domain. This allows the protocol to discriminate the two classes with greater accuracy.

Finally, these two examples also illustrate another reason why the QCDS protocol is suited for binary classification. Since the qubit measurement outcome is a random variable that follows a binomial probability distribution, the quantum sampling noise vanishes when the probability goes to either $0$ or $1$. By mapping displacements of one class to a qubit-excitation probability to $0$, and the other to $1$, the role of quantum sampling noise is reduced.

\subsection{Examples}
\label{app:quantum_computational_sensing_protocol:examples}

To illustrate the versatility of the QCDS protocol, we present the simulation and experiment data for different tasks (including the ones considered in Fig.~\ref{fig:main3}). For each task, we plot the distribution of the dataset, along with the performance in simulation and experiment. For the simulation, we consider the performance of the protocol both with perfect readout and with a finite readout fidelity of $95\%$ (which is approximately the experimental readout fidelity). This performance matches that of the experiment for low values of $N$. We also include the performance of the optimal cat-state protocol, in dashed lines, for the same scenarios. For clarity, we do not include the axes labels on the 2D plots of the function of the qubit-excitation probability. The scale is the same as that illustrated in the datasets, with a range: $|\boldsymbol{\alpha}| < 7.2$.

\begin{figure}[h!]
    \centering
    \includegraphics[width=\textwidth]{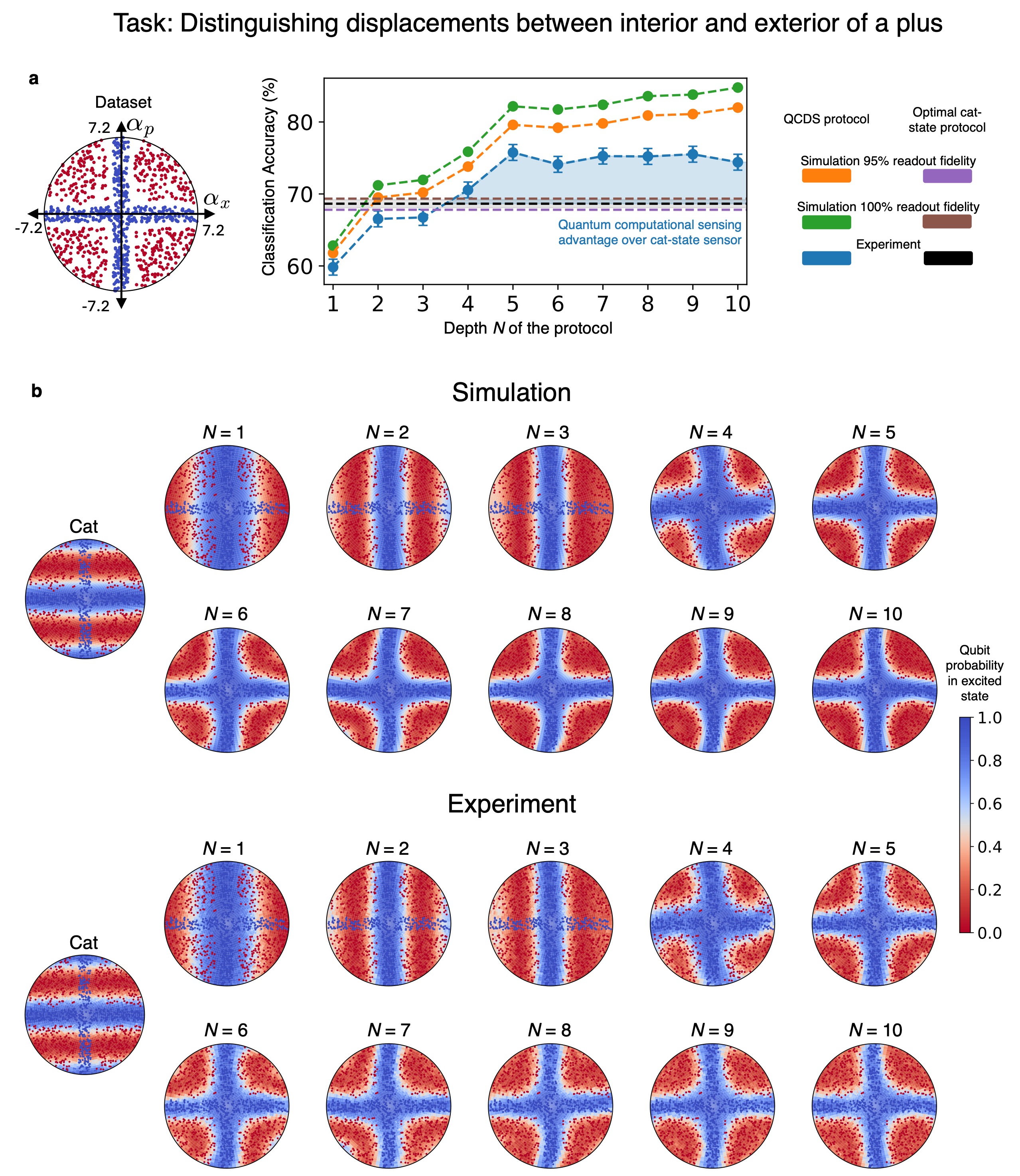}
    \caption{\textbf{Performance of the QCDS and cat-state protocols as a function of depth $N$ in distinguishing displacements between the interior and exterior of a plus symbol (same as Fig.~\ref{fig:main3}a).} \textbf{a.} Dataset and classification accuracy of the simulation and experiment as a function of $N$. \textbf{b.} Simulated and experimental plots of the qubit excitation probability as a function of the sensed displacement.}
    \label{fig:plus_detail}
\end{figure}
\clearpage

\clearpage

\clearpage

\begin{figure}[h!]
    \centering
    \includegraphics[width=\textwidth]{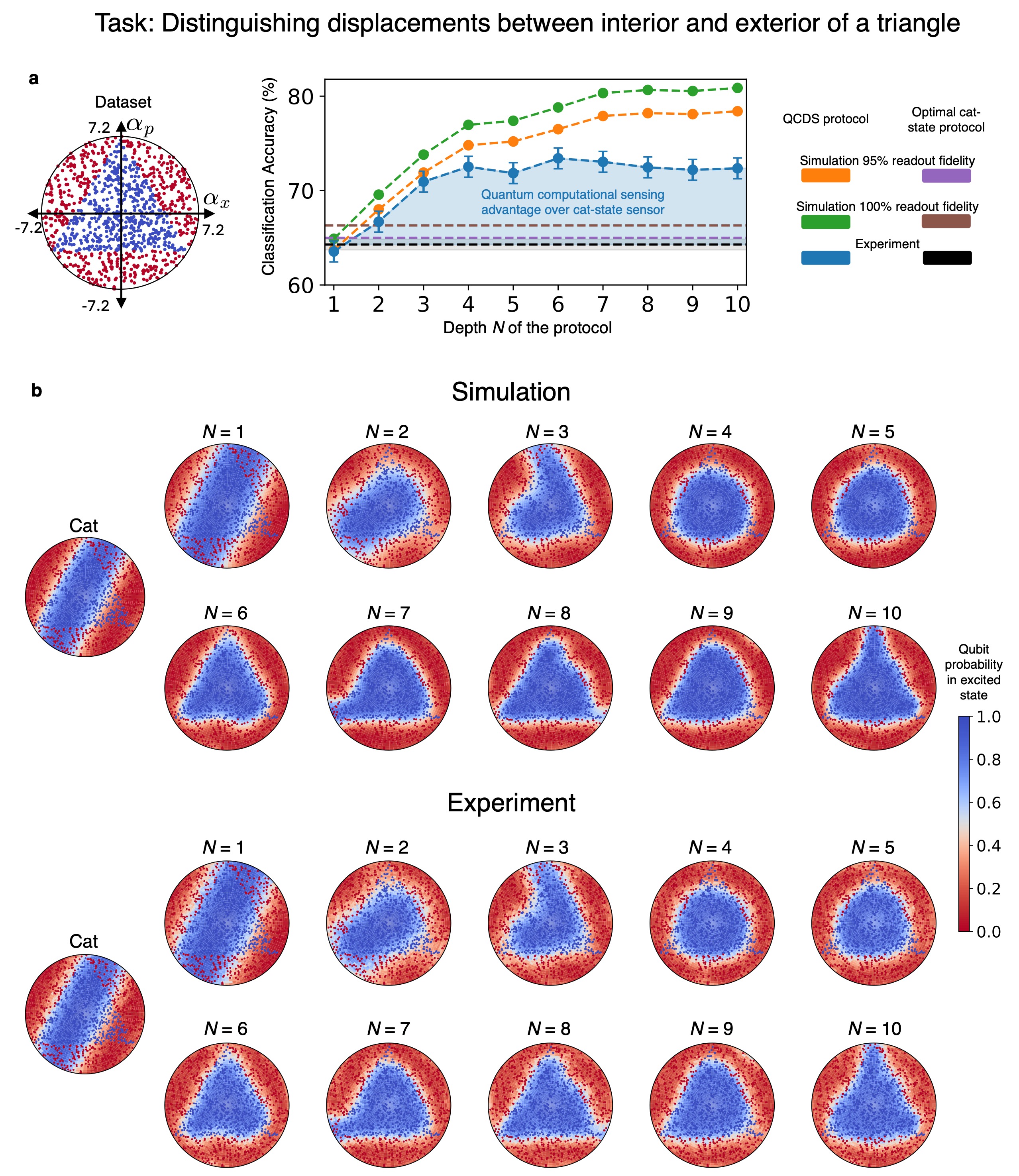}
    \caption{\textbf{Performance of the QCDS and cat-state protocols as a function of depth $N$ in distinguishing displacements between the interior and exterior of a triangle shape.} \textbf{a.} Dataset and classification accuracy of the simulation and experiment as a function of $N$. \textbf{b.} Simulated and experimental plots of the qubit excitation probability as a function of the sensed displacement.}
    \label{fig:triangle_detail}
\end{figure}
\clearpage

\begin{figure}[h!]
    \centering
    \includegraphics[width=\textwidth]{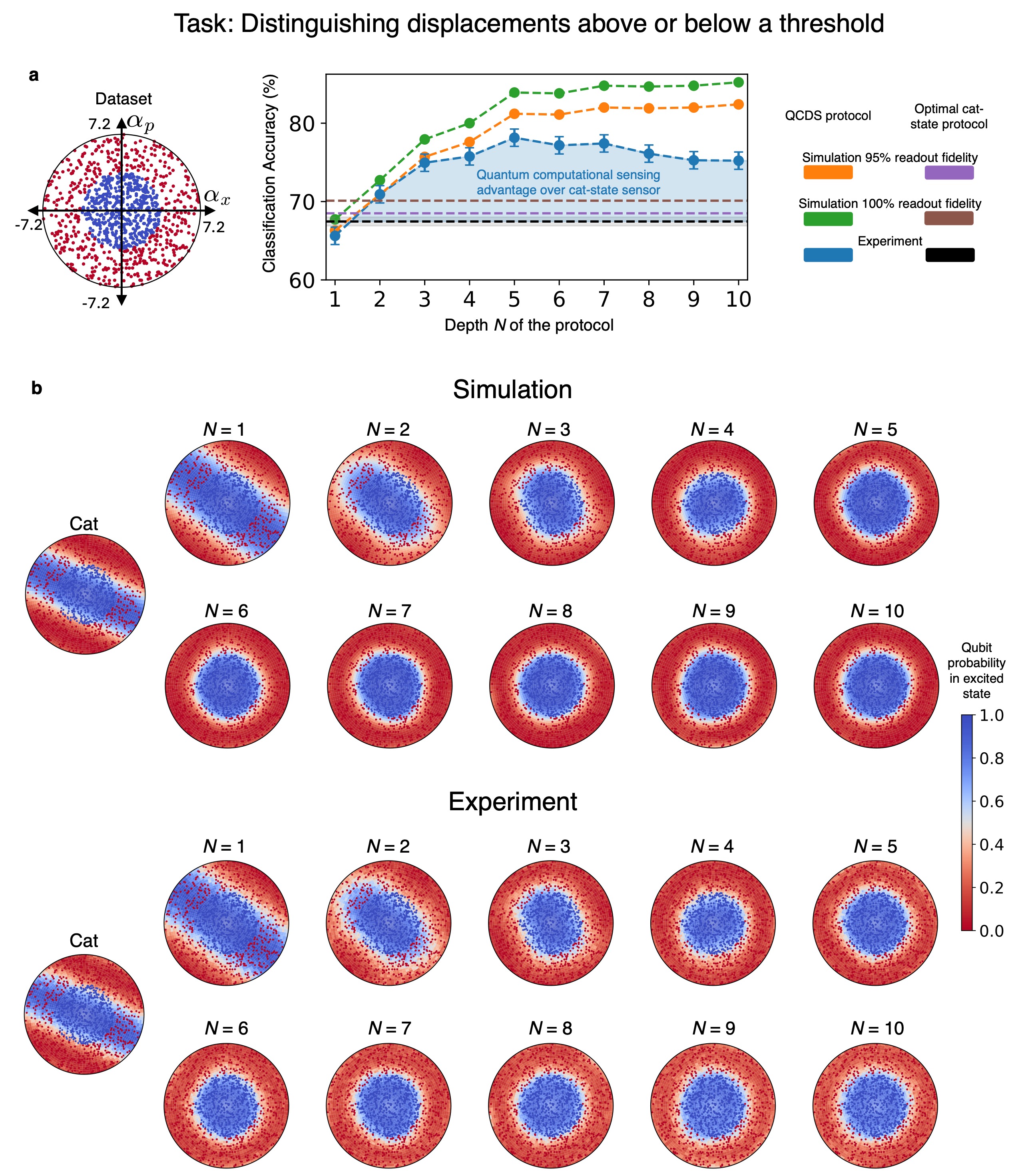}
    \caption{\textbf{Performance of the QCDS and cat-state protocols as a function of depth $N$ in distinguishing displacements above or below a threshold.} \textbf{a.} Dataset and classification accuracy of the simulation and experiment as a function of $N$. \textbf{b.} Simulated and experimental plots of the qubit excitation probability as a function of the sensed displacement.}
    \label{fig:threshold_detail}
\end{figure}
\clearpage

\begin{figure}[h!]
    \centering
    \includegraphics[width=\textwidth]{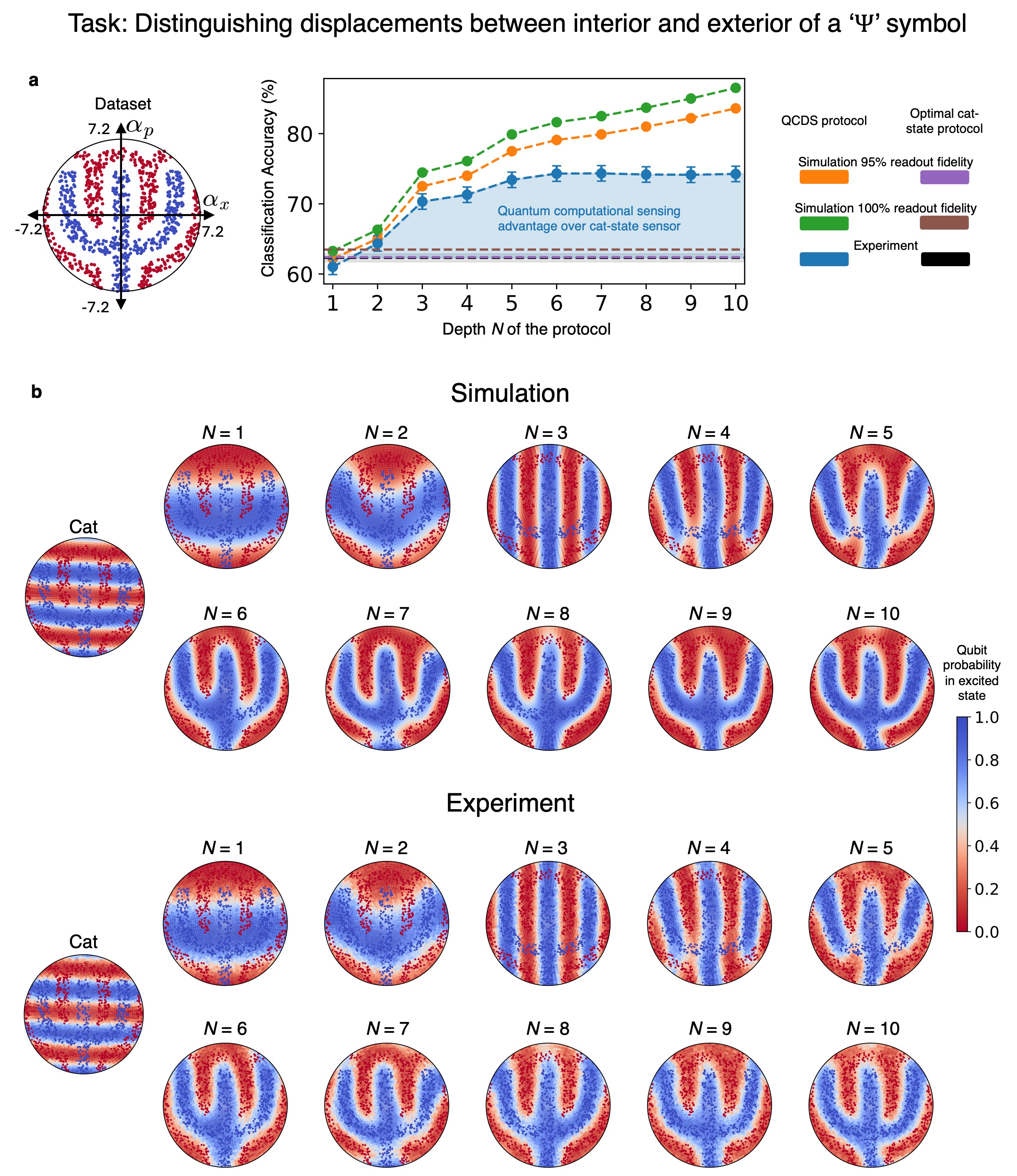}
    \caption{\textbf{Performance of the QCDS and cat-state protocols as a function of depth $N$ in distinguishing displacements between the interior and exterior of a distribution in the shape of $\Psi$.} \textbf{a.} Dataset and classification accuracy of the simulation and experiment as a function of $N$. \textbf{b.} Simulated and experimental plots of the qubit excitation probability as a function of the sensed displacement.}
    \label{fig:psi_detail}
\end{figure}
\clearpage

\begin{figure}[h!]
    \centering
    \includegraphics[width=\textwidth]{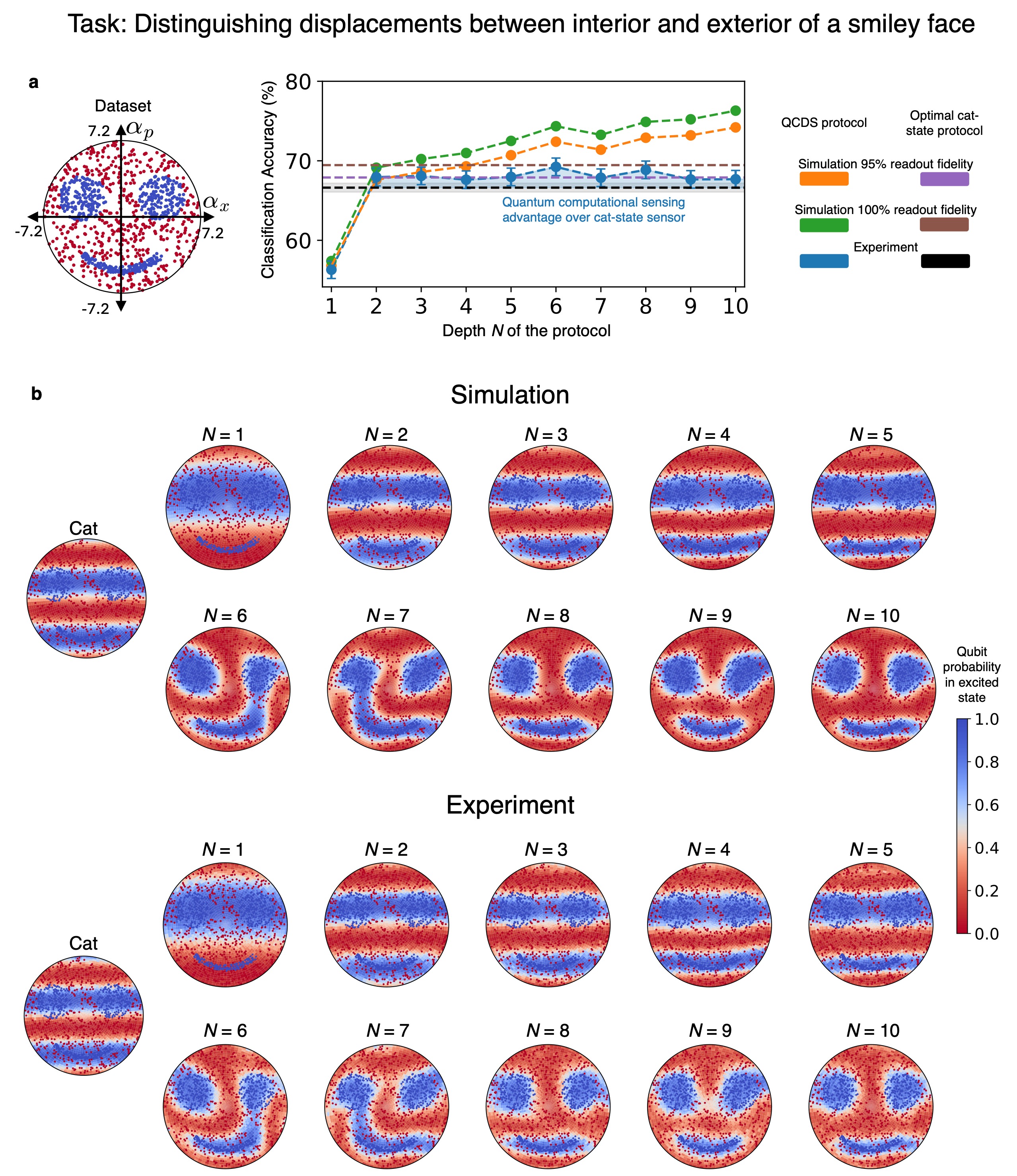}
    \caption{\textbf{Performance of the QCDS and cat-state protocols as a function of depth $N$ in distinguishing displacements between the interior and exterior of a distribution in the shape of a smiley face.} \textbf{a.} Dataset and classification accuracy of the simulation and experiment as a function of $N$. \textbf{b.} Simulated and experimental plots of the qubit excitation probability as a function of the sensed displacement.}
    \label{fig:smiley_detail}
\end{figure}
\clearpage

\begin{figure}[h!]
    \centering
    \includegraphics[width=\textwidth]{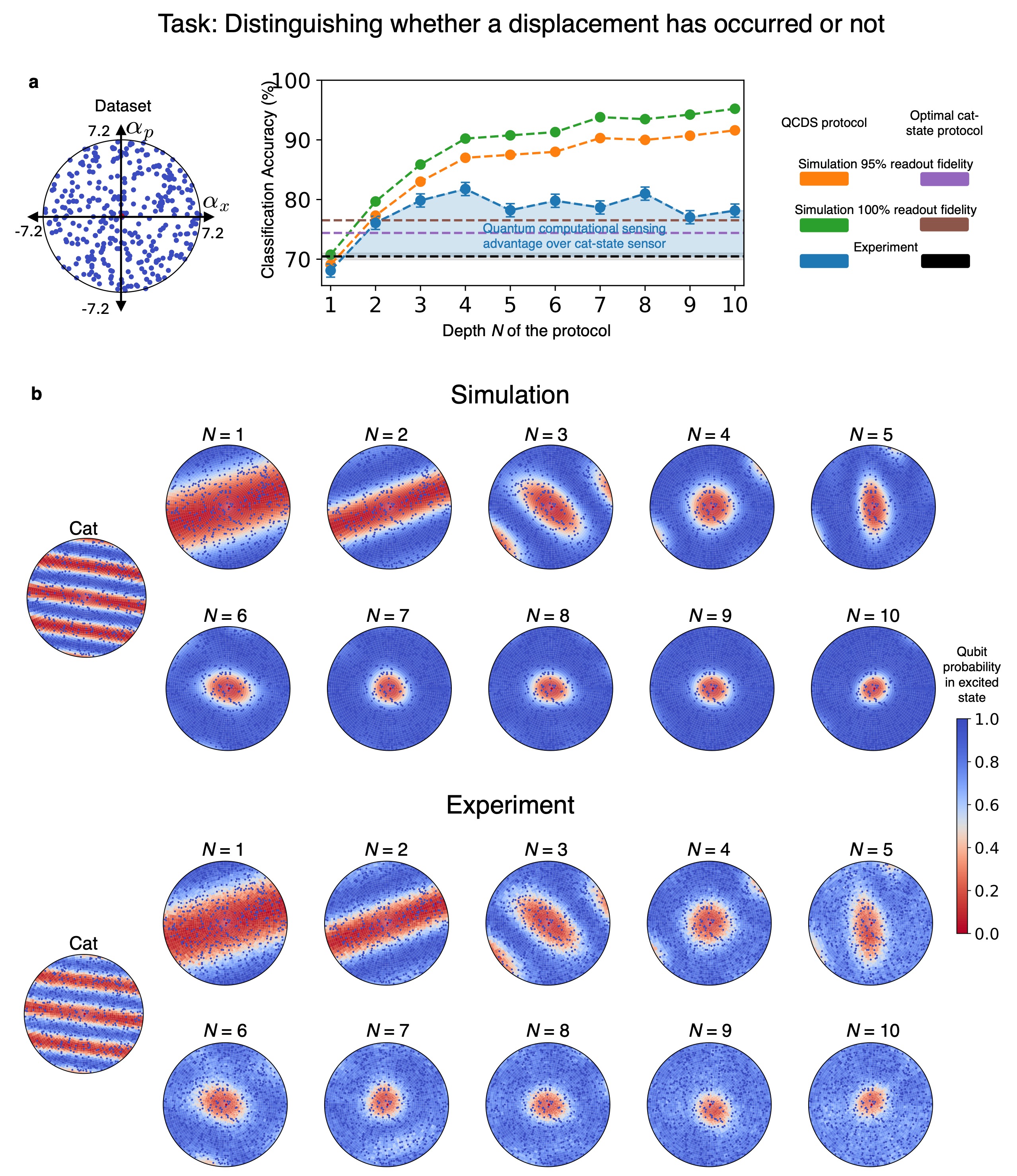}
    \caption{\textbf{Performance of the QCDS and cat-state protocols as a function of depth $N$ in distinguishing whether a displacement has occurred or not.} \textbf{a.} Dataset and classification accuracy of the simulation and experiment as a function of $N$. \textbf{b.} Simulated and experimental plots of the qubit excitation probability as a function of the sensed displacement.}
    \label{fig:zero_detail}
\end{figure}
\clearpage

\begin{figure}[h!]
    \centering
    \includegraphics[width=\textwidth]{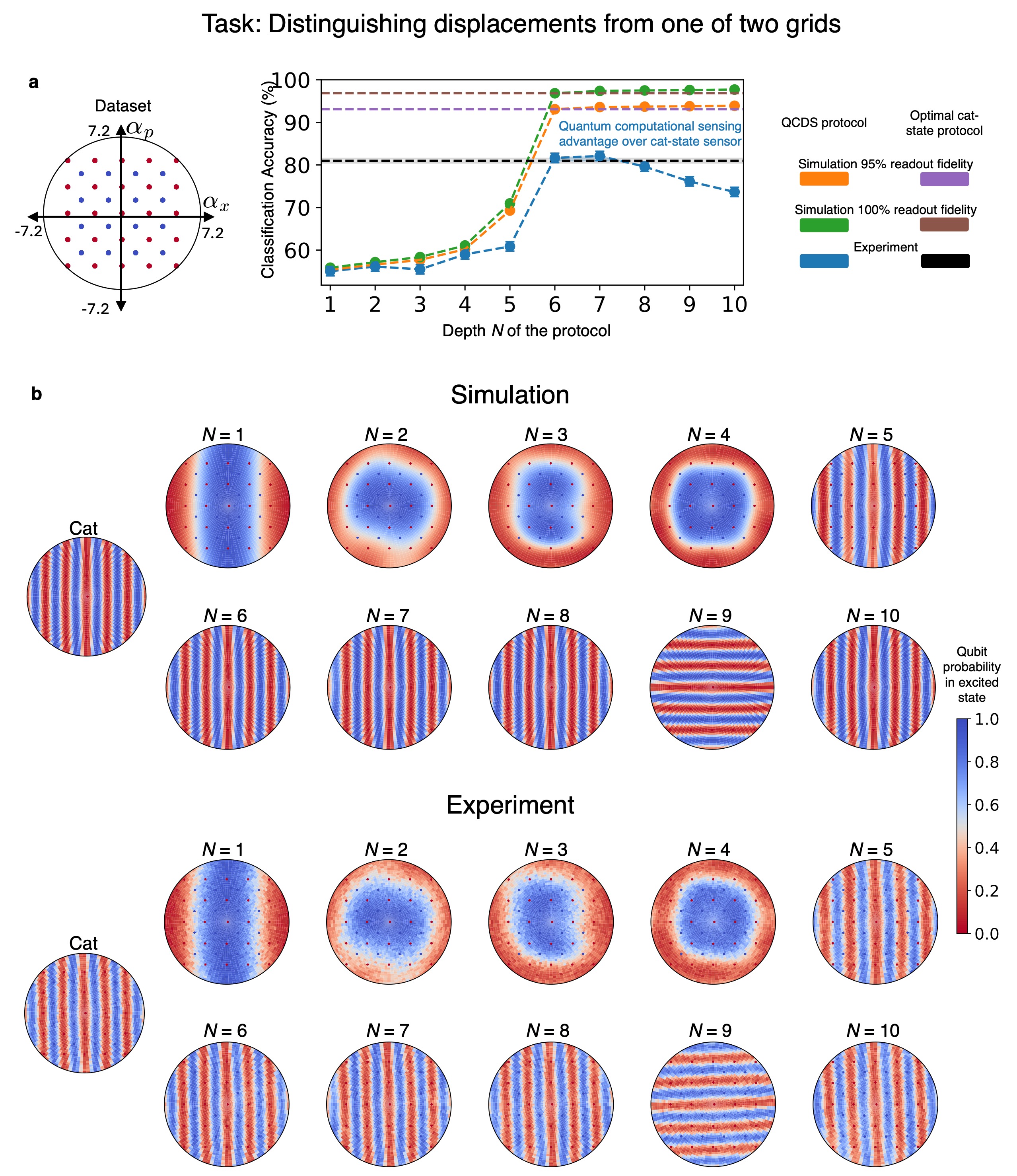}
    \caption{\textbf{Performance of the QCDS and cat-state protocols as a function of depth $N$ in distinguishing displacements from one of two distributions of a grid.} \textbf{a.} Dataset and classification accuracy of the simulation and experiment as a function of $N$. \textbf{b.} Simulated and experimental plots of the qubit excitation probability as a function of the sensed displacement.}
    \label{fig:checkerboard_detail}
\end{figure}
\clearpage

\begin{figure}[h!]
    \centering
    \includegraphics[width=\textwidth]{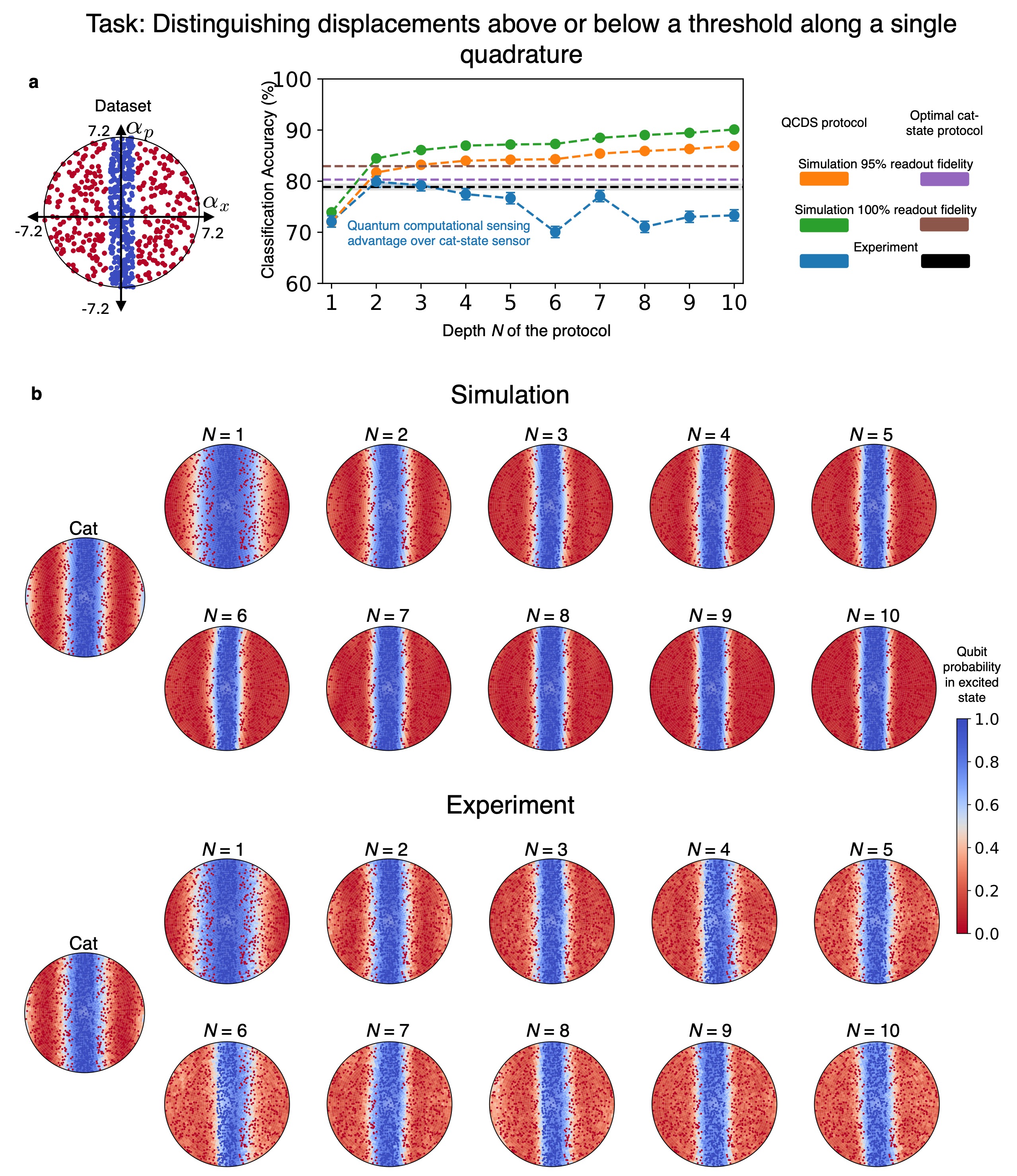}
    \caption{\textbf{Performance of the QCDS and cat-state protocols as a function of depth $N$ in distinguishing displacements above or below a threshold along the position quadrature.} \textbf{a.} Dataset and classification accuracy of the simulation and experiment as a function of $N$. \textbf{b.} Simulated and experimental plots of the qubit excitation probability as a function of the sensed displacement.}
    \label{fig:threshold_x_detail}
\end{figure}
\clearpage

\section{Constructing a trainable digital model}
\label{app:trainable_digital_model}

In this Appendix, we discuss the classical simulation of the protocol. We use this to optimize the parameters of the protocol for a particular classification task (see Appendix~\ref{app:training_procedure}). Hence, we aim to construct a faithful and efficient simulation of the experiment. Starting with the Hamiltonian in Eq.~(\ref{appeq:hsysq}) and setting $\hbar = 1$, it will prove convenient to express it in the basis of Pauli operators by using $\hat{\sigma}_z = 1-2\hat{q}^{\dagger}\hat{q}$. We also ignore higher-order nonlinear terms, which leaves us with the simplified Hamiltonian (in a rotating frame where bare qubit and storage cavity Hamiltonian terms can be removed):
\begin{align}
    \hat{\mathcal{H}} = \frac{\chi}{2}\hat{\sigma}_z\hat{a}^{\dagger}\hat{a} + \Omega(t)(\hat{\sigma}_x + i \hat{\sigma_y}) + \varepsilon(t)\hat{a}^{\dagger} + h.c..
\end{align}
The QCSD protocol shown in Fig.~\ref{fig:main1} of the main text is composed of three operations: arbitrary qubit rotations, echoed conditional displacements, and the sensing displacement. We discuss each in the following subsections.

\subsection{Arbitrary qubit rotations}
\label{app:trainable_digital_model:qubitrot}

When performing arbitrary qubit rotations, we consider the cavity drive signals to be turned off ($\varepsilon=0$). We define the qubit drive waveform as:
\begin{align}
    \Omega(t) = \frac{A \pi}{2} e^{i\phi}h_{\Omega}(t),
\end{align}
where $A$ defines the amplitude of the pulse, $\phi$ its phase, and $h_{\Omega}(t)$ its temporal waveform (see Appendix~\ref{app:quantum_computational_sensing_protocol:pulse_sequence}). For a pulse of length $t_p$, the pulse waveform is normalized such that $\int_{0}^{t_p} d\tau~h_{\Omega}(\tau) = 1$. The unitary operation defining arbitrary qubit rotations is therefore given by the time-ordered exponential:
\begin{align}
    {R}(\theta,\phi) = \mathcal{T}\exp\left\{ -i\int_{0}^{+t_p} d\tau~\frac{\chi}{2} \hat{\sigma}_z\hat{a}^{\dagger}\hat{a} + \frac{\theta}{2}h_{\Omega}(\tau)\left( \cos \phi~\hat{\sigma}_x + \sin\phi~\hat{\sigma}_y \right) \right\},
\end{align}
where $\theta=A\pi$. The normalization is chosen such that an amplitude of $A=1$ yields a qubit $\pi$-pulse.

To avoid having to compute this time-ordered exponential exactly, we use a first-order Trotter-Suzuki decomposition of the evolution over $N_t$ timesteps, 
\begin{align}
    {R}(\theta,\phi) \approx \prod_{j=1}^{N_t} \exp \left\{-i\frac{\chi}{2}\hat{\sigma}_z\hat{a}^{\dagger}\hat{a}\Delta_t \right\} \exp \left\{-i\frac{\theta}{2}h_{\Omega}(j\Delta t)\left( \cos \phi~\hat{\sigma}_x + \sin\phi~\hat{\sigma}_y \right)\Delta_t \right\}
\end{align}
where $N_t = \lfloor t_p/\Delta_t \rfloor$. We have found that a relatively small value of $N_t$ (around $5$) suffices for the Trotterized dynamics to converge to the exact evolution, due to the small value of the bare interaction strength.

\subsection{Echoed conditional displacement (ECD) gate}
\label{app:trainable_digital_model:ecd}

The ECD gate is composed of three elementary operations. The first operation applies a coherent drive $\varepsilon(t)$ to the oscillator, whose action can be denoted as a unitary operation $U_{\varepsilon}$. This is followed by a $\pi$-pulse on the qubit defined as $R_{\pi} \equiv R(\theta=\pi,\phi=0)$ (see Appendix~\ref{app:trainable_digital_model:qubitrot}). Lastly, we apply a second coherent drive $\varepsilon'(t)$ to the oscillator, described by $U_{\varepsilon'}$. In the rest of this Appendix subsection, we will show that this sequence of operations realizes a conditional displacement,
\begin{align}
    {\rm ECD}(\beta) = e^{-i\phi_g'-i\phi_e}\hat{D}(\alpha_g(t_f))\hat{D}^{\dagger}(\alpha_e(t_0))|g\rangle \langle e| + e^{-i\phi_e'-i\phi_g}\hat{D}(\alpha_e(t_f)) \hat{D}^{\dagger}(\alpha_g(t_0)) |e\rangle \langle g|,
    \label{appeq:ecd}
\end{align}
with a conditional displacement amplitude $\beta_j$ that depends on $(\varepsilon,\varepsilon')$ (similar to the protocol in Ref.~\cite{eickbusch2022fast}).

\subsubsection{Analyzing dynamics under oscillator drive using a conditional displaced frame}

The Hamiltonian governing evolution of the qubit-oscillator system during the application of an arbitrary coherent drive $\varepsilon(t)$ is given by:
\begin{align}
    \hat{\mathcal{H}}_{\varepsilon}(t) = \frac{\chi}{2} \hat{a}^\dagger \hat{a} \hat{\sigma}_z + \varepsilon^*(t) \hat{a} + \varepsilon(t) \hat{a}^\dagger.
    \label{appeq:heps}
\end{align}

To simplify the evolution of the quantum computational sensor under $\hat{\mathcal{H}}_{\epsilon}(t)$, it will prove convenient to choose a conditional displaced frame that is able to account exactly for the contribution to dynamics due to the (large) drive. We introduce the unitary conditional displacement operator:

\begin{align}
\hat{\Pi} = D(\alpha_g) |g\rangle \langle g| + D(\alpha_e) |e\rangle \langle e|.
\end{align}

States transformed under this operator are given by: $|\tilde{\psi}\rangle = \hat{\Pi}^\dagger|\psi\rangle$. Our task is now to derive the evolution equation, namely the displaced Schr\"odinger's equation, for states in this transformed frame:
\begin{align}
    \frac{d}{dt} |\tilde{\psi}\rangle = \frac{d}{dt} (\hat{\Pi}^\dagger |\psi\rangle) =  \dot{\hat{\Pi}}^\dagger |\psi\rangle + \hat{\Pi}^\dagger \frac{d}{dt} |\psi\rangle.
\end{align}
Introducing the transformed state in the first term, and using Schr\"odinger's equation in the original frame for the second term, we find:
\begin{align}
    \frac{d}{dt} |\tilde{\psi}\rangle &= \dot{\hat{\Pi}}^\dagger\hat{\Pi} |\tilde{\psi}\rangle + \hat{\Pi}^\dagger (-i \hat{\mathcal{H}} |\psi\rangle) \nonumber \\
    &= \dot{\hat{\Pi}}^\dagger\hat{\Pi} |\tilde{\psi}\rangle + ( -i\hat{\Pi}^\dagger \hat{\mathcal{H}}\hat{\Pi} ) |\tilde{\psi}\rangle \nonumber \\
    \implies \frac{d}{dt} |\tilde{\psi}\rangle &= -i\underbrace{\hat{\Pi}^\dagger \hat{\mathcal{H}}\hat{\Pi}}_{\hat{\tilde{\mathcal{H}}}} |\tilde{\psi}\rangle + \dot{\hat{\Pi}}^\dagger\hat{\Pi} |\tilde{\psi}\rangle
    \label{appeq:sch1}
\end{align}
where we have introduced the displacement-transformed Hamiltonian $\hat{\tilde{\mathcal{H}}}$, as well as the dynamical term that arises due to the time-dependent nature of the transformation. To simplify the above Schr\"odinger's equation in the displaced frame, we start with the second term, and note that:

\begin{align}
    \frac{d}{dt} {D}(\alpha) = \left[ \dot{\alpha} \hat{a}^\dagger -  \dot{\alpha}^* \hat{a} + \frac{1}{2} \dot{\alpha}^* \alpha - \frac{1}{2} \dot{\alpha}\alpha^* \right] D(\alpha),
\end{align}
where $\dot{\alpha} \equiv \frac{d\alpha}{dt}$. Therefore, we obtain:

\begin{align}
    \dot{\hat{\Pi}}^\dagger\hat{\Pi} = \left[ \dot{\alpha}_g^* \hat{a} - \dot{\alpha}_g \hat{a}^\dagger + \frac{1}{2} \dot{\alpha}_g^* \alpha_g - \frac{1}{2} \dot{\alpha}_g\alpha_g^* \right] |g\rangle \langle g| + \left[ \dot{\alpha}_e^* \hat{a} - \dot{\alpha}_e \hat{a}^\dagger + \frac{1}{2} \dot{\alpha}_e^* \alpha_e - \frac{1}{2} \dot{\alpha}_e\alpha_e^* \right] |e\rangle \langle e|,
\end{align}
where the displacement operator contributions cancel. The other term we must simplify in Eq.~(\ref{appeq:sch1}) is the displaced frame Hamiltonian $\hat{\tilde{\mathcal{H}}}$. Suppressing time labels for clarity, we note that: 

\begin{align}
\hat{\tilde{\mathcal{H}}} = \frac{\chi}{2} \left(\hat{\Pi}^\dagger \hat{a}^\dagger \hat{\Pi}\right)\left(\hat{\Pi}^\dagger \hat{a}\hat{\Pi}\right)\left( \hat{\Pi}^\dagger \hat{\sigma}_z \hat{\Pi}\right) + \varepsilon^* \left(\hat{\Pi}^\dagger \hat{a}\hat{\Pi}\right) + \varepsilon \left(\hat{\Pi}^\dagger \hat{a}^\dagger \hat{\Pi}\right),
\end{align}
where we have made use of the fact that the conditional displacement operator is a unitary operator. To proceed, we must calculate the displacement-transformed versions of the operators above: the three terms in round brackets, which eventually gives:

\begin{align}
    \hat{\tilde{\mathcal{H}}} =  ~~& \left[ \frac{\chi}{2}\sigma_z \left( a^\dagger a + \alpha_g a^\dagger + \alpha_g^* a + |\alpha_g|^2 \right) + \varepsilon^*(a + \alpha_g) + \varepsilon (a^\dagger + \alpha_g^*) \right] |g\rangle \langle g| \nonumber \\
    ~+&\left[   \frac{\chi}{2}\sigma_z \left( a^\dagger a + \alpha_e^* a^\dagger + \alpha_e a + |\alpha_e|^2 \right) + \epsilon^*(a + \alpha_e) + \epsilon (a^\dagger + \alpha_e^*) \right] |e\rangle \langle e|.
\end{align}

These results allow us to simplify Schr\"odinger's equation in the displaced frame, Eq.~(\ref{appeq:sch1}):

\begin{align}
    \frac{d}{dt}|\tilde{\psi}\rangle &= -i \left\{ \left[+\frac{\chi}{2} \hat{a}^\dagger \hat{a} + \left[\hat{a}^{\dagger}\left( +\frac{\chi}{2} \alpha_g + \varepsilon -i\dot{\alpha}_g \right) + h.c.\right] + \frac{\chi}{2}|\alpha_g|^2 +  \varepsilon^*\alpha_g + \varepsilon\alpha_g^* + \frac{i}{2} \big(\dot{\alpha}_g^* \alpha_g -  \dot{\alpha}_g\alpha_g^* \big) \right] |g\rangle \langle g|  \right\}|\tilde{\psi}\rangle \nonumber \\ 
    &~~~-i\left\{ \left[-\frac{\chi}{2} \hat{a}^\dagger \hat{a} + \left[\hat{a}^{\dagger}\left( -\frac{\chi}{2} \alpha_e + \varepsilon -i\dot{\alpha}_e \right) + h.c.\right] - \frac{\chi}{2}|\alpha_e|^2 +  \varepsilon^*\alpha_e + \varepsilon\alpha_e^* + \frac{i}{2} \big(\dot{\alpha}_e^* \alpha_e -  \dot{\alpha}_e\alpha_e^* \big) \right] |e\rangle \langle e|\right\}|\tilde{\psi}\rangle,
    \label{appeq:sch2}
\end{align}
where we have collected terms and have used $\hat{\sigma}_z|g\rangle \langle g| = |g\rangle \langle g|$ and $\hat{\sigma}_z|e\rangle \langle e| = -|e\rangle \langle e|$. As is standard, we can now choose the evolution of the displacement parameters $\alpha_g(t),\alpha_e(t)$ to cancel any terms in the Hamiltonian that are linear in the storage cavity creation and annihilation operators, thereby accounting exactly for the displacement due to the drive. This leads to the set of equations of motion:
\begin{subequations}
\begin{align}
    \dot{\alpha}_g &= -i \frac{\chi}{2} \alpha_g - i \varepsilon, \label{appeq:odesG} \\
    \dot{\alpha}_e &= +i \frac{\chi}{2} \alpha_e - i \varepsilon.
    \label{appeq:odesE}
\end{align}
\end{subequations}
To solve these equations, we additionally require a set of initial conditions determined by the initial state $\ket{\psi(t_0)}$, as we will discuss shortly. Before doing so, we can use the constraints imposed by Eqs.~(\ref{appeq:odesG}),~(\ref{appeq:odesE}), substituting the above into Eq.~(\ref{appeq:sch2}), which has two effects. First, this cancel terms linear in the creation and annihilation operators (implying that displacement terms are accounted for), and secondly the remaining time derivative terms are simplified. This simplifies to

\begin{align}
    \frac{d}{dt}|\tilde{\psi}\rangle &= -i \left\{ \left[+\frac{\chi}{2} \hat{a}^\dagger \hat{a} + \frac{1}{2}\varepsilon^*\alpha_g + \frac{1}{2}\varepsilon\alpha_g^* \right] |g\rangle \langle g|  + \left[-\frac{\chi}{2} \hat{a}^\dagger \hat{a} +  \frac{1}{2}\varepsilon^*\alpha_e + \frac{1}{2}\varepsilon\alpha_e^* \right] |e\rangle \langle e|\right\}|\tilde{\psi}\rangle.
    \label{appeq:sch}
\end{align}
The remaining interactions are a dispersive interaction in this frame, whose role can be echoed out in the full ECD implementation, and a drive-induced state-dependent phase. We can now integrate Schr\"odinger's equation to solve for the evolution of the displaced frame state of the qubit-cavity system, 
\begin{align}
    |\tilde{\psi}(t)\rangle = \left[ e^{-i\tau\frac{\chi }{2}\hat{a}^{\dagger}\hat{a}}e^{-i\phi_g}|g\rangle \langle g| + e^{+i\tau\frac{\chi}{2}\hat{a}^{\dagger}\hat{a}}e^{-i\phi_e}|e\rangle \langle e| \right]|\tilde{\psi}(t_0)\rangle 
\end{align}
where we have introduced the drive-induced phases:
\begin{align}
    \phi_j(t) &= \int_{t_0}^t d\tau~\left( \frac{1}{2}\varepsilon^*\alpha_j + \frac{1}{2}\varepsilon\alpha_j^* \right),~~~~~~j\in \{g,e\}.
    \label{appeq:drivePhase}
\end{align}
Finally, we can return to the lab frame, by undoing the conditional displacement transformation:

\begin{align}
    |\psi(t)\rangle =  \left[ \hat{D}(\alpha_g(t)) e^{-i\tau\frac{\chi }{2}\hat{a}^{\dagger}\hat{a}}e^{-i\phi_g}\hat{D}^{\dagger}(\alpha_g(t_0)) |g\rangle \langle g| + \hat{D}(\alpha_e(t)) e^{+i\tau\frac{\chi}{2}\hat{a}^{\dagger}\hat{a}}e^{-i\phi_e}\hat{D}^{\dagger}(\alpha_e(t_0))|e\rangle \langle e| \right]  |\psi(t_0)\rangle.
    \label{appeq:schsol}
\end{align}

\subsubsection{Initial conditions for displaced frame evolution}

Writing down Eq.~(\ref{appeq:schsol}) requires knowledge of the displacement trajectories $\alpha_j(t),~j \in \{g,e\}$, obtained by integrating  Eqs.~(\ref{appeq:odesG}),~(\ref{appeq:odesE}). To do so, we require initial conditions. The aim of the conditional displacement transformation is to center the quantum state to have vanishing quadrature expectation values for each qubit manifold; formally, we demand that the transformed state $|\tilde{\psi}(t_0)\rangle$ satisfies:
\begin{align}
    \langle \tilde{\psi}(t_0) | (\hat{a} |g\rangle \langle g|) |\tilde{\psi}(t_0)\rangle = \langle \tilde{\psi}(t_0) | (\hat{a} |e\rangle \langle e|) |\tilde{\psi}(t_0)\rangle = 0 .
    \label{appeq:initialDisp}
\end{align}
Evaluating this condition for the $\ket{g}$ sector first, we find (suppressing time labels for clarity):

\begin{align}
    \langle \tilde{\psi} | (\hat{a} |g\rangle \langle g|) |\tilde{\psi}\rangle = 0 
    &= \langle \psi | \left( \hat{D}(\alpha_g) \hat{a} \hat{D}^\dagger (\alpha_g) |g\rangle \langle g| \right) | \psi \rangle \nonumber \\
    &= \langle \psi | \left( \hat{a} |g\rangle \langle g| \right) | \psi \rangle - \alpha_g \langle \psi |g\rangle \langle g| \psi \rangle,
\end{align}
where we have used the standard action of the displacement operator: $\hat{D}(\alpha)\hat{a}\hat{D}^{\dagger}(\alpha) = \hat{a}-\alpha$. The above yields (restoring time labels):
\begin{align}
    \alpha_g(t_0) = \frac{\langle \psi(t_0) | (\hat{a} |g\rangle \langle g|) | \psi(t_0) \rangle}{|\langle g | \psi(t_0) \rangle|^2}.
\end{align}
We can analogously obtain: 
\begin{align}
    \alpha_e(t_0) = \frac{\langle \psi(t_0) | \left(\hat{a}|e\rangle \langle e|\right) | \psi(t_0) \rangle}{|\langle e | \psi(t_0) \rangle|^2},
\end{align}
which together specify the initial conditions required to evolve Eqs.~(\ref{appeq:odesG}),~(\ref{appeq:odesE}).

\subsubsection{Implementing an echoed conditional displacement}

Having described the evolution of the quantum computational sensor under a single coherent cavity drive, we proceed to describe the echoed conditional displacement (ECD) gate, which is composed of two cavity drives with a qubit $\pi$-pulse applied in between. We define the ECD gate over a time $t \in [t_0,t_f]$ as being composed of the following three operations:
\begin{enumerate}
    \item From $t_0$ to $t_m^-$ a storage drive $\varepsilon(t)$ is applied.
    \item From $t_m^-$ to $t_m^+$ a qubit $\pi$-pulse is applied.
    \item Finally, from $t_m^+$ to $t_f$ a second storage drive $\varepsilon'(t)$ is applied. We typically consider $\varepsilon'(t) = -\varepsilon(t)$.
\end{enumerate}

For the first storage cavity drive, the evolution is given by Eq.~(\ref{appeq:schsol}), reproduced below:
\begin{align}
    |\psi(t_m^-)\rangle =  U_{\varepsilon}|\psi(t_0)\rangle = -i \left[ \hat{D}(\alpha_g(t_m^-)) e^{-i\tau\frac{\chi }{2}\hat{a}^{\dagger}\hat{a}}e^{-i\phi_g}\hat{D}^{\dagger}(\alpha_g(t_0)) |g\rangle \langle g| + \hat{D}(\alpha_e(t_m^-)) e^{+i\tau\frac{\chi}{2}\hat{a}^{\dagger}\hat{a}}e^{-i\phi_e}\hat{D}^{\dagger}(\alpha_e(t_0))|e\rangle \langle e| \right]  |\psi(t_0)\rangle.
\end{align}
Upon completion of the first storage drive, a qubit $\pi$-pulse is applied. Using the definition of arbitrary qubit rotations $R(\theta,\phi)$ from Appendix~\ref{app:trainable_digital_model:qubitrot}, we define the qubit $\pi$ pulse required for the ECD gate by setting $A=1$. In the limit of a short pulse time $t_p$ ($=t_m^+-t_m^-$), $R_{\pi} \simeq -i\hat{\sigma}_x$. The state after the first conditional displacement operation and the qubit $\pi$-pulse is therefore approximately given by:
\begin{align}
    |\psi(t_m^+)\rangle =  R_{\pi}|\psi(t_m^-)\rangle = -i \left[ \hat{D}(\alpha_g(t_m^-)) e^{-i\tau\frac{\chi }{2}\hat{a}^{\dagger}\hat{a}}e^{-i\phi_g}\hat{D}^{\dagger}(\alpha_g(t_0)) |e\rangle \langle g| + \hat{D}(\alpha_e(t_m^-)) e^{+i\tau\frac{\chi}{2}\hat{a}^{\dagger}\hat{a}}e^{-i\phi_e}\hat{D}^{\dagger}(\alpha_e(t_0))|g\rangle \langle e| \right]  |\psi(t_0)\rangle.
    \label{appeq:statetmp}
\end{align}
The ECD gate is completed by applying the second conditional displacement operation under drive $\varepsilon'(t)$. Starting from the state at the conclusion of the preceding qubit $\pi$-pulse, the state at the end of the second storage drive is again given by Eq.~(\ref{appeq:schsol}):
\begin{align}
    |\psi(t_f)\rangle =  \left[ \hat{D}(\alpha_g(t_f)) e^{-i\tau\frac{\chi }{2}\hat{a}^{\dagger}\hat{a}}e^{-i\phi_g'}\hat{D}^{\dagger}(\alpha_g(t_m^+)) |g\rangle \langle g| + \hat{D}(\alpha_e(t_f)) e^{+i\tau\frac{\chi}{2}\hat{a}^{\dagger}\hat{a}}e^{-i\phi_e'}\hat{D}^{\dagger}(\alpha_e(t_m^+))|e\rangle \langle e| \right]  |\psi(t_m^+)\rangle.
    \label{appeq:statetf}
\end{align}
where the displacements $\alpha_g(t),\alpha_e(t)$ for $t\in[t_m^+,t_f]$ are governed by Eqs.~(\ref{appeq:odesG}),~(\ref{appeq:odesE}) respectively now with the drive $\varepsilon'(t)$ and with initial conditions $\alpha_g(t_m^+),\alpha_e(t_m^+)$ determined by $|\psi(t_m^+)\rangle$. Likewise, the phase parameters $\phi_g',\phi_e'$ are defined by Eqs.~(\ref{appeq:drivePhase}) with drive $\varepsilon'(t)$ and the aforementioned displacement trajectories $\alpha_g(t),\alpha_e(t)$ for $t\in[t_m^+,t_f]$. To determine the action of the composite ECD gate, we can combine Eq.~(\ref{appeq:statetmp}) and Eq.~(\ref{appeq:statetf}). To do so, we first compute the displacements $\alpha_g(t_m^+)$ and $\alpha_e(t_m^+)$. The first term is given by:

\begin{align}
    \alpha_g(t_m^+) &= \frac{\langle \psi(t_m^+) | (\hat{a} |g\rangle \langle g|) | \psi(t_m^+) \rangle}{|\langle g | \psi(t_m^+) \rangle|^2} \nonumber \\
    &= \frac{ \langle\tilde{\psi}(t_0)|\left(\hat{a}e^{-i\frac{\chi}{2}\tau}+\alpha_e(t_m^-)\right) |e\rangle \langle e|| \tilde{\psi}(t_0) \rangle}{ |\langle e | \psi(t_0) \rangle|^2 } \nonumber \\
    \implies \alpha_g(t_m^+) &= \alpha_e(t_m^-).
    \label{appeq:dispgpi}
\end{align}
An analogous calculation yields:
\begin{align}
    \alpha_e(t_m^+) = \alpha_g(t_m^-).
    \label{appeq:dispepi}
\end{align}
Eqs.~(\ref{appeq:dispgpi}),~(\ref{appeq:dispgpi}) follow from the fact that the qubit $\pi$-pulse maps $\ket{g} \leftrightarrow \ket{e}$, so the qubit-state-conditioned displacement before and after the pulse is similarly swapped. Using these results, we can finally combine Eq.~(\ref{appeq:statetmp}) and Eq.~(\ref{appeq:statetf}) to obtain the final state after the full ECD gate:

\begin{align}
    |\psi(t_f)\rangle = -i&\left[ \hat{D}(\alpha_g(t_f))e^{-i\phi_g'-i\phi_e}\hat{D}^{\dagger}(\alpha_e(t_0))|g\rangle \langle e| + \hat{D}(\alpha_e(t_f)) e^{-i\phi_e'-i\phi_g}\hat{D}^{\dagger}(\alpha_g(t_0)) |e\rangle \langle g| \right] |\psi(t_0)\rangle.
\end{align}

Note that the dispersive-shift-dependent term cancels out due to the echo. The residual phase that is accumulated is the drive-induced phase, which depends on both $\varepsilon,\varepsilon'$. Ignoring the irrelevant global phase, the full evolution  we have considered implements an operation that we can define as the ${\rm ECD}$ gate, via $|\psi(t_f)\rangle = {\rm ECD}|\psi(t_0)\rangle$, which takes the form:
\begin{align}
    {\rm ECD}(\varepsilon,\varepsilon') = e^{-i\phi_g'-i\phi_e}\hat{D}(\alpha_g(t_f))\hat{D}^{\dagger}(\alpha_e(t_0))|g\rangle \langle e| + e^{-i\phi_e'-i\phi_g}\hat{D}(\alpha_e(t_f)) \hat{D}^{\dagger}(\alpha_g(t_0)) |e\rangle \langle g|.
\end{align}

\subsubsection{Drive-induced displacement and accrued phases: analytic solutions}

The prior subsection shows that the conditional displacement accumulated during the {\rm ECD} gate is determined entirely by the trajectories $\alpha_{g,e}(t)$. For the case where the dynamics of these trajectories can be faithfully described by the linear ODEs in Eqs.~(\ref{appeq:odesG}),~(\ref{appeq:odesE}), we can formally integrate these equations to obtain the formal solutions:
\begin{subequations}
    \begin{align}
        \alpha_g(t) &= \alpha_g(0) e^{-i \frac{\chi}{2} t} - i \int_{0}^{t} d\tau \, \varepsilon(\tau) e^{-i \frac{\chi}{2} (t - \tau)}, \\
        \alpha_e(t) &= \alpha_e(0) e^{+i \frac{\chi}{2} t} - i \int_{0}^{t} d\tau \, \varepsilon(\tau) e^{+i \frac{\chi}{2} (t - \tau)}.
    \end{align}
\end{subequations}
For convenience we define $\varepsilon(\tau) = |\varepsilon|e^{i\phi_{\varepsilon}} f(\tau)$, where $f(\tau)$ is a real-valued temporal waveform, following which the solutions simplify to 
\begin{subequations}
    \begin{align}
        \alpha_g(t) &= \alpha_g(0) e^{-i \frac{\chi}{2} t} - i|\varepsilon|e^{i\phi_{\varepsilon}} \int_{0}^{t} d\tau \, f(\tau) e^{-i \frac{\chi}{2} (t - \tau)}, \\
        \alpha_e(t) &= \alpha_e(0) e^{+i \frac{\chi}{2} t} - i|\varepsilon|e^{i\phi_{\varepsilon}} \int_{0}^{t} d\tau \, f(\tau) e^{+i \frac{\chi}{2} (t - \tau)}.
    \end{align}
\end{subequations}
We note that the integral terms are independent of initial conditions. This allows us to write down the trajectories after time $t$ compactly as:
\begin{subequations}
    \begin{align}
        \alpha_g(t) &=  e^{-i \frac{\chi}{2} t} \left[ \alpha_g(0) - i|\varepsilon|e^{i\phi_{\varepsilon}} \mathcal{I}_1(t) \right] \label{appeq:trajAG}, \\
        \alpha_e(t) &= e^{+i \frac{\chi}{2} t} \left[ \alpha_e(0) - i|\varepsilon|e^{i\phi_{\varepsilon}} \mathcal{I}_1^*(t) \right]  \label{appeq:trajAE}.
    \end{align}
\end{subequations}
where we have introduced the integral of the drive waveform as:
\begin{align}
    \mathcal{I}_1(t) = \int_{0}^{t} d\tau \, f(\tau) e^{+i \frac{\chi}{2} \tau}
\end{align}

Having calculated the trajectories, we can now calculate the drive-induced phases given by Eq.~(\ref{appeq:drivePhase}). For concreteness, we consider the phase accrued conditioned on qubit state $\ket{g}$:

\begin{align}
    \phi_g(t) &= \int_{0}^t d\tau~\left( \frac{1}{2}\varepsilon^*\alpha_g + \frac{1}{2}\varepsilon\alpha_g^* \right) \nonumber \\
&= \frac{1}{2}|\varepsilon| e^{-i \phi_{\varepsilon}}\alpha_{g}(0)  \int_{0}^{t} d\tau \ f(\tau) e^{-i \frac{\chi}{2} \tau} - \frac{i}{2}|\varepsilon|^2 \int_{0}^{t} d\tau \ f(\tau)\mathcal{I}_1(\tau)e^{-i \frac{\chi}{2} \tau}  + c.c..
\end{align}
We note that the first integral term is simply $\mathcal{I}_1^*(t)$. The second term again depends only on the drive waveform and not on initial conditions, so we can define:
\begin{align}
    \mathcal{I}_2(t) = \int_{0}^{t} d\tau \ f(\tau)\mathcal{I}_1(\tau)e^{-i \frac{\chi}{2} \tau}  = \int_{0}^{t} d\tau \int_{0}^{\tau} d\tau' \ e^{-i\frac{\chi}{2}(\tau-\tau')} f(\tau)f(\tau').
\end{align}

From Eq.~(\ref{appeq:drivePhase}), the accrued phase conditioned on the qubit $\ket{e}$ state can be obtained from the above by making the substitution $g\to e$, $\chi \to -\chi$, with the latter being equivalent to the substitution $\mathcal{I}_1 \to \mathcal{I}_1^*, \mathcal{I}_2 \to \mathcal{I}_2^*$. We therefore obtain analytic forms of the accrued drive-induced phases:
 \begin{subequations}
    \begin{align}
    \phi_g(t) = \frac{|\varepsilon|}{2} \left( \alpha_g(0) \mathcal{I}_1^*(t)e^{-i\phi_{\varepsilon}} + \alpha_g^*(0) \mathcal{I}_1(t)e^{+i\phi_{\varepsilon}} \right) - \frac{i}{2}|\varepsilon|^2 \left( \mathcal{I}_2(t)-\mathcal{I}_2^*(t) \right) \label{appeq:drivePhaseAG} \\
    \phi_e(t) = \frac{|\varepsilon|}{2} \left( \alpha_e(0) \mathcal{I}_1(t)e^{-i\phi_{\varepsilon}} + \alpha_e^*(0) \mathcal{I}_1^*(t)e^{+i\phi_{\varepsilon}} \right) - \frac{i}{2}|\varepsilon|^2 \left( \mathcal{I}_2^*(t)-\mathcal{I}_2(t) \right)
    \label{appeq:drivePhaseAE}
\end{align}
\end{subequations}
Eqs.~(\ref{appeq:trajAG}),~(\ref{appeq:trajAE}) and  Eqs.~(\ref{appeq:drivePhaseAG}),~(\ref{appeq:drivePhaseAE}) provide analytic solutions for the drive-induced displacements and phases that we use in our digital model. For ${\rm ECD}$ gates with a fixed waveform $f(\tau)$, the integrals $\mathcal{I}_j$ need only be performed once, thereby circumventing the cost of integrating Eqs.~(\ref{appeq:odesG}),~(\ref{appeq:odesE}) for every displacement pulse of the ${\rm ECD}$ gate evolution in our digital model. 

\subsection{Sensing operation}

We implement the sensing displacement using a pulse $\varepsilon_s(t)$, of the pulse shape used in the calibration of the AWG amplitude to displacement (see Appendix~\ref{app:quantum_computational_sensing_protocol:pulse_sequence}), with the amplitude and phase determined by the dataset. Due to the analog nature of the pulse, this operation imparts a small conditional displacement, orthogonal to the direction of the unconditional displacement. In this regime, the trajectories $\alpha_{g,e}(t)$ can be approximated as:

\begin{subequations}
    \begin{align}
        \alpha_g(t) &= \alpha_g(t_0) e^{-i \frac{\chi}{2} t} - i \int_{t_0}^{t} d\tau \, \varepsilon_s(\tau) e^{-i \frac{\chi}{2} (t - \tau)} \approx \alpha_g(t_0) - i \int_{t_0}^{t} d\tau \, \varepsilon_s(\tau), \\
        \alpha_e(t) &= \alpha_e(t_0) e^{+i \frac{\chi}{2} t} - i \int_{t_0}^{t} d\tau \, \varepsilon_s(\tau) e^{+i \frac{\chi}{2} (t - \tau)} \approx \alpha_e(t_0) - i\int_{t_0}^{t} d\tau \, \varepsilon_s(\tau). 
    \end{align}
\end{subequations}
In this limit, we see that $\alpha_g(t)-\alpha_g(t_0) \approx \alpha_e(t)-\alpha(t_0) \equiv \alpha$, where
\begin{align}
    \alpha = - i \int_{t_0}^{t} d\tau \, \varepsilon_s(\tau).
\end{align}
Since the integral of the drive waveform is non-vanishing, $\alpha$ will be non-zero. This is not the case for the drive waveforms $\varepsilon(t)$ used in the ${\rm ECD}$ gate, by design: there, the aim is to impart conditional displacements only. The drive-induced phase also becomes equal for both states and subsequently becomes an overall phase that can be dropped. Under the assumption $\chi \tau \ll 1$, using Eq.~(\ref{appeq:schsol}) the result of the sensing operation can be approximately written as:
\begin{align}
    |\psi(t)\rangle &\approx  \left[ \hat{D}(\alpha_g(t)) \hat{D}^{\dagger}(\alpha_g(t_0)) |g\rangle \langle g| + \hat{D}(\alpha_e(t)) \hat{D}^{\dagger}(\alpha_e(t_0))|e\rangle \langle e| \right]  |\psi(t_0)\rangle \nonumber \\
    &\approx \left[\hat{D}(\alpha)|g\rangle \langle g| + \hat{D}(\alpha)|e\rangle \langle e| \right]|\psi(t_0)\rangle \approx \hat{D}(\alpha)|\psi(t_0)\rangle, 
\end{align}
which is equivalent to an approximate unconditional displacement. We emphasize again that we do not make this simplifying assumption in our simulation, and capture the conditional aspects of our sensing operation to reduce the simulation-reality gap.

\clearpage
\section{Training procedure}
\label{app:training_procedure}

In this section, we discuss the procedure for the training of the parameters of the QCDS protocol. The simulation, written using the PyTorch framework~\cite{paszke2019pytorch}, is run on GPUs. During the forward pass, we simulate the pulses using the description in Appendix~\ref{app:trainable_digital_model}. In the first epoch, for the ECD gate, we use the Euler method with 200 timesteps to simulate the semi-classical action of the oscillator-mode pulse, to compute the integrals described in Appendix~\ref{app:trainable_digital_model:ecd}. This speeds up the computation for all subsequent epochs, since we fix the magnitude of the conditional displacement (and only vary the phase). For the qubit rotation pulse (Appendix~\ref{app:trainable_digital_model:qubitrot}), we break the simulation into 5 timesteps, and directly compute the exponential of the Hamiltonian which includes both qubit drive term and the cross-Kerr interaction. While the effect of the cross-Kerr interaction is generally insignificant for a single pulse, the total effect for a large-$N$ circuit is non-negligible. For the action of the sensing operation, we use semi-classical simulation of the pulse, for all epochs of the simulation. Since this takes place once per simulation (rather than the multiple pulses of the ECD gate), this does not substantially increase the runtime of the entire simulation. Finally, the protocol concludes with computing the qubit probability in the excited state, from which the classification accuracy and loss function can be directly computed (see Appendix~\ref{app:quantum_computational_sensing_protocol}). For these simulations, we truncate the bosonic mode to the first $50$ Fock levels, sufficiently large to ensure that the truncation does not play a significant role.

The parameters of the simulation, such as the value of the cross-Kerr interaction and the scaling factor between the AWG amplitude and the displacement on the oscillator mode (see Appendix~\ref{app:system_hamiltonian_and_parameters}) are initially chosen to be that of the experiment. We then tune these values so that the simulation matches the experimental results of the cat-state sensing protocol illustrated in Fig.~\ref{fig:cat}a. The parameters to match the experiment are slightly different, within $5\%$ of the experimental values. We attribute this simulation-reality gap to the approximations made in our simulation. For instance, we do not take into account the semi-classical loss of the oscillator, and the higher order terms of the Hamiltonian of Eq.~(\ref{appeq:hsysq}). Such terms can play a role in the exact effect of the ECD gate, which was experimentally analyzed in Ref.~\cite{eickbusch2022fast}. We approximate them by a renormalization of the cross-Kerr and displacement scaling factor, which is captured by our calibration to the cat-state experiment.

\begin{figure}[h!]
    \centering
    \includegraphics[width=0.8\textwidth]{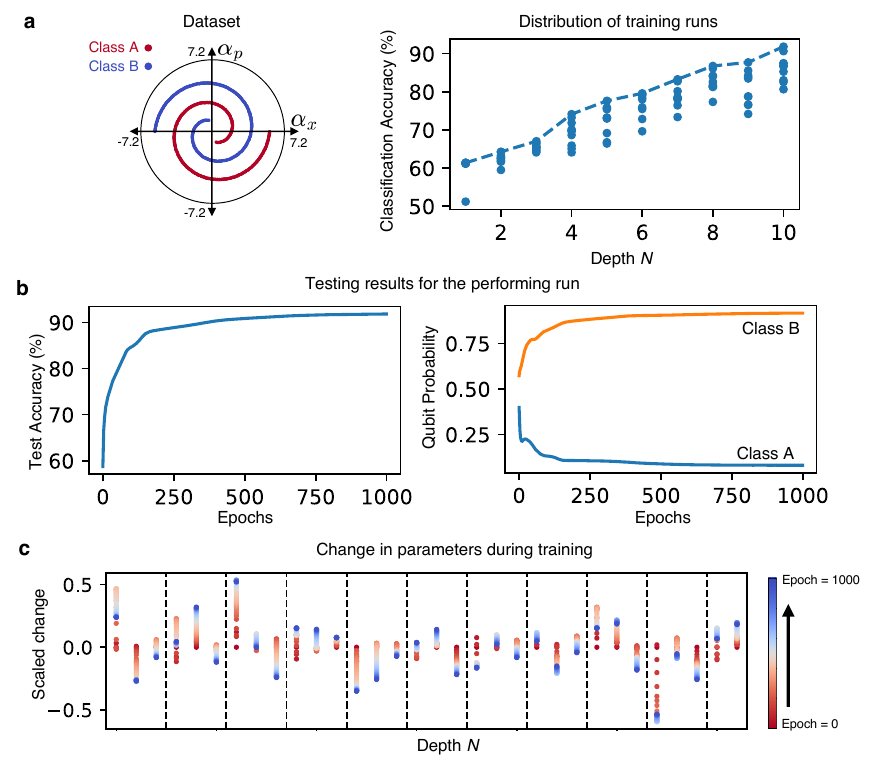}
    \caption{\textbf{Results of the training procedure in simulation.} \textbf{a.} Distribution of the final classification accuracy over different runs for a given depth $N$ for the dataset shown (equivalent to Fig.~\ref{fig:main3}c). Different runs, characterized by the randomly chosen initial parameters, give different final accuracy. We then choose the results of the best-performing run. \textbf{b.} Left: Classification accuracy Right: Average qubit probability in the excited state as a function of epochs. The training aims to push the two probabilities apart. \textbf{c.} Relative change in the parameters of the protocol over training. Each layer has three parameters, demarcated by the dashed lines. The scaled change represents the difference with respect to the initial value of the parameter chosen. The parameters associated with phases are scaled down by $2\pi$ to match the scale of the amplitude (which is between $0$ and $1$).}
    \label{fig:training}
\end{figure}

We randomly instantiate the parameters of the protocol. The phases of the qubit rotation pulse and echoed conditional displacement gate are chosen uniformly randomly from the entire interval: $[0,2\pi)$. The qubit rotation amplitude is chosen uniformly from the interval $[0,1]$, where an amplitude of $1$ performs a $\pi$ pulse. We clamp the value of the amplitude to be within this range during training. The parameters are updated using the conventional method of backpropagation, using an Adam optimizer with a learning rate of $5 \times 10^{-3}$. We train for $1000$ epochs for datasets of size $512$, and $250$ epochs for datasets of size $2048$. We generally notice the performance of the protocol plateaued by this point. The maximum classification accuracy varies from run to run, depending on the initial parameters. This spread generally tends to increase with circuit depth of the protocol $N$. We run $10$ instances for the tasks of Fig.~\ref{fig:main3}, and those in Appendix~\ref{app:quantum_computational_sensing_protocol:examples}, and select the best-performing protocol. For the tasks in Fig.~\ref{fig:main4}, we choose the best of $20$ runs. In Fig.~\ref{fig:training}, we plot the results of the training for the task considered in Fig.~\ref{fig:main3}c (see also Fig.~\ref{fig:training}a), and also show the variation between training runs from different parameter initializations. In Fig.~\ref{fig:training}b, we plot the simulation results for the best-performing training run for $N = 10$. In Fig.~\ref{fig:training}c, we plot the distribution of the parameters of the protocol over epochs.

During the training procedure across the several tasks considered in this work, we generally notice the following:
\begin{enumerate}
    \item The classification accuracy improves quickly at the start (for the first 100 epochs), and then slowly plateaus (for the last 900 epochs). This is characterized by the corresponding separation in the qubit excitation probability for the two classes. In some training runs, the slow plateaus in classification accuracy are interspersed by sudden jumps where the classification accuracy increases by several percentage points. This corresponds to the same behavior in the loss, since it is mathematically equivalent to the negative of the classification accuracy.

    \item The classification accuracy of the converged parameters varies across different realizations of the initial parameter. This spread increases with the depth $N$. For depths $N < 3$, there is hardly any variation in the classification accuracy across several runs. However, for the largest depths $N = 10$, the difference between the best- and worst- performing runs can be greater than $10\%$.

    \item There is a maximum learning rate beyond which the protocol does not train effectively. The value of $5 \times 10^{-3}$ lies well within the range for effective training while remaining sufficiently large to avoid excessively long training times.
\end{enumerate}

While we have not performed a systematic analysis (which is a potentially interesting avenue of future work), our observations hint at potential features of the loss landscape in the parameter space of the protocol. For instance, the variation in the converged performance for large $N$ protocols hints that the loss landscape might consist of many local minima, which the protocol might end up converging to. By running the training procedure several times with randomly initialized parameters, we improve the likelihood of converging to a minimum close to the global minimum. Furthermore, the two different scales of the variation of the loss hints that the landscape might consist of vast regions in parameter space with similar loss values. In such situations, the gradient of the loss with respect to the parameters can become small.

\clearpage
\section{Benchmarking QCDS against conventional quantum sensing protocols}
\label{app:benchmark}

\subsection{Overview}
\label{app:benchmark:overview}

In this Appendix section we provide details of various displacement sensing baselines considered. Table~\ref{table:baseline_overview} summarizes the benefits and drawbacks of these displacement sensing baselines, along with our QCDS protocol, when used for the task of binary classification of displacements. When the quantum computational displacement sensor achieves a better performance (quantified by the classification accuracy in our instance) over conventional sensors, we term that as quantum computational-sensing advantage (QCSA). Whether QCSA can be achieved, and how much, depends on the task. For our analysis, we primarily focus on the task introduced in Fig.~\ref{fig:main4}, regarding the classification of displacements originating from two arms of a spiral distribution. We sweep the winding number $W$, which quantifies the number of turns of the spirals of the task.

\begin{table}[h!]
    \centering  
    \setlength{\tabcolsep}{2pt}            
    \renewcommand{\arraystretch}{1.2}      

    \begin{tabular}{|l|l|l|}
    \hline
    \textbf{Sensor} & \textbf{Benefits} & \textbf{Drawbacks} \\ 
    \hline\hline 

    \begin{tabular}{@{}l@{}}
        Quantum computational\\
        displacement sensor
    \end{tabular} & 
    \begin{tabular}{@{}l@{}}
        1. Directly outputs the class prediction \\
        2. Performance increases with circuit \\
        ~~~depth $N$
    \end{tabular} &
    \begin{tabular}{@{}l@{}}
        1. Quantum circuit requires to be\\
        ~~~trained for each task \\
        2. Performance limited by qubit-readout \\
        ~~~fidelity
    \end{tabular}\\ 
    \hline
    \hline

    Cat-state sensor & 
    \begin{tabular}{@{}l@{}}
        1. Sensitive to small displacements \\
        2. Simple to implement in experiment
    \end{tabular} &
    \begin{tabular}{@{}l@{}}
        1. Sensitive to single-component of \\
        ~~~displacement \\
        2. Sensitivity -- dynamic range tradeoff \\
        3. Performance limited by qubit-readout \\
        ~~~fidelity
    \end{tabular}\\ 
    \hline

    Compass-state sensor & 
    \begin{tabular}{@{}l@{}}
        1. Sensitive to small displacements \\
        2. Sensitive to both components of \\
        ~~~displacement
    \end{tabular} &
    \begin{tabular}{@{}l@{}}
        1. Limited expressive capacity \\
        2. Performance limited by qubit-readout \\
        ~~~fidelity
    \end{tabular}\\ 
    \hline

    \begin{tabular}{@{}l@{}}
        Phase-preserving amplifier and\\
        classical postprocessing backend
    \end{tabular} &
    \begin{tabular}{@{}l@{}}
        1. High expressive capacity \\
        2. Sensitive to both components of \\
        ~~~displacement
    \end{tabular} &
    \begin{tabular}{@{}l@{}}
        1. Limited by the fundamental noise \\ 
        ~~~introduced by the amplifier\\
        2. Classical postprocessing \\
        ~~~required to be trained for each task
    \end{tabular}\\ 
    \hline

    \begin{tabular}{@{}l@{}}
        Squeezed probe-state,\\
        phase-sensitive amplifier and \\
        classical postprocessing backend
    \end{tabular} &
    \begin{tabular}{@{}l@{}}
        1. Sensitive to small displacements \\
        2. High expressive capacity of the \\
        ~~~single-component displacement
    \end{tabular} &
    \begin{tabular}{@{}l@{}}
        1. Sensitive to single-component of \\
        ~~~displacement \\
        2. Classical postprocessing \\
        ~~~required to be trained for each task
    \end{tabular}\\ 
    \hline

    \begin{tabular}{@{}l@{}}
        GKP probe-state,\\
        quantum phase estimation and \\
        classical postprocessing backend
    \end{tabular} &
    \begin{tabular}{@{}l@{}}
        1. High expressive capacity \\
        2. Sensitive to both components of \\
        ~~~displacement
    \end{tabular} &
    \begin{tabular}{@{}l@{}}
        1. Limited dynamic range \\
        2. Classical postprocessing \\
        ~~~required to be trained for each task
    \end{tabular}\\ 
    \hline

    \begin{tabular}{@{}l@{}}
        Two-mode squeezed probe-state,\\
        phase-sensitive amplifiers and \\
        classical postprocessing backend
    \end{tabular} &
    \begin{tabular}{@{}l@{}}
        1. High expressive capacity \\
        2. Performance increases with \\
        ~~~two-mode squeezing
    \end{tabular} &
    \begin{tabular}{@{}l@{}}
        1. Hard to implement in experiment \\
        ~~~with high efficiency \\
        2. Classical postprocessing \\
        ~~~required to be trained for each task
    \end{tabular}\\ 
    \hline

    \end{tabular}
    \caption{\textbf{Benefits and drawbacks of different protocols for classifying displacements.} The first row is the quantum computational displacement sensor introduced in this work, while the rest are examples of conventional displacement sensing protocols}
    \label{table:baseline_overview}
\end{table}

\begin{figure}[h!]
    \centering
    \includegraphics[width=\textwidth]{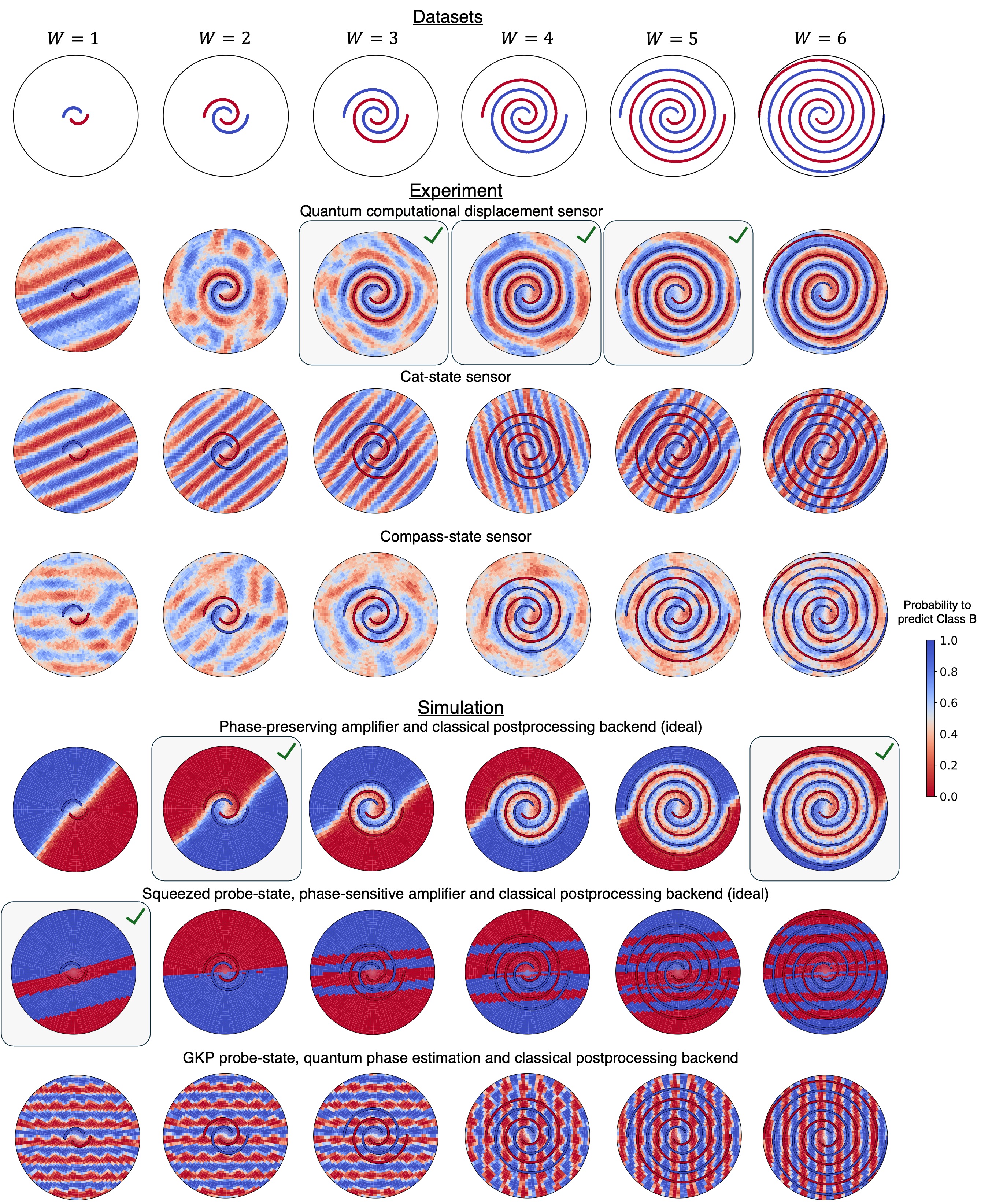}
    \caption{\textbf{Response of the various protocols as a function of displacement for the spirals task considered in Fig.~\ref{fig:main4}.} The response is defined as the probability for the protocol to predict class B given it received a displacement at that point. Highlighted check-marked boxes indicate the best performing protocol at that winding number (including comparing against ideal amplifier protocols).}
    \label{fig:spiralsN}
\end{figure}

In Fig.~\ref{fig:spiralsN}, we plot the response of various protocols for different values of the winding number. For clarity, we do not include the axis labels on the 2D plots of the function of the qubit-excitation probability. The scale is the same as that of Fig.~\ref{fig:main4}, with a range: $|\boldsymbol{\alpha}| < 8.7$. As the winding number increases, so does the task complexity. For each winding number, we highlight the protocol which achieves the highest classification accuracy. For three of the six winding numbers plotted, the quantum computational sensor achieves the best performance, outperforming even the protocols involving theoretically ideal amplifiers. The response is the probability of the protocol for prediction class B. When this value is either $0$ or $1$, that signifies that the protocol predicts the class label deterministically. When the value is not these values, it signifies the role of the quantum sampling noise.

For example, consider the protocol involving a phase-preserving amplifier to measure both components of the displacement (fourth row in Fig.~\ref{fig:spiralsN}). For displacements near the boundary of the two classes, quantum noise (equivalent to a total noise of a single photon in the ideal case) randomly produces different estimates for the same sensed displacement. Due to this noise, the output of the classical postprocessing backend is also inherently stochastic. Hence the probability to predict class B, for instance, takes on a value between $0$ and $1$.

On the other hand, some protocols do not introduce any quantum noise. For the case of the infinitely squeezed probe-state, along with ideal phase-sensitive amplifier, the output is deterministic for a given sensed displacement (see the fifth row in Fig.~\ref{fig:spiralsN}). Therefore the output of the classical postprocessing backend is also deterministic. However, since this protocol is only sensitive to a single-component, it misclassifies several displacement data points of the datasets. Finally, for the protocols which output a single-bit of information (which are protocols with a single-qubit measurement, and the ones performed in experiment), the probability to predict class B is equivalent to the probability of the qubit to be measured in the excited state.

The training of the parameters of the protocol (either that of the quantum circuit and/or the classical postprocessing) are performed with the goal to maximize the classification accuracy in a single shot. Therefore (and is visually apparent in Fig.~\ref{fig:spiralsN}), we forego the requirement that the estimate of $\mathcal{F}(\boldsymbol{\alpha})$ is unbiased for all value of $\boldsymbol{\alpha}$ in the dataset. Different scenarios and tasks (such as the many shot regime, estimation and so on) can be addressed by a corresponding choice in the loss function to minimize. To maximize performance for each protocol, we take into account the inherent stochasticity during training. For example, in the training of the classical postprocessing MLP for the phase-preserving amplifier protocol, we produce samples of the outcomes of the quantum measurement: $(\tilde{\alpha}_x, \tilde{\alpha}_p)$. By minimizing the loss function, the MLP learns to approximate the function $\mathcal{F}_{\rm QS}(\tilde{\alpha}_x, \tilde{\alpha}_p)$ to match as close as possible to $\mathcal{F}(\alpha_x, \alpha_p)$, where $(\alpha_x, \alpha_p)$ is the true displacement sensed by the quantum sensor. Importantly, the functional form of $\mathcal{F}_{\rm QS}$ will not in general match $\mathcal{F}$, but is rather whatever the function which optimizes accuracy~\cite{khan2025quantum_novel}.

In the next subsections, we discuss the simulation details of the implementation of these protocols. Furthermore, for protocols compatible with the quantum computational-sensing protocol, we also implement them in experiment. In the simulations of protocols incompatible with our protocol, we consider both the theoretical best-possible case and the expected result in experiments using state-of-the-art experiments. We train the classical postprocessor (along with any parameters of the quantum sensing protocol) for $5000$ epochs, by which the performance has saturated. For protocols with inherent stochasticity in the simulation, we average the accuracy over the last $100$ epochs to obtain an unbiased estimate of the classification accuracy.

\clearpage
\subsection{Quantum computational displacement sensor}
\label{app:benchmark:quantum_computational_sensor}

\begin{figure}[h!]
    \centering
    \includegraphics[width=\textwidth]{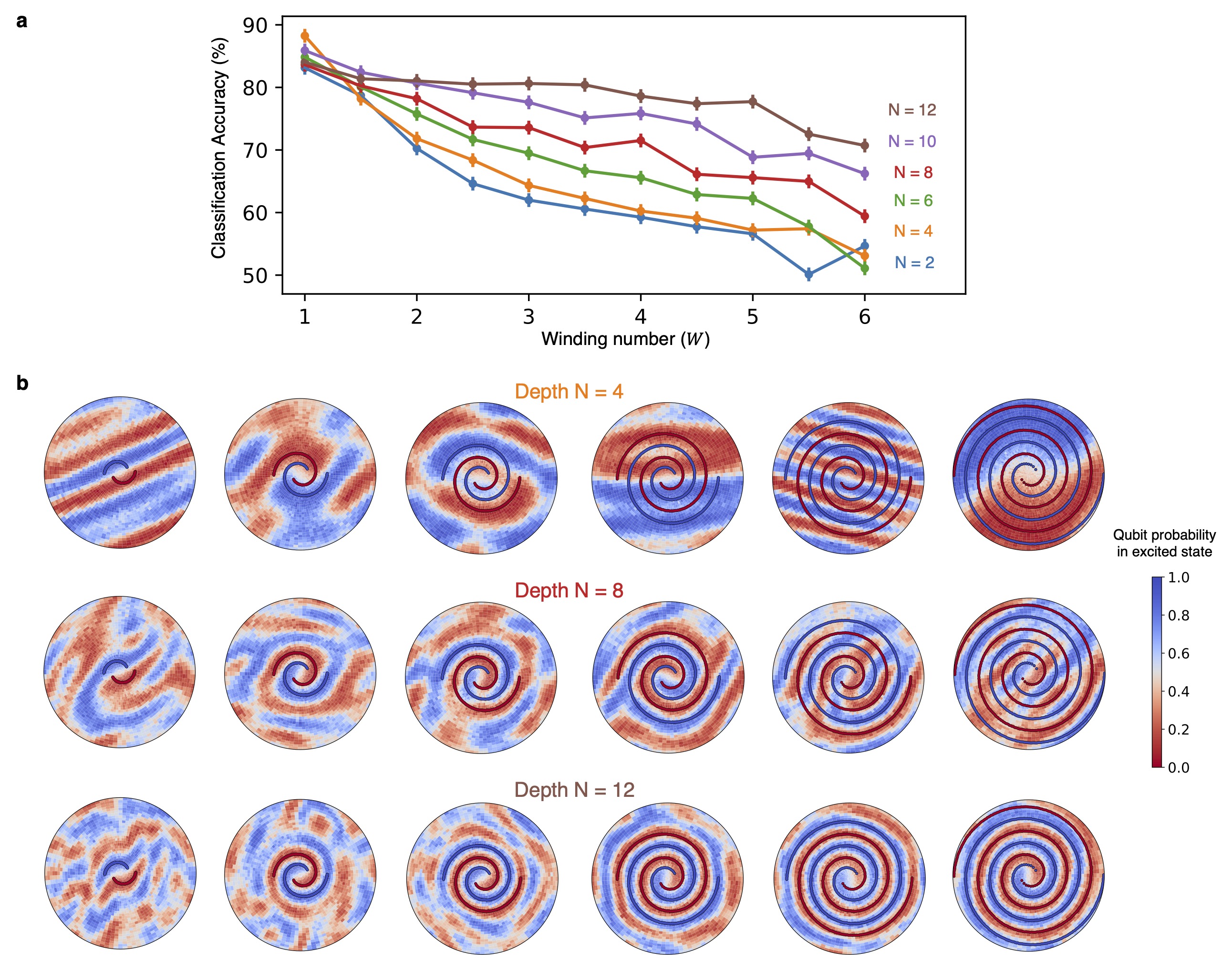}
    \caption{\textbf{Performance of the quantum computational sensor for the spirals task.} \textbf{a.} Experiment classification accuracy as a function of the winding number for different depths of the protocol. As the circuit depth increases, the performance tends to increase. \textbf{b.} The qubit-excitation probability as a function of the sensed displacement, showing how the ability of the protocol to effectively learn the distribution at higher depths.}
    \label{fig:spiralsN_qsp}
\end{figure}

In this subsection, we discuss the implementation of our quantum computational sensor for this task. In Fig.~\ref{fig:spiralsN_qsp}a, we plot the experimental classification accuracy of the protocols as a function of the circuit depth $N$. The best performance, across $N$, is plotted in Fig.~\ref{fig:main4}. For our experiment, this is usually the $N = 12$ protocol. However, for small winding numbers, lower depths performed better. This is because, as illustrated in Fig.~\ref{fig:spiralsN_qsp}b, the task-complexity is less for small winding numbers. In this situation, the additional expressive capacity of large $N$ protocols is not useful. Rather, the high-contrast of the shorter-time protocol of small $N$ perform better. On the other hand, for large winding numbers, only the largest depths achieve a reasonable classification accuracy. For each depth and winding number, the best-performing protocol out of $20$ simulations is chosen.

\clearpage
\subsection{Cat-state sensor}
\label{app:benchmark:cat_state_sensor}

\begin{figure}[h!]
    \centering
    \includegraphics[width=0.75\textwidth]{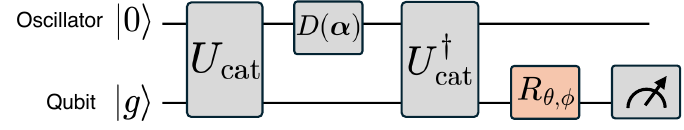}
    \caption{\textbf{Circuit diagram for the cat-state sensing protocol, which we demonstrate in experiment.}}
    \label{fig:circuit_cat}
\end{figure}

The cat-state sensing protocol measures a single component of the sensed displacement by mapping it onto the qubit phase~\cite{sinanan2024single}. This protocol can be realized for a $N = 1$ version of the QCDS protocol. In general, this proceeds by preparing the qubit in the $\ket{+}$ state. Then, an echoed conditional displacement gate entangles the qubit with the cavity. Therefore, $U_{\rm cat} = R(\theta=\pi, \phi=0)\rm{ECD}(\boldsymbol{\beta})$. This state is sensitive to displacements orthogonal to the axis of the cat-state. This arises from the geometric phase from the action of two displacement operators, which is proportional to the cross-product of the vectors of the displacement. We then implement an echoed conditional displacement in the opposite axis, which disentangles the qubit from the oscillator state. This imparts a phase on the qubit, proportional to the size of the cat $\beta$, and the orthogonal sensing displacement $\alpha_{\perp}$ (associated with the area enclosed in the phase space of the oscillator). The cat-state protocol therefore maps a displacement sensing task to a phase sensing Ramsey task. A final qubit rotation allows the measurement to be maximally sensitive to the phase. The circuit diagram is illustrated in Fig.~\ref{fig:circuit_cat}.

\begin{figure}[h!]
    \centering
    \includegraphics[width=0.8\textwidth]{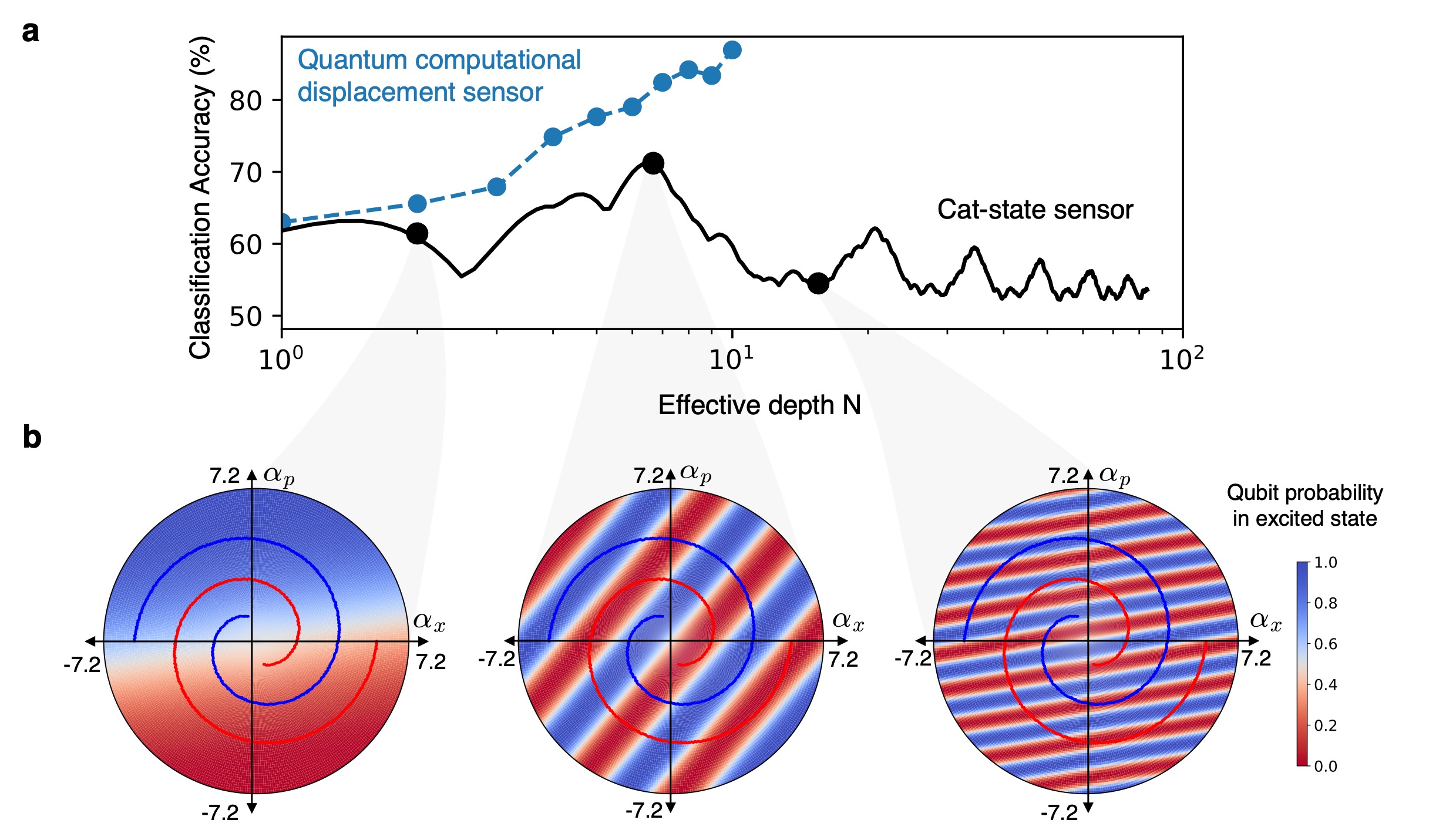}
    \caption{\textbf{Simulated performance of the cat-state sensing protocol as a function of cat-size.} \textbf{a.} We consider the performance of the ideal cat-state sensing protocol in simulation for the task of distinguishing displacements from arms of a spirals distribution, considered in Fig.~\ref{fig:main3}c. The effective depth is defined as the conventional depth for the QCDS protocol. For the cat-state protocol, it is an effective proxy for the cat-state size, where we assume each depth implements an ECD gate of fixed magnitude of $\beta = 0.24$ (as in experiment). The performance does not generally increase with the cat-state size. \textbf{b.} Response of the qubit probability in the excited state as a function of the sensing displacement for the cat-state protocol, which visually shows why larger cat states do not result in improved classification accuracy.}
    \label{fig:cat_sweep}
\end{figure}

When the axis of displacement is known, such a protocol is efficient in sensing small displacements on the order of $1/\beta$. Consequently, the sensitivity can be increased by producing cat-states of larger size $\beta$, such that it effectively amplifies the action of a small displacement $\alpha_\perp$ onto an $O(1)$ phase $\sim \beta\alpha_\perp$, which can be measured with a larger signal-to-noise ratio. This enables a larger Fisher information than is possible with a standard phase-preserving amplifier, as we experimentally observed in Fig.~\ref{fig:cat}. However, this protocol is not efficient for our tasks of single-shot binary classification. This is because of two reasons. First, the protocol is only sensitive to a single component of the displacement. This places a fundamental upper bound on the performance we expect to achieve, based on the task. Second, the tradeoff between the dynamic-range and sensitivity of the protocol. The tasks we consider have datasets with a support which is not small. Larger cat sizes do not necessarily increase performance; rather, there exists an optimal cat-size. We illustrate this in Fig.~\ref{fig:cat_sweep}, where we consider the simulated classification accuracy as a function of the cat-size (in units of the size of the conditional displacement of our protocol) for the task in Fig.~\ref{fig:main3}c. 

The classification accuracy is non-monotonic as a function of the cat-size. The oscillations arise due to the nature of the dataset, depending on when the fringe separation matches the length scale of the arm-to-arm separation of the spirals. To highlight this, we plot the simulated response of the protocol. For small cat-size, the protocol is only able to capture large-wavelength features of the dataset. For large cat-size, the fringes do not necessarily line up with that of the dataset. The optimal size produces fringes which match the dataset length-scale (in this case the arm-to-arm separation). As a comparison, we plot the simulated performance of the quantum computational displacement sensor for this task, which tends to increase with $N$.

In all our benchmarks, we compare against the best performing cat-state protocol. To do this, we set the cat size and orientation to be trainable parameters. We do this by restricting the circuit to be a depth $N = 1$ protocol, but where the magnitude and phase of the conditional displacement amplitude $\beta$ is trained (along with the parameters of the final qubit rotation). However, in general, the trained protocol will not be implemented robustly since we have calibrated the protocol for a fixed magnitude ECD gate. However, this state can be equivalently realized by multiple ECD gates. Therefore, to realize a high-fidelity protocol, we map the single ECD gate to multiple fixed-size ECD gates (of the size used in our protocol). This is done using the parameters described in Appendix~\ref{app:quantum_computational_sensing_protocol:cat_state}. The last ECD gate has a tunable amplitude, which allows this mapping to generate any continuous size cat size.

Similar to the quantum computational sensing protocol, we train for $10$ different realizations for the tasks of Fig.~\ref{fig:main3} and those in Appendix~\ref{app:quantum_computational_sensing_protocol:examples}. We train for $20$ different realizations for the tasks of Fig.~\ref{fig:main4}. We notice that, due to the small number of trainable parameters, the spread in performance of the different realizations is small.

\clearpage
\subsection{Compass-state sensor}
\label{app:benchmark:compass_state_sensor}

\begin{figure}[h!]
    \centering
    \includegraphics[width=0.75\textwidth]{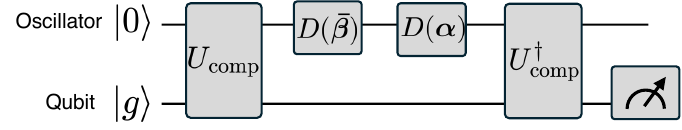}
    \caption{\textbf{Circuit diagram for the compass-state sensing protocol, which we demonstrate in experiment.}}
    \label{fig:circuit_compass}
\end{figure}

Another approach for conventional quantum sensing of displacements is using compass states~\cite{toscano_sub-planck_2006}, which belong to the family of generalized Schr\"odinger cat-states. The compass-state sensing protocol relies on the fact that a compass state that is displaced by $D(\alpha)$ is quasiorthogonal to the undisplaced compass state on a scale set by $1/|\alpha|$~\cite{toscano_sub-planck_2006}. Importantly, unlike cat states, compass states and other higher-order generalized Schr\"odinger cat states can be sensitive to displacements along an arbitrary axis. Therefore a scheme that measures the overlap of a possibly-displaced compass state and the original compass state can be used as a Heisenberg-limited sensor of small displacements along a general quadrature. Such a scheme for compass state sensing can be performed using a qubit-cavity sensor~\cite{toscano_sub-planck_2006}, with a protocol illustrated in Fig.~\ref{fig:circuit_compass}. In particular, the compass state protocol is given by:
\begin{align}
    \ket{\psi_f(\alpha)} = U_{\rm comp}^{\dagger}~D(\alpha)D(\bar{\beta})~U_{\rm comp}\ket{\psi_0} .
    \label{appeq:compassprotocol}
\end{align}

This protocol starts from the sensor ground state $\ket{\psi_0} = \ket{0,g}$ and prepares a displaced compass state entangled with the qubit, using the operation $D(\bar{\beta})~U_{\rm comp}$, where:
\begin{align}
    D(\bar{\beta})~U_{\rm comp}\ket{\psi_0} = D(\bar{\beta})\left[ \frac{1}{\sqrt{2}}\ket{c_{4e}}\!\ket{e} +  \frac{1}{\sqrt{2}}\ket{c_{4g}}\!\ket{g} \right] \equiv D(\bar{\beta})\ket{\psi_{\rm comp}}
    \label{appeq:compassprobe}
\end{align}
and the states $\ket{c_{4e}},\ket{{c}_{4g}}$ are qubit-state-dependent compass states defined as:
\begin{subequations}
\begin{align}
    \ket{c_{4e}} &= \frac{1}{2}\left( \ket{i\beta} + \ket{-\beta} - \ket{-i\beta} + \ket{\beta} \right) \label{appeq:compassE} \\
    \ket{c_{4g}} &= \frac{1}{2}\left( \ket{i\beta} - \ket{-\beta} - \ket{-i\beta} - \ket{\beta} \right) 
    \label{appeq:compassG}
\end{align}
\end{subequations}
The compass state size is determined by the parameter $\beta$. Following the sensing displacement, the compass state-preparation operation is undone to arrive at the final sensor state $\ket{\psi_f(\alpha)}$. Qubit state measurement on this final state can be shown to provide an estimate of $\alpha$ that has Heisenberg scaling with the compass state size $\beta$, for small displacements $\alpha$.

\subsubsection{Exact compass-state sensing protocol}

To determine the best-case performance of the compass state sensing protocol, we explore theoretically the performance when an analytic protocol is used for compass state preparation. In particular, we consider the unitary operation proposed in Ref.~\cite{dalvit_quantum_2006}:
\begin{align}
    U_{\rm comp} = R_{\frac{\pi}{2}}{D}_c(-i\beta)R_{\pi}~{D}_c(-\beta)~R_{\frac{\pi}{2}}~R_c\left( \frac{\pi}{4}\right){D}_c(2\beta e^{i\pi/4})R_{\frac{\pi}{2}},
    \label{appeq:ucompexact}
\end{align}
where we have introduced the single-qubit operations:
\begin{align}
    R_{\frac{\pi}{2}} &= \frac{1}{\sqrt{2}}\left( \mathbb{I}_2 + i~\hat{\sigma}_y \right) \quad\mathrm{and}\quad 
    R_{\pi} = i\hat{\sigma}_y 
\end{align}

and the conditional cavity displacement and conditional cavity rotation operations respectively:
\begin{align}
    {D}_c(\beta) &= {D}(\beta )|e\rangle\langle e| + |g\rangle\langle g| \quad\mathrm{and}\quad 
    R_c\left( \phi \right) = e^{i \phi \hat{\sigma}_x\hat{a}^{\dagger}\hat{a}}
\end{align}

Using Eq.~(\ref{appeq:ucompexact}) and Eq.~(\ref{appeq:compassprotocol}) for the compass state sensing protocol, it can be shown that the final sensor state takes the form:
\begin{align}
    \ket{\psi_f(\alpha)} =  \ket{\psi_e(\alpha)}\!\ket{e} + \ket{\psi_g(\alpha)}\!\ket{g}
\end{align}
where:
\begin{align}
    \ket{\psi_g(\alpha)} = ~&\frac{1}{4}\Bigg\{ f_1\ket{ (\bar{\beta}+2i\beta+\alpha)e^{-i\frac{\pi}{4}}}+f_2\ket{ (\bar{\beta}-2i\beta+\alpha)e^{-i\frac{\pi}{4}}}+f_3\ket{(\bar{\beta}+\alpha)e^{-i\frac{\pi}{4}}}   \nonumber \\
    &~~+ f_4\ket{ (\bar{\beta}+2\beta+\alpha)e^{+i\frac{\pi}{4}}}+f_5\ket{ (\bar{\beta}-2\beta+\alpha)e^{+i\frac{\pi}{4}}}+f_6\ket{(\bar{\beta}+\alpha)e^{+i\frac{\pi}{4}}}  \Bigg\}
\end{align}
with the coefficients:
\begin{align}
    f_1 &= f_2 = - \exp \left( \frac{1}{2}(\alpha\bar{\beta}^* - \alpha^*\bar{\beta} ) \right) \nonumber \\
    f_3 &= 2\exp \left( \frac{1}{2}(\alpha\bar{\beta}^* - \alpha^*\bar{\beta} ) \right) \cos \left( \beta\alpha^*+\beta^*\alpha + \beta\bar{\beta}^* + \beta^*\bar{\beta} \right) \\
    f_4 &= f_5 = +\exp \left( \frac{1}{2}(\alpha\bar{\beta}^* - \alpha^*\bar{\beta} ) \right)  \\
    f_6 &= 2 \exp \left( \frac{1}{2}(\alpha\bar{\beta}^* - \alpha^*\bar{\beta} ) \right) \cos\left( -i( \beta\alpha^*-\beta^*\alpha + \beta\bar{\beta}^*-\beta^*\bar{\beta}) \right)
\end{align}

\begin{figure}[h!]
    \centering
    \includegraphics[width=0.8\textwidth]{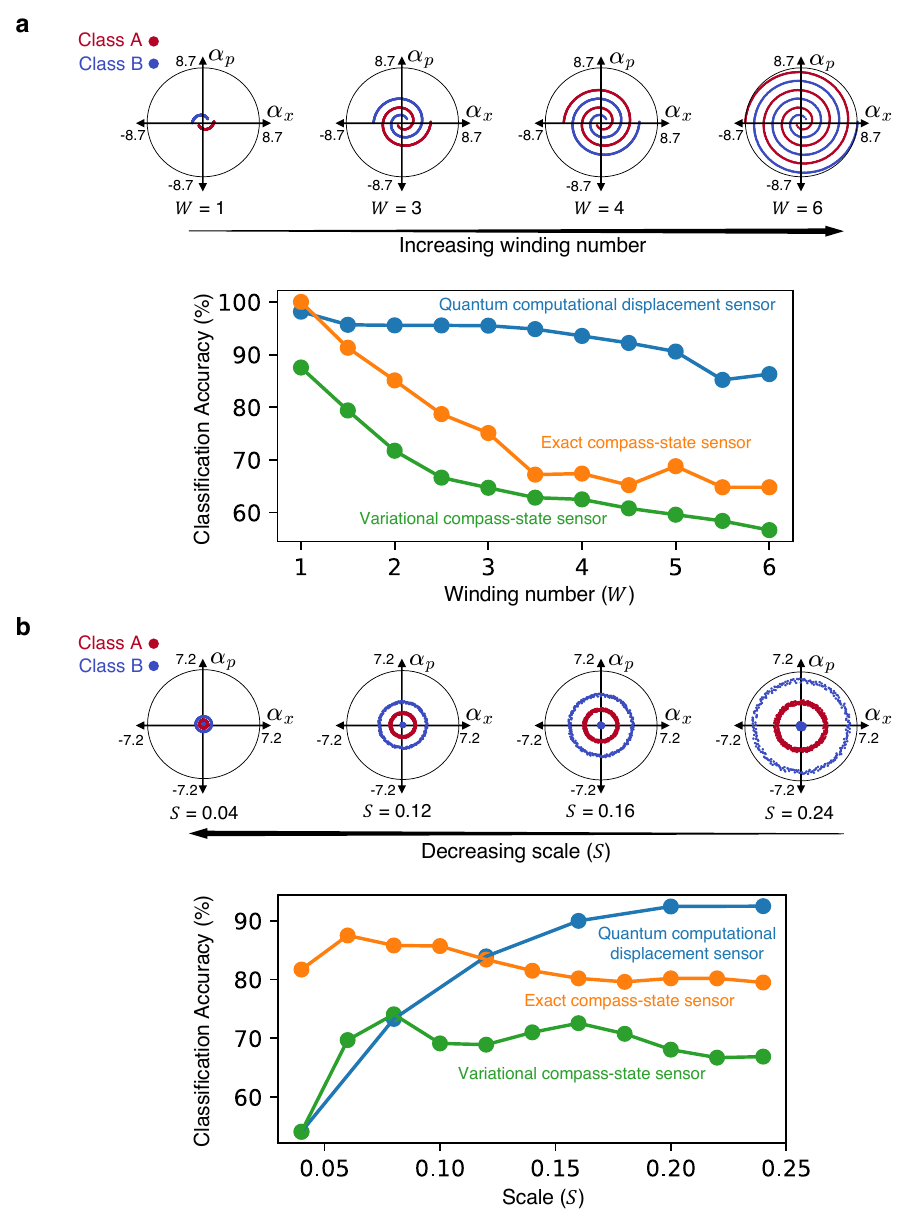}
    \caption{\textbf{Simulated performance of the compass state protocols.} \textbf{a.} Simulated classification accuracy as a function of the winding number for the task introduced in Fig.~\ref{fig:main4} for three different protocols: the exact compass-state sensing protocol, variational compass-state sensing protocol (compatible with experiment) and QCDS protocol. \textbf{b.} Performance for the same set of protocols, this time for the family of tasks described by a scale number. As the scale reduces, the tasks becomes more challenging to classify. For the variational compass-state sensor and the quantum computational sensor, the size of the conditional displacement per gate is fixed, which limits the performance of the protocol for small $S$ values (for finite $N$).}
    \label{fig:compass_performance}
\end{figure}

The probability of measuring the qubit in its ground state can be computed using:
\begin{align}
    p(\alpha|\beta,\bar{\beta}) = \langle \psi_f(\alpha) | g \rangle \langle g| \psi_f(\alpha) \rangle =  \langle \psi_g(\alpha) | \psi_g (\alpha) \rangle
\end{align}
which suffices to fully determine the response of the compass state sensor. The full analytic form of $p(\alpha|\beta,\bar{\beta})$ is unwieldy; however it can be easily computed numerically to determine the response of the compass state sensor as a function of the sensing displacement $\alpha$, following which the classification accuracy can be obtained for any classification task. As mentioned before, compass state sensing is designed for estimating small displacements $\alpha$. For binary classification tasks such as those we are considering here, the optimal choices of $\beta, \bar{\beta}$ are not a priori known. Using the analytic form of the compass state sensor response, we use training to obtain the optimal parameters. The resulting simulated performance for these compass state sensing protocols optimized for binary classification are shown in Fig.~\ref{fig:compass_performance} (in orange).

\subsubsection{Variational compass-state sensing protocol (compatible with experiment)}

We also leverage the programmability of our qubit-oscillator platform to perform the compass state protocol defined by Eq.~(\ref{appeq:compassprotocol}) in experiment. The exact compass state protocol of Eq.~(\ref{appeq:ucompexact}) cannot be implemented in our experiment due to the unavailability of a native conditional rotation operation $R_c$. However, since we do have access to ${\rm ECD}$ gates and single qubit rotations, which are known to enable universal control over the oscillator state~\cite{eickbusch2022fast}, our platform can be used to generate the compass states required for large enough depth $N$. To do so, we adapt our trainable digital model to now optimize the probe-state-preparation unitary $U$ to prepare the specific probe state defined by Eq.~(\ref{appeq:compassprobe}), namely an entangled superposition of compass states, with the specific states defined by Eqs.~(\ref{appeq:compassE}),~(\ref{appeq:compassG}). Formally, our training procedure determines the parameters $\{\theta^{\rm opt}_i,\phi^{\rm opt}_i,\beta^{\rm opt}_i\}$ that yield:
\begin{align}
    U_{\rm comp}\ket{\psi_0} = \prod_{i=0}^{N-1} R(\theta^{\rm opt}_i,\phi^{\rm opt}_i){\rm ECD}(\beta^{\rm opt}_i)\ket{\psi_0} \approx \ket{\psi_{\rm comp}}.
\end{align}
These optimal parameters are determined by minimizing $1-\mathcal{F}$, where $\mathcal{F} = |\langle \psi_{\rm comp}|U_{\rm comp}\ket{\psi_0}|^2$ is the fidelity between the prepared state and the target entangled compass state~\cite{eickbusch2022fast}:
\begin{align}
    \{\theta^{\rm opt}_i,\phi^{\rm opt}_i,\beta^{\rm opt}_i\} = \underset{\{\theta_i,\phi_i,\beta_i\}}{\rm argmin}~|1-\mathcal{F}|.
\end{align}
Note that a displaced compass state can be prepared by applying the unconditional displacement $D(\bar{\beta})$ following the application of $U_{\rm comp}$, and is therefore not included in the training procedure. We train the parameters of the protocol for a circuit depth of $N = 10$ for a range of compass state sizes up to $\beta = 3$.

Using this state preparation scheme, we are able to experimentally perform the compass state sensing protocol for binary classification tasks. In doing so, one must again ask what the optimal $\beta,\bar{\beta}$ values are that characterize the optimal compass state protocol for a given task. Since we do not possess an exact analytic form of the compass state sensor response for this variationally-optimized state preparation protocol, we determine these optimal choices of $\beta$ and $\bar{\beta}$ via a grid search, which is feasible since only a relatively small number of parameters need to be optimized. The resulting parameters determine the optimal compass state sensing protocols that we implement in experiment. The resulting classification performance is shown in Fig.~\ref{fig:compass_performance} (in green) for simulation and Fig.~\ref{fig:main4} of the main text for experiment.

\clearpage
\subsection{Phase-preserving amplifier and classical postprocessing backend}
\label{app:benchmark:phase_preserving_amplifier}

\begin{figure}[h!]
    \centering
    \includegraphics[width=0.75\textwidth]{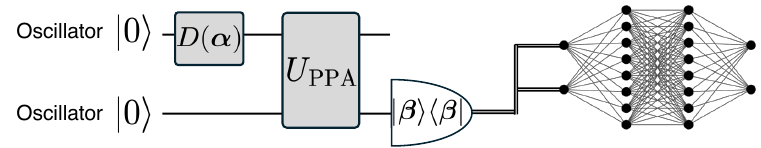}
    \caption{\textbf{Circuit diagram for using a phase-preserving amplifier followed by a classical postprocessor}. We simulate the phase-preserving amplifier as a quantum sensor baseline using both ideal and experimental parameters.}
    \label{fig:circuit_heterodyne}
\end{figure}

A conventional sensing strategy is to perform heterodyne measurement of the oscillator, which provides unbiased estimates of its position and momentum $(\widetilde{\alpha}_x, \widetilde{\alpha}_p)$ in a single shot~\cite{clerk2010introduction}. The advantage of this technique is its simplicity: based on the estimate, further postprocessing tasks can be performed, as illustrated in Fig.~\ref{fig:circuit_heterodyne}. This is unlike the previous two protocols (and in general those involving a qubit which is measured), which have a limited dynamic range. However, the drawback is that the estimates come with unavoidable noise. This noise sets a bound on the minimum error of outputs from downstream postprocessing. In our context, this results in a non-zero classification error, since displacements from the two classes could produce the same output estimate from the phase-preserving amplifier. In this case, there is no other way to know which class the displacement originated from. 

\subsubsection{Theoretical description}
The quantum protocol can be described by the unitary:
\begin{align}
    \Udec = \exp \{-i\hat{\mathcal{H}}_{\rm L} \},
\end{align}
The protocol for an ideal (lossless) phase-preserving amplifier can be described by a sensing bosonic mode $\hat{a}$ coupled to an additional (readout) mode $\hat{b}$, with both modes initialized in the vacuum state $\ket{\psi_0} = \ket{0,0}$. Following the sensing operation which imparts a displacement $D(\alpha)$ on mode $\hat{a}$, the two modes are evolved under the Hamiltonian $\hat{\mathcal{H}}_{\rm L}$ for a linear phase-preserving quantum amplifier:
\begin{align}
    \hat{\mathcal{H}}_{\rm L} = i\chi \left( \hat{a}^{\dagger}\hat{b}^{\dagger} - \hat{a}\hat{b} \right),
\end{align}
where $\chi$ defines the amplifier interaction strength.

A simple approach to evaluate expectation values with respect to the quantum state of the phase-preserving linear amplifier is in the Heisenberg representation. In this case, expectation values are always computed with respect to the initial amplifier state, while time evolution is encoded in the dynamics of operators. In this case, the final operator for the amplifier is given in terms of the initial operators, depending on the gain $\mathcal{G}$:
\begin{align}
    \hat{b} = \sqrt{\mathcal{G}}~\hat{b}_0 + \sqrt{\mathcal{G}-1}~\hat{a}^{\dagger}_0 
    \label{appeq:heisppa}
\end{align}
where $\sqrt{\mathcal{G}} = \cosh \chi$~\cite{scully_quantum_1997}. In the ideal case, we consider the limit of infinite gain (so that the noise contribution from subsequent amplifiers are negligible). The outputs are the quadratures of the $\hat{b}$ mode, which is classical in nature:
\begin{equation}
    \hat{X}_b = \sqrt{\mathcal{G}}~\hat{X}_{b_0} + \sqrt{\mathcal{G}-1}~\hat{X}_{a_0}  \quad \rm and \quad
    \hat{P}_b = \sqrt{\mathcal{G}}~\hat{P}_{b_0} - \sqrt{\mathcal{G}-1}~\hat{P}_{a_0}.
\end{equation}
In the ideal case, the signal mode $\hat{b}_0$ is in a coherent state, while the idler mode $\hat{a}_0$ is in vacuum. In this case, the variance of the quadratures are: $\langle(\Delta \hat{X})^2\rangle = \langle(\Delta \hat{P})^2\rangle = 1/2$. The output modes have the variance:
\begin{equation}
    \langle(\Delta \hat{X}_b)^2\rangle = \langle(\Delta \hat{P}_b)^2\rangle = \mathcal{G}\frac{1}{2} + (\mathcal{G}-1)\frac{1}{2}.
\end{equation}
In the limit of infinite gain, the above expression tends to $G$, equivalent to a $1$ when referred to the input (that is, the noise to be added to the input signal to produce equivalent signal-to-noise ratio). Realistic amplifiers have higher noise figures. In the microwave domain, tremendous progress has been made to approach the theoretical-best case scenario of $1$ photon worth of noise. As a baseline for state-of-the-art performance, we use the experiment parameters of the first phase of the HAYSTAC experiment~\cite{zhong2018results, al2017design}. In this experiment, the quantum efficiency of the measurement chain is carefully optimized since this directly limits the sensitivity of the experiment in detecting small displacements. The noise added varies over time and the frequency of measurement, but we use the lower end of the range: 2.3 photons of total noise added~\cite{zhong2018results}.

\begin{figure}[h!]
    \centering
    \includegraphics[width=0.7\textwidth]{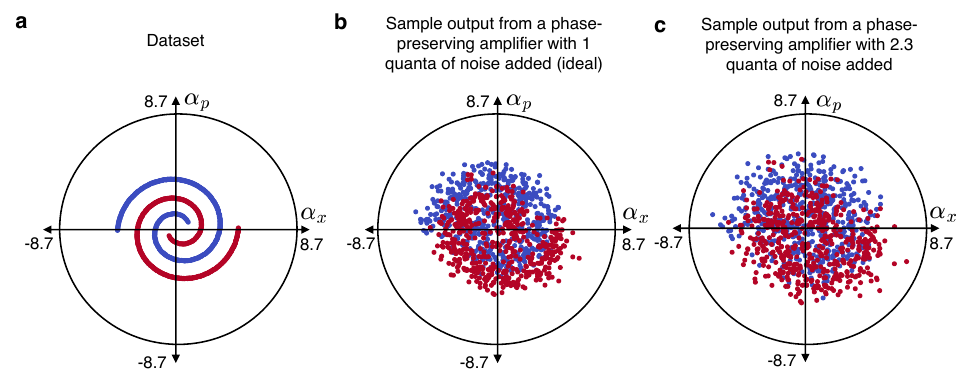}
    \caption{\textbf{Effect of the noise added by phase-preserving amplifiers on the dataset.} \textbf{a.} Distribution of the dataset for the two classes of the task considered in Fig.~\ref{fig:main4}, with winding number $W=3$. If the values of the displacement data points were available classically, a classical neural network can achieve essentially $100\%$ classification accuracy. \textbf{b.} Single realization of the noisy output of an ideal phase-preserving amplifier. The task can now be thought of as performing classification on this new dataset. Due to the noise, the two distributions of displacements now overlap, and hence cannot be distinguished in a single sample. \textbf{c.} Same as \textbf{b.}, but with the parameters of the HAYSTAC experiment~\cite{zhong2018results}.}
    \label{fig:ppa_noise}
\end{figure}

\subsubsection{Classical postprocessing}

To convert the noisy estimates into a prediction, we use a multilayer perceptron (MLP). As can be seen in Fig.~\ref{fig:ppa_noise}, once the sensed displacement is amplified, the displacement probability distributions of the two classes overlap, which sets a maximum upper bound on the classification accuracy (which depends on the parameters of the amplifier). In our simulation, we consider the performance of an ideal phase-preserving amplifier and one with the parameters of the amplifier used in the HAYSTAC experiment~\cite{zhong2018results}. During the training of the MLP, we redraw new samples from the theoretical Gaussian distribution of the noise added by the amplifier. This allows the MLP to train noise-aware, which increases its performance by reducing the chance to overfit to the training dataset. The MLP architecture we use is: $2 \to 64 \to 64 \to 2$, with the input dimension of $2$ representing the position and momentum quadratures of the output of the amplifier, and the output dimension of $2$ representing the two classes of the task. The predicted label is the index of the output with the larger value.

This architecture has sufficient expressive capacity to achieve $100\%$ classification accuracy on the original displacement dataset. We also notice that changing the architecture, either the width of the hidden layers, or the number of layers, does not yield a noticeable increase the performance. The errorbars in performance presented in Fig.~\ref{fig:main4} correspond to the difference in performance between the upper and lower end of the estimate of the sensed displacement (i.e., the scale factor discussed in Appendix~\ref{app:system_hamiltonian_and_parameters:characterization}).

\clearpage
\subsection{Squeezed probe-state, phase-sensitive amplifier and classical postprocessing backend}
\label{app:benchmark:squeezed_phase_sensitive_amplifier}

Another conventional sensing strategy which we use as a benchmark is one which only measures one component of the displacement. This protocol was implemented in the second phase of the HAYSTAC experiment, which provided a sensing advantage since the axis of the displacement is known beforehand (and can be tuned in experiment)~\cite{backes2021quantum, bai2025dark}. In this protocol, the initial state of the sensing mode is a squeezed vacuum state. Following the displacement, the low-noise quadrature is measured. This outputs a real number, an ideally-noiseless estimate. Similar to Appendix~\ref{app:benchmark:phase_preserving_amplifier}, an MLP is trained to convert this estimate into a class prediction.

\begin{figure}[h!]
    \centering
    \includegraphics[width=0.75\textwidth]{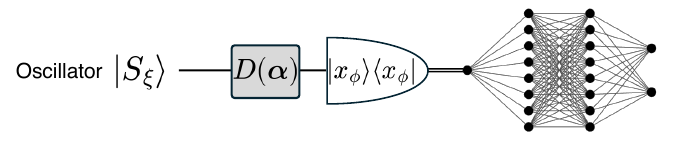}
    \caption{\textbf{Circuit diagram for using a squeezed probe-state, phase-sensitive amplifier followed by a classical post processor.} We simulate this protocol for both ideal (infinitely squeezed probe-state $\ket{S_{\infty}}$ and phase-sensitive amplifier) and experimental parameters.}
    \label{fig:circuit_squeezed}
\end{figure}

\subsubsection{Theoretical description}
The squeezed probe state required for this protocol can be generated by squeezing the vacuum state:
\begin{equation}
    \ket{\psi_{\rm probe}} =  \hat{S}(\xi) \ket{0} \equiv \ket{S_{\xi}},
\end{equation}
where $\hat{S}(\xi) = \exp{\left(\frac{1}{2}\left(\xi^* {a^\dagger}^2 - \xi a^2 \right) \right)}$ is the squeeze operator, and $\xi = r \exp{\left( i\phi \right)}$, where $r$ is the squeeze parameter and $\phi$ sets the orientation of the squeezed quadrature~\cite{scully_quantum_1997}. The squeeze parameter sets the amount of noise in this quadrature. For $\phi = 0$:
\begin{equation}
    \langle \hat{X}^2 \rangle = \frac{1}{2}e^{-2r} \quad {\rm and} \quad
    \langle \hat{P}^2 \rangle = \frac{1}{2}e^{2r}.
\end{equation}
Hence, for $r>0$ the noise on the position quadrature (in this case) can be reduced below the vacuum noise floor, at the expense of increasing the noise in the conjugate (momentum) operator. We consider two cases. In the ideal case, we assume there is no noise on the low-noise quadrature, which is achieved by infinite squeezing ($\ket{S_{\infty}}$ where $r\to\infty$) and an ideal phase-sensitive amplifier. In the realistic case, we consider the parameters of the HAYSTAC experiment, which achieves up to a $4~\rm dB$ reduction in noise, which corresponds to a squeezing strength of $r \simeq 0.46$.

\subsubsection{Classical postprocessing}

The optimal postprocessing strategy is based on maximum-likelihood, by assigning a class label to each possible value of the real-valued output of the squeezed mode. We approximate this strategy by using a MLP. Empirically, we found it easier to train the MLP by binning the real-valued output, and converting to a one-hot vector, increasing the input dimension. We chose a dimension of $32$ bins. We also append to this vector the real-valued output itself, providing a total input dimension of $33$. This appending allows us to also train the orientation of squeezing $\phi$, in order to maximize classification accuracy. Due to the simplicity of the task, we use a relatively small architecture: $33 \to 8 \to 2$. We found this architecture both easiest to train and achieve the highest possible classification accuracy. As seen in Fig.~\ref{fig:main4}, this protocol fails to obtain $100\%$ classification accuracy, even in the ideal limit. This is because the tasks considered in Fig.~\ref{fig:main4} (and more generally throughout this work), require information of both components of the displacement for effective classification. Errorbars in performance in Fig.~\ref{fig:main4}, only relevant for the simulation with HAYSTAC's experiment parameters, correspond to the upper and lower estimates of the sensed displacement.

\clearpage
\subsection{Gottesman--Kitaev--Preskill (GKP) probe-state, quantum phase estimation and classical postprocessing backend}
\label{app:benchmark:gkp_state}

A protocol can estimate both position and momentum beyond the limit set by the uncertainty principle by instead measuring their modular counterparts. In this subsection, we simulate the performance of a quantum sensing protocol based on using the GKP state.

\begin{figure}[h!]
    \centering
    \includegraphics[width=0.85\textwidth]{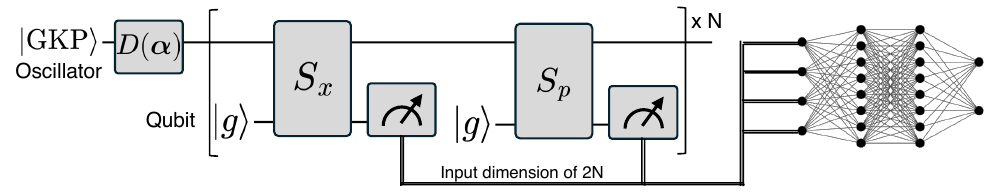}
    \caption{\textbf{Circuit diagram for using a GKP probe-state, quantum phase estimation followed by a classical post processor.} We simulate this protocol for the experimental parameters of Refs.~\cite{sivak2023real, valahu2025quantum}.}
    \label{fig:circuit_gkp}
\end{figure}

\subsubsection{Theoretical description}

Modular operators $\hat{x}_{[2\pi/l_x]}$ and $\hat{p}_{[2\pi/l_p]}$, where $\hat{q}_{m} = \hat{q}_m~\rm mod~m $, can commute with each other. One solution when this is possible is when $l_x = l_p = l_s= \sqrt{2\pi}$. This can be achieved by instantiating the oscillator in a GKP state, which is a simultaneous eigenstate of the shift operators $S_x = e^{-i l_x \hat{x}_{[2\pi/l_x]}} = e^{-i l_x \hat{x}}$ and $S_p = e^{-i l_p \hat{p}_{[2\pi/l_p]}} = e^{-i l_p \hat{p}}$. This enables simultaneous measurements of both quadratures beyond the SQL limit, with the only drawback the finite dynamic range of displacements which can be resolved~\cite{duivenvoorden2017single, labarca_quantum_2025}. This protocol is therefore useful for the quantum sensing of small displacements. In this work, where we use a convention of the quadrature representation, the periodicity is $\sqrt{\pi}$. Since the GKP states are eigenstates of the shift operators, and they commute with each other, they can be measured repeatedly using the quantum phase estimation algorithm. Ref.~\cite{valahu2025quantum} uses an ancilla qubit, similar to our setup, to measure these observables.

\begin{figure}[h!]
    \centering
    \includegraphics[width=0.5\textwidth]{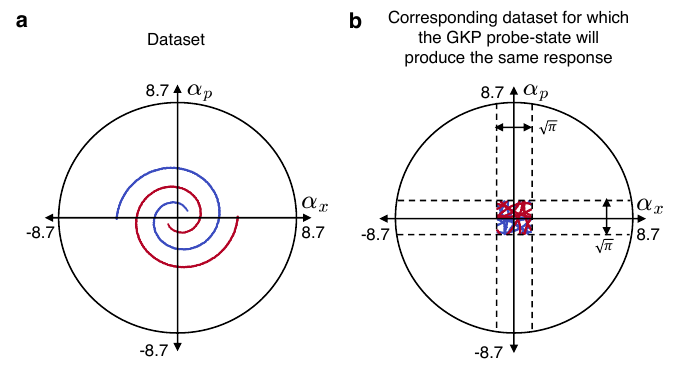}
    \caption{\textbf{Effect of the modular nature of the response of the GKP probe-state.} \textbf{a.} Distribution of the dataset for the two classes of the task considered in Fig.~\ref{fig:main4}, with winding number $W=3$. \textbf{b.} Due to the modular response of the GKP probe-state, it cannot distinguish displacement modulo $\sqrt{\pi}$ (for instance) along both quadratures. We can equivalently think of this constraint as `wrapping' the dataset with this periodicity. This results in a much `harder' task, even with the reduced noise in estimation of both quadratures.}
    \label{fig:gkp_modular}
\end{figure}

\subsubsection{Simulating the expected performance of a superconducting circuit experiment}

Ref.~\cite{sivak2023real} experimentally demonstrated quantum error correction of GKP states with an ancilla qubit. The GKP state can be error corrected by measuring stabilizers and feedback operations. However, as theoretically considered in Ref.~\cite{labarca_quantum_2025}, instead of using the measurement outcomes of the stabilizer measurement for error correction, they can be used for metrology. Repeated measurements of the stabilizers provide an estimate of the displacement sensed by the GKP state. Therefore, while the experiment demonstration of Ref.~\cite{sivak2023real} doesn't explicitly perform displacement sensing, we use the simulation results of Ref.~\cite{labarca_quantum_2025} to estimate the experiment's performance. The simulations in Ref.~\cite{labarca_quantum_2025} generate the probability distribution of the bitstring outcomes for a sequence of noiseless successive measurements, as a function of the displacement, for a set of GKP sizes $\Delta$. We consider $\Delta = 0.35$, which is close to the experimental value of Ref.~\cite{sivak2023real}. From this we consider the situation of $10$ rounds of noiseless measurements, which corresponds to $20$ bits in total ($10$ bits for each quadrature). At $10$ rounds, the amount of information extracted is close to the Holevo bound (for this size of GKP state). Therefore, our simulation yields a considerable overestimate of the expected performance of what the experiment would likely achieve. We then sample from this probability distribution (given the input displacement value) in simulation to stochastically generate a 20-bit long sequence (using the multinomial samples in PyTorch~\cite{paszke2019pytorch}). This is then processed by a MLP of architecture: $20 \to 64 \to 64 \to 2$ to perform prediction. 

As is evident from the modular response in Fig.~\ref{fig:main4} and Fig.~\ref{fig:gkp_modular}, the GKP protocol is unable to achieve a high classification accuracy for the tasks considered. However, we do note, the modular nature which wraps the dataset as illustrated in Fig.~\ref{fig:gkp_modular}, usually will not have overlaps. Therefore, in principle, for a sufficiently efficient protocol, the two classes can be distinguished without error. However, it is straightforward to expect such a protocol will require GKP states of many photons and high fidelity in the quantum phase estimation measurement -- beyond the reach of experimental platforms. Furthermore, in the unlucky scenario, the modular nature might result in overlaps of the displacement distributions -- in which case, they cannot be distinguished.

\subsubsection{Simulating the expected performance of a trapped ion experiment}

Ref.~\cite{valahu2025quantum} experimentally performs displacement sensing, where the collective bosonic mode of trapped ions are prepared in the GKP state. Their experiment achieved approximately a $5~\rm dB$ reduction in the total mean-squared error compared to the theoretical heterodyne limit. We follow the theoretical description of the protocol of Ref.~\cite{valahu2025quantum} for our simulations. In their experiment, the best performance was achieved for two rounds of measurements, corresponding to $4$ bits of measurements, two for each quadrature. The probability of outcome for each qubit measurement can be expressed as:

\begin{equation}
    P_{x/p} = \frac{1}{2} + \frac{\eta_{x/p}}{2}\cos{\left( \alpha_{x/p} l_s / \sqrt{2} + \theta_{x/p} \right)},
\end{equation}
where $P_x$ and $P_p$ represent the probability of measurement when measuring displacement $(\alpha_x, \alpha_p)$ along the position and momentum quadratures respectively. $\eta_x$ and $\eta_p$ are the corresponding visibility parameters, which takes into account experimental imperfections such as qubit readout error which reduces the information obtained per round of measurement. The experiment approximates both $\eta_x$ and $\eta_p$ as the same value, equivalent to $\eta$ with a value of $0.72$, which we also set in simulation. We train the offset angles $\theta_x$ and $\theta_p$, independent for each layer. We do this by using the straight-through estimator on the backward pass to back propagate through the binomial sampler. This allows the simulation to find the best possible parameters for the particular task. The bitstrings are processed by an MLP of architecture: $4 \to 64 \to 64 \to 2$ to map the bitstrings to a class prediction. As usual, changing the hyperparameters do not yield any significant improvements.

\begin{figure}[h!]
    \centering
    \includegraphics[width=0.7\textwidth]{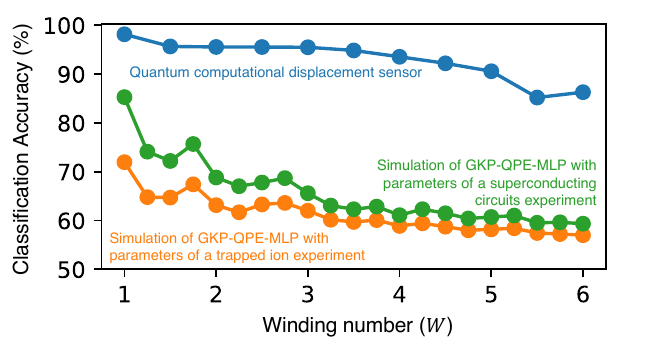}
    \caption{\textbf{Simulated performance of the GKP probe-state, QPE and classical postprocessing protocol.} Simulated classification accuracy as a function of the winding number for the task introduced in Fig.~\ref{fig:main4} for the two GKP experiments and QCDS protocol.}
    \label{fig:gkp_performance}
\end{figure}

\clearpage
\subsection{Two-mode squeezed probe state, phase-sensitive amplifiers and classical postprocessing backend}
\label{app:benchmark:two_mode_squeezed_phase_sensitive_amplifier}

Two-mode squeezing (TMS) provides a resource to enable sub-vacuum-noise estimation of non-commuting quadratures of a sensed displacement~\cite{braunstein_dense_2000, steinlechner_quantum-dense_2013}. In this section we provide theoretical details of this TMS scheme for displacement sensing, and of the classical postprocessing backend used to process the estimated displacement for classification tasks. We also provide results of the comparison between our QCDS protocol and the TMS scheme for binary classification tasks, showing how QCDS can provide an increased efficiency with probe state photon number. This is relevant since experimentally realizing TMS can be challenging, especially for the scenario of the displacement sensing of a single cavity mode considered in this work~\cite{eberle2013stable, qiu2023broadband, liu2022noise}.

\begin{figure}[h!]
    \centering
    \includegraphics[width=0.75\textwidth]{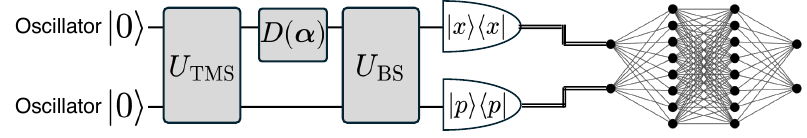}
    \caption{\textbf{Circuit diagram for using a two-mode squeezed probe-state, phase-sensitive amplifiers and classical postprocessing backend}. We simulate this protocol as a quantum sensor baseline with varying amounts of two-mode squeezing.}
    \label{fig:circuit_tms}
\end{figure}

\subsubsection{Theoretical description}

We consider a quantum sensor comprised of two bosonic modes $\hat{a}$ and $\hat{b}$. The quantum sensor is initialized in the product ground state $\ket{\psi_0} = \ket{0, 0}$. Then, a two-mode squeezed vacuum state is constructed via the action of the two-mode squeezing operator~\cite{scully_quantum_1997}:
\begin{align}
    \ket{\psi_{\rm probe}} = U_{\rm TMS}\ket{\psi_0} = \exp \left( r \hat{a}^{\dagger}\hat{b}^{\dagger} - r \hat{a}\hat{b} \right)\ket{\psi_0}.
    \label{appeq:tms}
\end{align}
Starting from this probe state, we consider the use of an interferometric setup for displacement sensing~\cite{braunstein_dense_2000}, under which the final state of the sensor is defined by:
\begin{align}
    \ket{\psi_f} = U_{\rm BS}D(\alpha)U_{\rm TMS} \ket{\psi_0} = U_{\rm BS}D(\alpha)\ket{\psi_{\rm probe}}
\end{align}
Here $D(\alpha)$ is as defined in the main text, and $U_{\rm BS}$ defines a balanced beam splitter, which takes the form:
\begin{align}
    U_{\rm BS} = \exp \left( -\frac{\pi}{4}(\hat{a}^{\dagger}\hat{b} - \hat{b}^{\dagger}\hat{a} )  \right).
    \label{appeq:ubs}
\end{align}

We would like to determine the statistical properties of measurements made on the final state $\ket{\psi_f}$ of the quantum sensor, from which estimates of the sensed displacement $\alpha$ can be obtained. For a Gaussian circuit such as this one, it is sufficient to evaluate first and second-order moments of measured observables, and to perform this calculation in the Heisenberg representation. To begin, we define the expectation value with respect to the quantum sensor probe state as $\langle \hat{f} \rangle = \langle \psi_{\rm probe}|\hat{f}| \psi_{\rm probe}\rangle$. Then, the specific choice of initial state leads to the first-order moments:
\begin{align}
    \langle \hat{a} \rangle = \langle \hat{b} \rangle = 0. 
    \label{appeq:tmsstats1}
\end{align}
The properties of two-mode squeezing yield the following self- and cross-moments for the two bosonic modes:
\begin{align}
    \langle \hat{a}^{\dagger}\hat{a} \rangle &= \langle \hat{b}^{\dagger}\hat{b} \rangle = \sinh^2 r \\
    \langle \hat{a}^2 \rangle &= \langle \hat{b}^2 \rangle = 0 \\
    \langle \hat{a}\hat{b} \rangle &= \cosh r \sinh r\\
    \langle \hat{a}^{\dagger}\hat{b} \rangle &= 0 
    \label{appeq:tmsstats2}
\end{align}
The total number of photons in the probe state is therefore $\bar{n} = 2\sinh^2 r$. We introduce canonical position quadratures for the two modes, defined as $\hat{x}_a = \frac{1}{\sqrt{2}}(\hat{a}+\hat{a}^{\dagger})$, $\hat{x}_b = \frac{1}{\sqrt{2}}(\hat{b}+\hat{b}^{\dagger})$. We have chosen the phase of the two-mode squeezing interaction in Eq.~(\ref{appeq:tms}) such that the particular joint quadrature $\hat{X}_- = \frac{1}{\sqrt{2}}\left(\hat{x}_a - \hat{x}_b \right)$ of modes $\hat{a}$ and $\hat{b}$ is squeezed; this is easily verified by computing $\avg{\hat{X}_-^2}$, which is given by:
\begin{align}
    \avg{\hat{X}_-^2} &= \frac{1}{4} \langle \left( \hat{a}^2 + \hat{a}\hat{a}^{\dagger} - \hat{a}\hat{b} - \hat{a}\hat{b}^{\dagger} + \hat{a}^{\dagger}\hat{a} + \hat{a}^{\dagger 2} - \hat{a}^{\dagger}\hat{b} - \hat{a}^{\dagger}\hat{b}^{\dagger} - \hat{b}\hat{a} - \hat{b}\hat{a}^{\dagger} + \hat{b}^2 + \hat{b}\hat{b}^{\dagger}  - \hat{b}^{\dagger}\hat{a} - \hat{b}^{\dagger}\hat{a}^{\dagger}  + \hat{b}^{\dagger}\hat{b} + \hat{b}^{\dagger 2} \right) \rangle \nonumber \\
    &= \frac{1}{2}\left( 1 + 2\sinh^2 r - 2\cosh r \sinh r \right)   \nonumber \\
    &= \frac{1}{2}e^{-2r}
    \label{appeq:tmsstatsX2}
\end{align}
Therefore, for $r>0$, the variance of the joint quadrature $\hat{X}_-$ is reduced below its vacuum value. Introducing the conjugate quadratures $\hat{p}_a = -\frac{i}{\sqrt{2}}(\hat{a}-\hat{a}^{\dagger})$, $\hat{p}_b = -\frac{i}{\sqrt{2}}(\hat{b}-\hat{b}^{\dagger})$, and the corresponding joint quadrature $\hat{P}_+ = \frac{1}{\sqrt{2}}\left(\hat{p}_a+\hat{p}_b\right)$, we can easily verify that $\avg{\hat{P}_+^2} = \frac{1}{2}e^{-2r}$ also. The respective conjugate quadratures $\hat{X}_+ = \frac{1}{\sqrt{2}}\left( \hat{x}_a + \hat{x}_b \right)$ and $\hat{P}_- = \frac{1}{\sqrt{2}}\left(\hat{p}_a-\hat{p}_b\right)$ will instead be amplified in accordance with the uncertainty principle, as may be readily verified.

It is often convenient to parametrize the number of photons in the initial two-mode squeezed state---which is the irreducible resource that determines the utility of this sensor~\cite{braunstein_squeezing_2005}---in terms of a squeezing value ${\rm S(dB)}$ in dB. Formally, the amount of squeezing characterized in this way is obtained by comparing the variance of the squeezed quadrature relative to a vacuum field quadrature:
\begin{align}
    S(r)~{\rm [dB]} = 10 \log_{10} \left( \frac{\frac{1}{2}}{\frac{1}{2}e^{-2r}} \right) =  10 \log_{10} \left( e^{2r} \right) \simeq 8.7r~{\rm [dB]}
    \label{appeq:sqdef}
\end{align}

To determine the statistical properties of measurements made on the final sensor state in the Heisenberg representation, we can compute the evolution of operators $\hat{a}_0,\hat{b}_0$ under the protocol defined by Eq.~(\ref{appeq:tms}). The expectation value for an arbitrary operator $\hat{O}$ can be given by:
\begin{align}
    \avg{\hat{O}}_f = \langle \psi_f | \hat{O} |\psi_f \rangle = \bra{\psi_{\rm probe}}D^{\dagger}(\alpha)U_{\rm BS}^{\dagger}\hat{O}U_{\rm BS}D(\alpha) \ket{\psi_{\rm probe}} 
\end{align}

We can determine the action of the beam splitter operation $U_{\rm BS}$ by noting that the evolution of the operators $\hat{a}$ and $\hat{b}$ under  $U_{\rm BS}$ is formally determined by the Heisenberg equations of motion:
\begin{align}
    \dot{\hat{a}} = -\frac{\pi}{4}\hat{b} \quad {\rm and} \quad
    \dot{\hat{b}} = +\frac{\pi}{4}\hat{a}
\end{align}
which yield the solution:
\begin{align}
    \hat{a}(t) = e^{ +i(\pi/4)t} \hat{A} + e^{ -i(\pi/4)t}\hat{B}
\end{align}
Using initial conditions $\hat{a}(0) = \hat{a}$ and $\dot{\hat{a}}(0) = -\frac{\pi}{4}\hat{b}$, we find:
\begin{align}
    \hat{A} + \hat{B} = \hat{a} \quad {\rm and} \quad
    -i\hat{A} + i\hat{B} = \hat{b}
\end{align}
which yields $\hat{A} = \frac{1}{2}(\hat{a}+i\hat{b}), \hat{B} = \frac{1}{2}(\hat{a}-i\hat{b})$. We therefore finally obtain:
\begin{align}
    \hat{a}(t) = \cos (\pi t/4)~\hat{a} - \sin (\pi t/4)~\hat{b} 
\end{align}
Setting $t=1$, we have:
\begin{align}
    \hat{a}_{\rm BS} = \frac{1}{\sqrt{2}}\hat{a} - \frac{1}{\sqrt{2}}\hat{b}
\end{align}
Since $\dot{\hat{a}} = -\frac{\pi}{4}\hat{b}$, we find:
\begin{align}
    \hat{b}(t) = \sin (\pi t/4)~\hat{a} + \cos (\pi t/4)~\hat{b} 
\end{align}
and again, setting $t=1$, we obtain:
\begin{align}
    \hat{b}_{\rm BS} = \frac{1}{\sqrt{2}}\hat{a} + \frac{1}{\sqrt{2}}\hat{b}
\end{align}

In the Heisenberg representation, we are interested in obtaining the evolved  operators:
\begin{align}
    \hat{a}_f &\equiv D^{\dagger}(\alpha)U_{\rm BS}^{\dagger}~\hat{a}~U_{\rm BS}D(\alpha) = D^{\dagger}(\alpha)\left( \frac{1}{\sqrt{2}}\hat{a} - \frac{1}{\sqrt{2}}\hat{b} \right)D(\alpha)  = \frac{1}{\sqrt{2}}\hat{a} - \frac{1}{\sqrt{2}}\hat{b} + \frac{1}{\sqrt{2}}\alpha
    \label{appeq:af}
\end{align}
where we have used the standard property of the displacement operator $D^{\dagger}(\alpha)\hat{a}D(\alpha) = \hat{a} + \alpha$. We similarly obtain:
\begin{align}
    \hat{b}_f &\equiv D^{\dagger}(\alpha)U_{\rm BS}^{\dagger}~\hat{b}~U_{\rm BS}D(\alpha)  = D^{\dagger}(\alpha)\left( \frac{1}{\sqrt{2}}\hat{a} + \frac{1}{\sqrt{2}}\hat{b} \right)D(\alpha)  = \frac{1}{\sqrt{2}}\hat{a} + \frac{1}{\sqrt{2}}\hat{b} + \frac{1}{\sqrt{2}}\alpha
\end{align}
We can now calculate the statistical properties of measurements made upon conclusion of the displacement sensing protocol. We therefore see that both mode operators have a dependence on the sensed displacement $\alpha$. We can now compute appropriate observables of both modes to construct estimators for both components of the sensed displacement. In particular, using Eq.~(\ref{appeq:af}) the $\hat{x}_{af}$ quadrature of mode $\hat{a}_f$ is given by:
\begin{align}
    \hat{x}_{af} = \frac{1}{\sqrt{2}}\left(\hat{a}_f + \hat{a}_f^{\dagger} \right) = \frac{1}{\sqrt{2}}\left(\hat{x}_a - \hat{x}_b + \sqrt{2}\alpha_x\right) = \hat{X} + \alpha_x
\end{align}
Therefore, the $\hat{x}_{af}$ quadrature can be used to estimate $\alpha_x$, while its noise properties are determined by the two-mode-squeezed quadrature $\hat{X}$. Similarly, we see that the $\hat{p}_b$ quadrature is given by:
\begin{align}
    \hat{p}_{bf} = \frac{-i}{\sqrt{2}}\left(\hat{b}_f - \hat{b}_f^{\dagger} \right) = \frac{-i}{\sqrt{2}}\left(\hat{p}_a + \hat{p}_b + \sqrt{2}\alpha_p\right) = \hat{P} + \alpha_p
\end{align}
Using Eqs.~(\ref{appeq:tmsstatsX2}), we can immediately write down:
\begin{align}
    \avg{\hat{x}_a}_f &= \avg{\hat{X}} + \alpha_x = \alpha_x \nonumber \\
    \Delta \hat{x}_a^2 &= \avg{\hat{x}_a^2}_f - \avg{\hat{x}_a}_f^2 =  \avg{\hat{X}^2} + \alpha_x^2 - \alpha_x^2 = \frac{1}{2}e^{-2r}
    \label{appeq:tmsvarx}
\end{align}
An analogous calculation can be used to show that the $\hat{p}_{b}$ quadrature provides an estimate of $\alpha_p$, so that
\begin{align}
    \avg{\hat{p}_b}_f &= \avg{\hat{P}} + \alpha_p = \alpha_p \nonumber \\
    \Delta \hat{p}_b^2 &= \avg{\hat{p}_b^2}_f - \avg{\hat{p}_b}_f^2 =  \avg{\hat{P}^2} + \alpha_p^2 - \alpha_p^2 = \frac{1}{2}e^{-2r}
    \label{appeq:tmsvary}
\end{align}

As a result, local homodyne measurements of the $\hat{x}_a$ and $\hat{p}_b$ quadratures provide unbiased estimators of the position and momentum quadratures of the sensed displacement respectively, both with a squeezed variance determined by the two-mode squeezing parameter $r$. 

\subsubsection{Classical postprocessing}

Like some of the other protocols, we use a multi-layer perceptron (MLP) to perform classification based on the output of the TMS protocol. The MLP we use has the structure $2\rightarrow 64 \rightarrow 64 \rightarrow 64 \rightarrow 2$. We have verified that MLPs that are deeper (i.e. have more layers) or larger (i.e. have more neurons) do not lead to a qualitative change in classification performance for the TMS scheme.

\subsubsection{Simulation results}

We now consider the classification accuracy of this protocol in simulation for the task considered in Fig.~\ref{fig:main4}. In addition, we also consider the effects of experimental non-idealities, which will limit the overall efficiency of this protocol. In superconducting circuit implementations, the main experimental imperfection is the quantum efficiency $\eta$ of the amplifiers used for the generation of two-mode squeezed states and homodyne measurements~\cite{krantz2019quantum, eberle2013stable, qiu2023broadband, liu2022noise}. The effect of these imperfections can be modeled by a beamsplitter operation with an idler mode, followed by an ideal amplifier. We therefore simulate the performance of this protocol as a function of the amount two-mode squeezing $S$ (measured in dB) and the quantum efficiencies $\eta_1$ (for the amplifier generating the two-mode squeezed state) and $\eta_2$ (for the amplifier performing the measurement). At input, the variance of the field is:
\begin{equation}
    \sigma_{\mathrm{input}}^2 = \frac{1}{2} \eta_1 10^{-S/10} + \frac{1}{2} (1-\eta_1).
\end{equation}
The effect of the beam splitter interaction on the output not only introduces noise, but also reduces the strength of the signal. The variance of the measurement signal is:
\begin{align}
    \sigma_{\mathrm{output}}^2 &= \eta_2 \sigma_{\mathrm{input}}^2 + \frac{1}{2} (1-\eta_2) \\
    &=  \frac{1}{2} \eta_1 \eta_2 10^{-S/10} + \frac{1}{2} (1-\eta_1) \eta_2 + \frac{1}{2} (1-\eta_2),
\end{align}
while the signal is scaled by: $\alpha_{\mathrm{output}} = \sqrt{\eta_2} \alpha$, where $\alpha$ is the sensed displacement. Together, we can define the signal-to-noise ($\mathrm{SNR}$) quantity:
\begin{equation}
    \mathrm{SNR}(S, \eta) = \frac{\alpha_{\mathrm{output}}}{\sigma_{\mathrm{output}}} = \frac{\sqrt{\eta} \alpha}{\sqrt{\frac{1}{2} \eta^2 10^{-S/10} + \frac{1}{2} (1-\eta) \eta + \frac{1}{2} (1-\eta)}},
\end{equation}
where we have made the assumption that both amplifiers have equal efficiency $\eta_1 = \eta_2 = \eta$. When $\eta < 1$, this imposes an upper limit on the maximum value of $\mathrm{SNR}$, even in the limit of infinite two-mode squeezing.

\begin{figure}[h!]
    \centering
    \includegraphics[width=\textwidth]{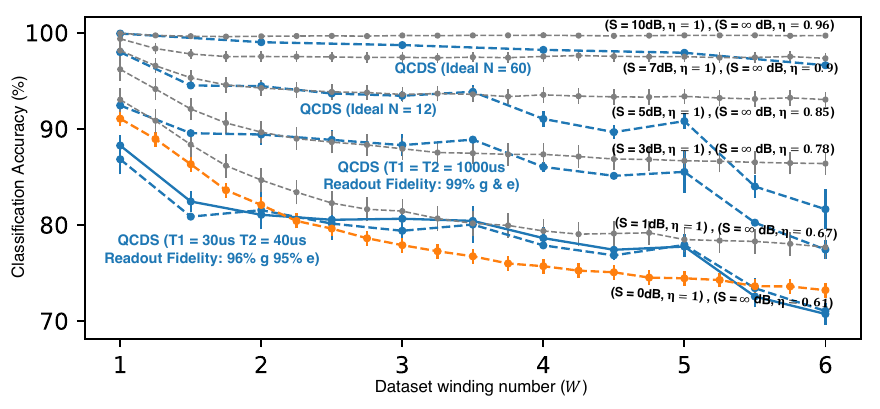}
    \caption{\textbf{Performance of the QCDS and two-mode squeezed-state protocols for the task defined in Fig.~\ref{fig:main4}.} We plot the classification accuracy obtained for different levels of squeezing and efficiency. We choose different values of two-mode squeezing, at perfect efficiency $\eta$. At $0~\rm dB$ of squeezing, this is equivalent to the ideal limit of the heterodyne (PPA-MLP - plotted in orange) protocol. For each value, there is a corresponding efficiency $\eta < 1$ which cannot obtain a higher classification accuracy. In addition, we plot the accuracy of the QCDS protocol under several regimes. First we consider the ideal regime for two different values of $N$. Next, we simulate the performance of the protocol under different experimental imperfections such as finite qubit coherence times T1 and T2, and readout fidelity conditioned on the ground and excited states. The parameters associated with our experiment match well with the experimental results (shown in blue solid), demonstrating that our simulation indeed capture most the effects of the relevant experimental imperfections.}
    \label{fig:tms_real}
\end{figure}

In Fig.~\ref{fig:tms_real}, we plot the simulation performance of the two-mode squeezed-state protocol under different values of $S$ and $\eta$. As a comparison, we plot the performance of the QCDS protocol (both in ideal simulations and simulations which model decoherence and experimental non-idealities). When assuming a qubit coherence times of $100~\rm us$ and readout fidelity of $99\%$, the expected performance exceeds the maximum accuracy obtainable with an efficiency $\eta = 0.78$. While the parameters of the former (i.e., QCDS) protocol is within the regime of state-of-the-art parameters, the parameters of the latter (i.e., TMS-MLP) protocol are experimentally challenging. Our simulation results highlight the prospects of the QCDS protocol to achieve larger QCSA under current trends in improvement in superconducting hardware.

\subsubsection{Photon number efficiency}

As derived in Eqs.~(\ref{appeq:tmsvarx}),~(\ref{appeq:tmsvary}), the dense coding scheme using TMS enables measurement of both quadratures of a displacement received in a bosonic mode with arbitrary precision, limited only by the degree of squeezing available. An appropriate comparison against our QCDS protocol then becomes one of resource efficiency: for the same resources, can the QCDS protocol outperform the TMS protocol (since the TMS protocol can achieve an arbitrarily good performance given enough resources). 

The resource we consider, which is typical for many sensing applications, is the probe state photon number: the average photon number in the sensor prior to receiving the sensing displacement. For the TMS protocol, the photon number $\bar{n}_{\rm TMS} = 2\sinh^2r$, which is directly related to the degree of TMS $S(r)$ via Eq.~(\ref{appeq:sqdef}). We use the analogous quantity $\bar{n}_{\rm QCDS}$ of our QCDS protocol to provide a detailed analysis of the relative performance of these two protocols for performing displacement classification tasks.

The task we consider is the spirals classification task from the main text. However, in addition to varying the winding number $W$ of the task, we will also allow an overall variation of the displacement scale $d$, which amounts to receiving the displacement operation $D(d\alpha)$. For simulations and experiments in the main text, $d=1.0$. In this section, we consider smaller displacement scales, down to $d=0.1$. Reducing the displacement scale increases the task difficulty and requires increasing the sensing resources. 

For a range of winding numbers and displacement scales, we optimize QCDS protocols for binary classification of the spirals dataset. We consider depth $N=60$ QCDS protocols. Our optimization proceeds in two stages. We first train the protocols as detailed in Appendix~\ref{app:training_procedure}, which constitutes an unconstrained loss minimization: the only constraint on the photon numbers in the QCDS protocol is imposed by the Hilbert space cutoff. 

We then instantiate the QCDS protocol with the best performing model yielded by this first training procedure, and subsequently perform a compression stage: training is performed with a modified loss function introducing a probe state photon number penalty: $\mathcal{L}_{\rm com} = \mathcal{L} + \lambda~ \bar{n}_{\rm QCDS}$, The probe state photon number $\bar{n}_{\rm QCDS}$ is computed at every training epoch, and the penalty $\lambda$ is adjusted dynamically (over epochs) to ensure the validation accuracy is not reduced too drastically as $\bar{n}_{\rm QCDS}$ is compressed. We find that this additional compression stage is particularly useful at smaller displacement scales $d$, where the unconstrained training procedure can sometimes favor protocols with large probe state photon numbers with marginal gains in classification accuracy.

\begin{figure}[h!]
    \centering
    \includegraphics[width=\textwidth]{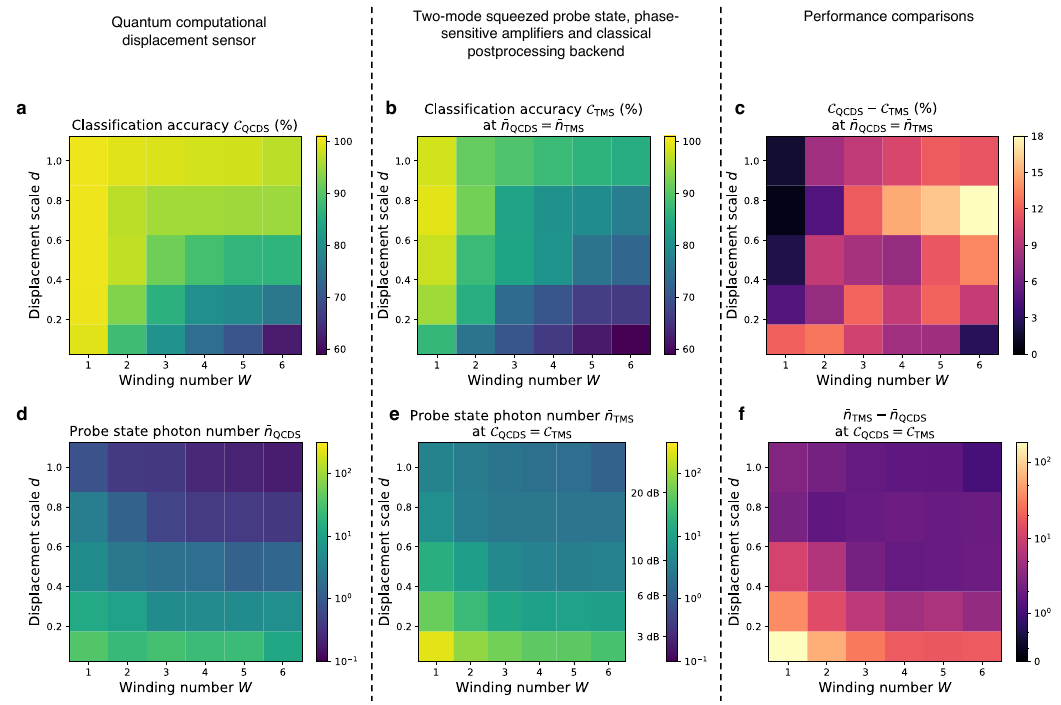}
    \caption{\textbf{Simulation performance of the QCDS protocol and the protocol involving two-mode squeezed state as a function of the winding number $W$ and scale of the dataset $d$}. $d=1$ corresponds to the task considered in Fig.~\ref{fig:main4}. We compare the photon number of the probe-state requirements of the two protocols. \textbf{a.} Classification accuracy $\mathcal{C}_{\mathrm{QCDS}}$ of the QCDS protocol at depth $N=60$. The classification accuracy tends to reduce with increasing $W$ and decreasing $d$, which makes the task harder. \textbf{b.} Classification accuracy $\mathcal{C}_{\mathrm{TMS}}$ of the two-mode squeezed state based protocol, when it has access to the same probe-state photons as the QCDS. \textbf{c.} Difference in accuracy. We observe that the QCDS protocol achieves a higher classification accuracy across the entire parameter space with equivalent photon number. \textbf{d.} Photon numbers of the probe-state generated by the parameters of the QCDS protocol for the performance listed in \textbf{a.} \textbf{e.} Photon number required by the two-mode squeezed-state based protocol to achieve the same classification accuracy as \textbf{a.}, also labeled by the corresponding squeezing strength. \textbf{f.} Difference in photon numbers between the two protocols. As before, across the entire parameter space, two-mode squeezed state based protocol requires more photons than the QCDS protocol to achieve equivalent classification accuracy. The results of \textbf{c.} and \textbf{f.} empirically highlight the photon-number efficiency of the QCDS protocol.}
    \label{fig:tms_performance}
\end{figure}

In Fig.~\ref{fig:tms_performance}, we perform a detailed comparison of the performance of the QCDS protocol and the TMS protocols. Fig.~\ref{fig:tms_performance}(a) shows the classification accuracy achieved by our optimized QCDS protocol $\mathcal{C}_{\rm QCDS}$, as a function of spirals winding number and displacement scale. Recall that our experiment corresponds to a displacement scale $d=1$. For the smaller displacement scales down to $d=0.1$ we are able to explore in simulation, we note that the engineered QCDS probe state photon number $\bar{n}_{\rm QCDS}$ increases, as shown in Fig.~\ref{fig:tms_performance}(d). 

We compare QCDS performance against the TMS protocol in two related ways. We can first ask: if the TMS protocol is allowed the same number of probe state photons as the optimized QCDS protocol, what classification accuracy $\mathcal{C}_{\rm TMS}$ can it achieve? This result is shown in Fig.~\ref{fig:tms_performance}(b); for visibility, the difference between the QCDS and TMS performance is plotted directly in Fig.~\ref{fig:tms_performance}(c). Most importantly, we find that for $\mathcal{C}_{\rm QCDS} > \mathcal{C}_{\rm TMS}$ for all the tasks considered here. The achieved advantage we have observed can vary from a few percentage points to 18 percentage points; we note that the extent of advantage depends on several factors, including the task difficulty and the training variability.

We can also ask the complementary question: what probe state photon number must the TMS protocol use to reach the same accuracy as the QCDS protocol, $\mathcal{C}_{\rm QCDS} = \mathcal{C}_{\rm TMS}$. Under this requirement, the probe state photon number needed by the TMS protocol is depicted in Fig.~\ref{fig:tms_performance}(e). Again to aid the comparison with the QCDS protocol, we plot the difference in probe state photon numbers in Fig.~\ref{fig:tms_performance}(f). Note that we always find $\bar{n}_{\rm TMS} > \bar{n}_{\rm QCDS}$ for the tasks considered here, so the QCDS protocol is more photon-number-efficient. Furthermore, the photon number requirements of the TMS protocol can be substantially larger than those of the QCDS protocol especially at smaller displacement scales.

Our results show that, at least for the spirals task considered in this work, across a range of task complexities (characterized by the spirals winding number and also by the extent of the displacement scale), the QCDS protocol is more efficient than the TMS protocol: it is able to reach a higher classification accuracy for the same probe state photon number. Equivalently, it requires a smaller probe-state photon number to reach the same classification accuracy.

The efficiency of the QCDS protocol comes with some important constraints, inherent to quantum computational sensing in general. The TMS protocol, like other conventional quantum sensing protocols, is task-agnostic: as it aims to estimate the raw sensed displacement, the protocol is unchanged regardless of whatever task one may eventually be interested in. The QCDS protocol, on the other hand, must be optimized for the specific task, which can be expensive. Nevertheless, this optimization allows the QCDS protocol to extract relevant information more efficiently.

\clearpage
\subsection{Additional examples}
\label{app:benchmark:additional_examples}

\begin{figure}[h!]
    \centering
    \includegraphics[width=0.8\textwidth]{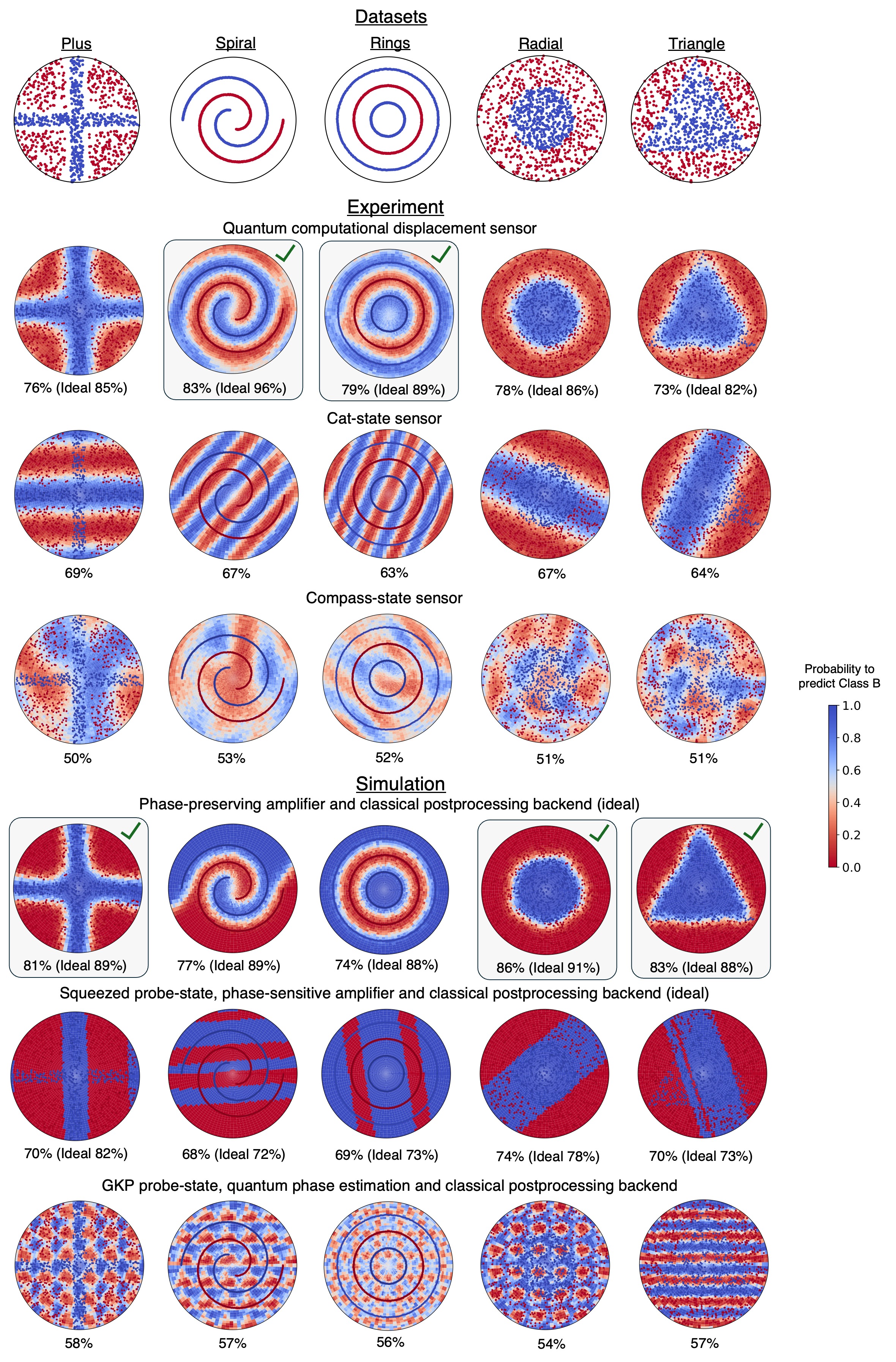}
    \caption{\textbf{Response of the various protocols as a function of displacement for the first five tasks considered in Appendix~\ref{app:quantum_computational_sensing_protocol:examples}.} The response is defined as the probability for the protocol to predict class B given it received a displacement at that point. Highlighted check-marked boxes indicate the best performing protocol (for simulation this corresponds to the performance at state-of-the-art experimental parameters).}
    \label{fig:examples_part_1}
\end{figure}

\begin{figure}[h!]
    \centering
    \includegraphics[width=0.8\textwidth]{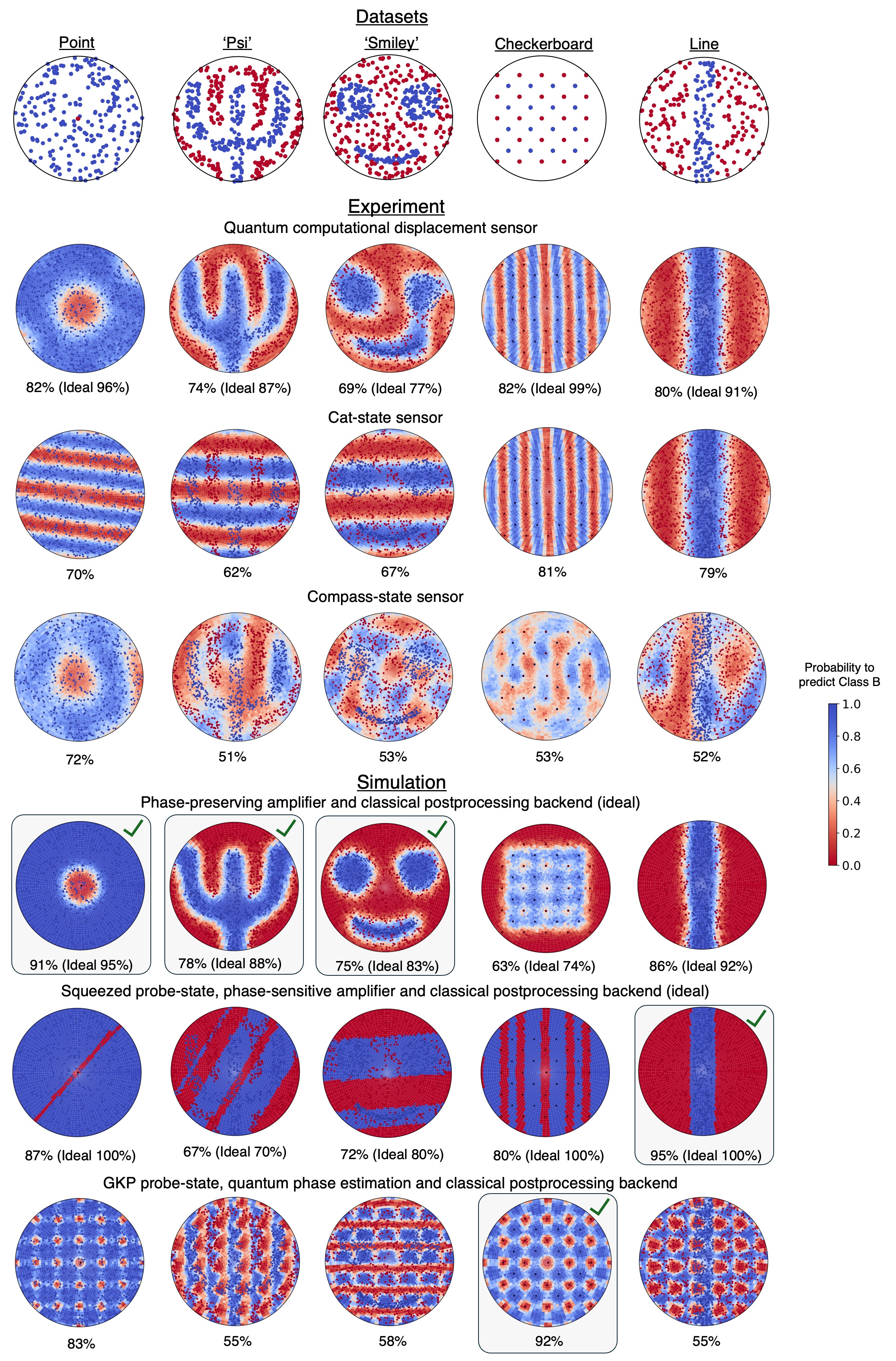}
    \caption{\textbf{Response of the various protocols as a function of displacement for the second five tasks considered in Appendix~\ref{app:quantum_computational_sensing_protocol:examples}.} The response is defined as the probability for the protocol to predict class B given it received a displacement at that point. Highlighted check-marked boxes indicate the best performing protocol (for simulation this corresponds to the performance at state-of-the-art experimental parameters).}
    \label{fig:examples_part_2}
\end{figure}

We compare the performance of the QCDS protocol, along with the set of quantum displacement sensing protocols considered so far, for several different tasks (see Figs.~\ref{fig:examples_part_1} and~\ref{fig:examples_part_2}). For three of the protocols, we also provide the accuracy under ideal conditions. For the amplifier-based protocols, this corresponds to $100\%$ quantum efficiency, and infinite squeezing. For the QCDS protocol, this corresponds to the accuracy achieved in simulation at depth $N = 10$ without taking into account errors due to decoherence or qubit readout fidelity. 

The best-performing protocol depends strongly on the structure of the dataset. This is due to the fact that each protocol is suited for different forms of sensing tasks. Below we discuss, for each task, the types of datasets for which they are suitable. However, we notice that the QCDS protocol achieves a competitive performance (that is, either the highest or close-to-the-highest classification accuracy) across all tasks. Furthermore, it outperforms the experimental quantum sensing protocols (cat-state and compass-state protocols) across all tasks.

\underline{\emph{Quantum computational displacement sensor}} -- Achieves the highest, or close to the highest, classification accuracy across all the tasks considered. This is due to the high expressivity, or space of functions, which can be realized. However, as the circuit depth increases, the effect of decoherence is enhanced, which limits the maximum performance which can be achieved. The tasks where the protocol achieves the highest classification accuracy are those which can be efficiently expressed by the protocol, such as 'spiral' and 'rings' tasks, but also inefficiently by other protocols. Since the QCDS protocol is designed particularly for single-shot sensing, this protocol outperforms the other two protocols (cat and compass) which use the same hardware of a qubit coupled to the sensing oscillator.

\underline{\emph{Cat-state sensor}} -- This protocol has a sinusoidal response to the displacement, with the periodicity, orientation and offset of the response trainable to match the dataset. Therefore this protocol is restricted to distinguishing the datasets by a single Fourier component. The cat-state sensor achieves high classification accuracy for tasks (such as the 'checkerboard' and 'line' tasks) where the maximum difference in a Fourier component is large between the two datasets.

\underline{\emph{Compass-state sensor}} -- This protocol is designed for estimating the magnitude of the displacement when the phase is unknown. Consequently, this protocol is not inherently optimal for single-shot displacement sensing setting tasks considered in this work. Therefore, the performance of this protocol is limited across almost all the tasks considered. The protocol achieves a relatively high classification accuracy for the 'point' task, since that inherently requires distinguishing weak displacements with unknown phase.

\underline{\emph{Phase-preserving amplifier and classical postprocessing backend}} -- Across all datasets, this protocol achieves relatively high classification accuracy. Due to the classical postprocessing, which we allow to utilize arbitrarily large expressivity, the response to the displacement closely matches the distribution of the dataset. The bottleneck in achieving ideal classification accuracy is the fundamental quantum noise introduced in performing heterodyne measurement via a phase-preserving amplifier. As discussed previously, this introduces additive Gaussian noise on top of the sensed signal. Consequently, the effective dataset viewed by the classical postprocessing is the convolution of the original dataset by a Gaussian kernel of the noise profile. In Fourier space, this is equivalent to a Gaussian kernel multiplying the distribution of the datasets. Importantly, high frequency components experience a low signal-to-noise ratio. If these components are required in accurately distinguishing the datasets, then the performance of this protocol will be limited. Therefore, the overall scale of the datasets sets the classification accuracy which can be achieved. For instance, datasets where the length scale is small (such as between the arms of the distributions in the 'spiral' tasks, between concentric rings in the 'rings' task and the points in the 'checkerboard' tasks) result in relatively low classification accuracy.

\underline{\emph{Squeezed probe-state, phase-sensitive amplifier and classical postprocessing backend}} -- This protocol achieves high accuracy when only a single component of the displacement is required to separate the two datasets. This protocol tends to outperform the cat-state sensor (and always does in the ideal scenario) since it can represent arbitrary functions of the single component of the displacement (while the cat-state sensor is restricted to sinusoidal responses). Therefore this protocol achieves low classification accuracy for the 'rings', 'spirals', and 'threshold' tasks. In the ideal scenario of noiseless measurement of a single quadrature, this protocol achieves perfect classification accuracy for 'line', 'point' and 'checkerboard' -- all which can be distinguished by a single component.

\underline{\emph{GKP probe-state, quantum phase estimation and classical postprocessing backend}} -- Due to the modular response of this protocol, it achieves relatively low classification accuracy for tasks where the dataset is on a scale much larger than the scale of periodicity. We designed the 'checkerboard' dataset to match the periodicity of the protocol. Therefore, this protocol achieves a high classification accuracy for this protocol (since we simulate the expected performance of a GKP probe-state with an experimentally achievable finite photon-number, the performance is below perfect classification accuracy). Furthermore, this protocol also achieves high classification accuracy on the 'point' task, due to its small support for one of the class.

\clearpage

\clearpage

\clearpage
\subsection{Resources}
\label{app:benchmark:resources}

In this subsection, we benchmark the various protocols considered via several resource metrics. Our choice of metrics is based on potential constraints one might face when applying these protocols for practical applications. For instance, the photon number of the probe-state is a prototypical example of a quantum resource. The larger the photon number a protocol requires to perform the task, the greater the requirements placed on the hardware (such as the coherence timescale of all the components). We discuss each metric in detail, discussing their relevance and the values for each protocol. We do note, however, that all the protocols are optimized to maximize classification accuracy (to benchmark our protocol against the best-possible performance of quantum sensing protocols on current quantum hardware). Therefore, the numbers listed in Table~\ref{table:resource} should not be understood as strict limits, but rather as a qualitative comparison across protocols.

\underline{\emph{Number of photons in the oscillator before sensing}} -- The larger this value is, the more likely the effects of decoherence (for instance, due to photon loss) will affect the performance of the corresponding protocol in experiment. Across all the tasks considered, we observe that the number of photons in the probe state for the protocols implemented in experiment (QCDS, CaS, CoS) is small, with the values of the QCDS protocol generally on the lower end. The photon number for the GKP-MLP protocol is based on the theoretical number based on the size of the grid state chosen~\cite{labarca_quantum_2025} which has been experimentally realized~\cite{sivak2023real}. Finally, the number of photons for amplifier based protocols is estimated from the corresponding squeezing value (if relevant). As discussed before, these numbers have not been optimized. For instance, one can train the QCDS protocol under the constraint of photon number. This can be done by directly incorporating the photon number in the loss function during training.

\underline{\emph{Duration of the protocol in microseconds}} -- This corresponds to the total duration for one round of sensing, including any state preparation, the sensing duration, and the final measurement. This is a useful metric when many sensing rounds will be performed, since this directly sets the repetition rate of the experiment. Therefore, the smaller this number, the more samples which can be collected within a given duration. For the QCDS protocol, this depends on the depth $N$ of best performance. The duration increases linearly with $N$. In our experiments, we wait a time of $4\mathrm{ms}$ between rounds of the experiment to allow the device to equilibrate to the ground state. We have not included this in the estimate since this value can be cut down significantly (which we did not perform in this work) using methods based on those experimentally demonstrated in Ref.~\cite{sivak2023real}. In a practical implementation, it is important to include this wait time. The experimental demonstrations of the amplifier-based protocols (PPA-MLP, PSA-MLP)~\cite{zhong2018results, al2017design, backes2021quantum, bai2025dark} do not have a strict definition for the duration of a single experiment run since the measurements are performed continuously in time. Therefore, the comparison depends on the particular application. For the purpose of the tasks considered in this work, we consider the time required to perform a single round of measurement for estimating the displacement of the oscillator used in this work. This corresponds roughly to the cavity T1, which sets the timescale the signal from the oscillator would 'leak out' into a transmission line attached to the amplifier. This timescale can be significantly reduced depending on the requirements of a task. Finally, the duration of the GKP-MLP protocol depends on the timescale for one round of error correction (which is also roughly the time for one round of phase estimation) and the number of rounds required. Ref.~\cite{sivak2023real} reports that $10$ rounds of error correction (one for each quadrature) are required to stabilize in the GKP manifold. Each round took roughly $10~\mathrm{us}$ In the assessment of this protocol, we consider $10$ rounds of phase estimation (one for each quadrature)~\cite{labarca_quantum_2025}. Therefore the expected timescale for one measurement round is roughly $200~\mathrm{us}$.

\underline{\emph{Number of entangling gates}} -- For the protocols involving an oscillator coupled to the qubit, this corresponds to the number of ECD gates. For the QCDS protocol, this is twice the depth of the protocol. For the CaS protocol, due to the simplicity of the protocol, only two ECD gates are required. We implement the compass-state protocol by variationally training a sequence of single-qubit rotations and ECD gates. The total number of ECD gates in the circuit ansatz was $20$. However, an experiment designed to efficiently implement the compass-state protocol (which cannot be achieved with the device parameters of our experiment) will be able to significantly reduce this number. For the GKP-MLP protocol, this corresponds to the total number of rounds of error correction and phase-estimation, each requiring 2 ECD gates. Finally, the TMS-MLP protocol required one entangling operation when the device can natively implement the two-mode squeezing Hamiltonian. However, the relevant metric in this situation is the maximum strength of the two-mode squeezing parameter which can be realized.

\underline{\emph{Number of trainable parameters}} -- This includes both the parameters used (if any) for the quantum protocol and parameters used (if any) for the classical postprocessing backend. As before, we emphasize that none of the protocols were optimized with this number in mind. For instance, the number of parameters used in the classical postprocessing could potentially be significantly reduced, and furthermore depending on the task. Since the number of trainable parameters for classical postprocessing is a economically cheap resource (for the range of values required in this work), we instead choose the postprocessing MLP which maximizes classification accuracy. Therefore, even increasing this number will not improve the performance of the protocol. For the QCDS protocol, the number of parameters is $3N + 2$. For the amplifier-based protocol and GKP-MLP, the number of parameters are essentially all in the classical postprocessing (since there are minimal quantum degrees of freedom (if any) which can be optimized).

\underline{\emph{Ancilla resource}} -- We list the requirement of what (if any) additional hardware is required to perform the protocol. QCDS, CaS, CoS and GKP-MLP all require an ancilla qubit. Therefore, the performance of this protocol depends not only on the coherence of the sensing oscillator, but also the ancilla qubit (placing additional hardware constraints). TMS-MLP requires an ancilla oscillator mode, which is entangled with the sensing oscillator.

\underline{\emph{Calibration overhead}} -- Not all protocols utilizing equivalent quantum hardware are equivalently feasible to run in experiment. Some protocols incur additional overhead, based on the calibrations required to implement them efficiently. This can place a constraint on practical tasks especially if the device drifts, requiring constant calibration. While the exact calibration overhead depends on the task and application, we divide the protocols into two broad categories. CaS, CoS, PPA-MLP and PSA-MLP all require relatively low levels of calibrations in superconducting circuits. This is due to their low degrees of freedom, which can be easily tuned up. QCDS, GKP-MLP and TMS-MLP usually incur high calibration overhead in superconducting circuits. For instance, the QCDS protocol requires an accurate simulation to train the parameters of the protocol. We do note that after designing the simulation, the simulation-reality gap remains small -- we achieved similar performance across tasks after a span of 6 months, requiring only simple tune-ups (single-qubit rotation pulses). GKP-MLP requires implementation of error correction, which incurs large overhead and is experimentally challenging. TMS-MLP requires producing two-mode squeezed states. While producing such states in superconducting circuits can be experimentally feasible, achieving below-SQL sensitivity can be challenging.

\begin{table}[H]
    \renewcommand{\arraystretch}{1.3}
    
    \begin{minipage}[t]{0.49\textwidth}
    \centering
    \begin{adjustbox}{width=\linewidth}
    \begin{tabular}{|c|c|c|c|c|c|c|c|}
    \hline
    
    Task & \multicolumn{7}{c|}{Sensing protocol} \\
    \hline

    & QCDS & CaS & CoS & PPA-MLP & PSA-MLP & GKP-MLP & TMS-MLP\\
    \hline

    \multicolumn{8}{|c|}{\rule[-1.5ex]{0pt}{5ex} \textbf{Number of photons in the oscillator before sensing}} \\
    \hline
    
    Spiral $W=3.5$ & 0.38 & 0.42 & 0.64 & \multirow{11}{*}{0} 
    &\multirow{11}{*}{\begin{tabular}{@{}c@{}}
    0.23 (4dB)\\
    $\infty$ (Ideal)
    \end{tabular}} 
    &\multirow{11}{*}{3.6} 
    &\multirow{11}{*}{\begin{tabular}{@{}c@{}}
    0.01 (1dB)\\
    0.12 (3dB) \\
    0.37 (5dB) \\
    $\infty$ (Ideal)
    \end{tabular}} \\
    \cline{1-4}

    Plus & 0.06 & 0.05 & 0.04 & & & &\\
    \cline{1-4}
    
    Spiral & 0.2 & 0.18 & 0.16 & & & &\\
    \cline{1-4}
    
    Rings & 0.17 & 0.23 & 0.64 & & & &\\
    \cline{1-4}
    
    Radial & 0.03 & 0.02 & 0.25 & & & &\\
    \cline{1-4}
    
    Triangle & 0.01 & 0.02 & 0.25 & & & &\\
    \cline{1-4}
    
    Point & 0.01 & 0.19 & 0.09 & & & &\\
    \cline{1-4}
    
    'Psi' & 0.1 & 0.1 & 0.16 & & & &\\
    \cline{1-4}
    
    'Smiley' & 0.06 & 0.06 & 0.25 & & & &\\
    \cline{1-4}
    
    Checkerboard & 0.45 & 0.45 & 0.81 & & & &\\
    \cline{1-4}
    
    Line & 0.04 & 0.04 & 0.16 & & & &\\
    \hline

    \multicolumn{8}{|c|}{\rule[-1.5ex]{0pt}{5ex} \textbf{Duration of the protocol in microseconds}} \\
    \hline
    
    Spiral $W=3.5$ & 11.46 & 6.18 & \multirow{11}{*}{9.8} & \multirow{11}{*}{$\sim$250} &\multirow{11}{*}{$\sim$250} & \multirow{11}{*}{$\sim$200} & \multirow{11}{*}{$\sim$250} \\
    \cline{1-3}

    Plus & 5.30 & 2.66 & & & & &\\
    \cline{1-3}
    
    Spiral & 9.70 & 4.42 & & & & &\\
    \cline{1-3}
    
    Rings & 9.70 & 5.30 & & & & &\\
    \cline{1-3}
    
    Radial & 5.30 & 2.66 & & & & &\\
    \cline{1-3}
    
    Triangle & 6.18 & 2.66 & & & & &\\
    \cline{1-3}
    
    Point & 4.42 & 4.42 & & & & &\\
    \cline{1-3}
    
    'Psi' & 6.18 & 3.54 & & & & &\\
    \cline{1-3}
    
    'Smiley' & 6.18 & 3.54 & & & & &\\
    \cline{1-3}
    
    Checkerboard & 7.06 & 6.18 & & & & &\\
    \cline{1-3}
    
    Line & 2.66 & 2.66 & & & & &\\
    \hline

    \multicolumn{8}{|c|}{\rule[-1.5ex]{0pt}{5ex}\textbf{Ancilla resource}} \\
    \hline

    All tasks & Qubit & Qubit & Qubit & None & None & Qubit & Oscillator\\
    \hline

    \end{tabular}
    \end{adjustbox}
    \end{minipage}
    \hfill
    \begin{minipage}[t]{0.49\textwidth}
    \centering
    \begin{adjustbox}{width=\linewidth}
    \begin{tabular}{|c|c|c|c|c|c|c|c|}
    \hline

    Task & \multicolumn{7}{c|}{Sensing protocol} \\
    \hline

    & QCDS & CaS & CoS & PPA-MLP & PSA-MLP & GKP-MLP & TMS-MLP\\
    \hline

    \multicolumn{8}{|c|}{\rule[-1.5ex]{0pt}{5ex} \textbf{Number of entangling gates}} \\
    \hline
    
    Spiral $W=3.5$ & 24 & \multirow{11}{*}{2} & \multirow{11}{*}{20} & \multirow{11}{*}{0} &\multirow{11}{*}{0} & \multirow{11}{*}{40} & \multirow{11}{*}{1} \\
    \cline{1-2}

    Plus & 10 & & & & & &\\
    \cline{1-2}
    
    Spiral & 20 & & & & & &\\
    \cline{1-2}
    
    Rings & 20 & & & & & &\\
    \cline{1-2}
    
    Radial & 10 & & & & & &\\
    \cline{1-2}
    
    Triangle & 12 & & & & & &\\
    \cline{1-2}
    
    Point & 8 & & & & & &\\
    \cline{1-2}
    
    'Psi' & 12 & & & & & &\\
    \cline{1-2}
    
    'Smiley' & 12 & & & & & &\\
    \cline{1-2}
    
    Checkerboard & 14 & & & & & &\\
    \cline{1-2}
    
    Line & 4 & & & & & &\\
    \hline
    
    \multicolumn{8}{|c|}{\rule[-1.5ex]{0pt}{5ex} \textbf{Number of trainable parameters}} \\
    \hline
    
    Spiral $W=3.5$ & 38 & \multirow{11}{*}{3} & \multirow{11}{*}{4} & \multirow{11}{*}{4482} &\multirow{11}{*}{291} & \multirow{11}{*}{5634} & \multirow{11}{*}{4482} \\
    \cline{1-2}

    Plus & 17 & & & & & &\\
    \cline{1-2}
    
    Spiral & 32 & & & & & &\\
    \cline{1-2}
    
    Rings & 32 & & & & & &\\
    \cline{1-2}
    
    Radial & 17 & & & & & &\\
    \cline{1-2}
    
    Triangle & 20 & & & & & &\\
    \cline{1-2}
    
    Point & 14 & & & & & &\\
    \cline{1-2}
    
    'Psi' & 20 & & & & & &\\
    \cline{1-2}
    
    'Smiley' & 20 & & & & & &\\
    \cline{1-2}
    
    Checkerboard & 23 & & & & & &\\
    \cline{1-2}
    
    Line & 8 & & & & & &\\
    \hline

    \multicolumn{8}{|c|}{\rule[-1.5ex]{0pt}{5ex} \textbf{Calibration overhead}} \\
    \hline

    All tasks & High & Low & Low & Low & Low & High & High\\
    \hline

    \end{tabular}
    \end{adjustbox}
    \end{minipage}

    \caption{\textbf{Resource estimates.} We characterize the QCDS protocol, along with all quantum sensing protocols considered for several metrics. In cases where the value is task dependent, we list their values for the task considered in Fig.~\ref{fig:main4} at $W = 3.5$ and for the $10$ tasks considered in Appendix~\ref{app:quantum_computational_sensing_protocol:examples}. Legend: QCDS -- quantum computational displacement sensor, CaS -- cat-state sensor, CoS -- compass-state sensor, PPA-MLP -- phase-preserving amplifier and classical postprocessing backend, PSA-MLP -- squeezed probe-state, phase-sensitive amplifier and classical postprocessing backend, GKP-MLP -- GKP probe-state, quantum phase estimation and classical postprocessing backend, TMS-MLP -- two-mode squeezed vacuum, phase-sensitive amplifiers and classical postprocessing backend.}
    \label{table:resource}
\end{table}

\newpage
\section{Potential quantum computational displacement sensors}
\label{app:future}

\subsection{QCDS based on displacement and SNAP gates}
\label{app:future:snap}

SNAP (selective number-dependent arbitrary phase) gates~\cite{heeres2015cavity} combined with displacement operations provide universal control over a bosonic mode~\cite{krastanov2015universal}. In this appendix section, we discuss how SNAP gates can be used to realize a quantum computational displacement sensor using a single bosonic mode. The QCDS protocol we propose has the same structure as our bosonic QSP-inspired protocol, although here we consider operations only in the Hilbert space of the single bosonic mode that is being used for sensing, which is the mode of operation of SNAP gates. Formally, starting from the initial sensor state $\ket{0}$, the protocol can be defined as:
\begin{align}
    \ket{\psi(\bm{\alpha})} = {\rm SNAP}({\bm{\zeta}}')U^{\dagger}(\vec{\bm{\zeta}},\vec{\bm{\beta}})D(\bm{\alpha})U(\vec{\bm{\zeta}},\vec{\bm{\beta}})\ket{0}.
    \label{appeq:qcdssnap}
\end{align}
The unitary $U$ takes the form $U(\vec{\bm{\zeta}},\vec{\bm{\beta}}) = \prod_{i=0}^{N-1} D(\bm{\beta}_i){\rm SNAP}({\bm{\zeta}}_i)$. where $D(\bm{\beta}_i)$ defines \textit{unconditional} displacements of the bosonic mode, while ${\rm SNAP}(\bm{\zeta}_i)$ is the SNAP gate. The SNAP gate is defined as is usual:
\begin{align}
    {\rm SNAP}(\bm{\zeta}) = \sum_{n=0}^{\mathcal{N}} e^{i m_n\zeta_n} \ket{n}\bra{n},
\end{align}
where $\bm{\zeta} = (\zeta_1,\ldots,\zeta_{\mathcal{N}})$ where $\mathcal{N}$ denotes the maximum Fock state number that the SNAP gate can be applied on. We note that SNAP gates can be realized in cQED using an auxiliary control qubit that is dispersively coupled to the bosonic mode; however at the conclusion of the gate the qubit and bosonic mode are disentangled, so that the system of interest remains the single bosonic mode that constitutes the quantum computational sensor. Following a final SNAP gate that leads to the sensor state $\ket{\psi(\bm{\alpha})}$, the QCDS protocol is completed by performing a measurement of the bosonic mode to extract information about the class label. For binary classification where only a single bit of information is ultimately desired, a natural choice is parity measurement of the final state of the bosonic mode. In practice, such a measurement can be implemented using the same control qubit that is used to implement the SNAP gate on the bosonic mode~\cite{wang2020efficient}.

We now discuss the various trainable parameters that are optimized in the QCDS protocol defined by Eq.~(\ref{appeq:qcdssnap}). We have introduced vectors of trainable parameters for each operation in the $N$ layers of the protocol, as before. Starting with the unconditional displacement operations, the vector of trainable parameters $\vec{\bm{\beta}} = (\bm{\beta}_0,\ldots,\bm{\beta}_{N-1})$ defines the complex-valued displacements applied, where $\bm{\beta}_i$ is the displacement for the $i$th layer. As with the ECD operation, we have assumed that $|\bm{\beta}_i|$ is fixed for all layers and equal to the ECD displacement magnitude, and the phase of the displacements is trainable.

For SNAP gates, we define the vector of trainable parameters $\vec{\bm{\zeta}} = ({\bm{\zeta}}_0,\ldots,{\bm{\zeta}}_{N-1})$, where $\bm{\zeta}_i$ defines the trainable phases applied in the $i$th layer. For this simulated implementation of the SNAP gate, we can assume control over $\mathcal{N}$ Fock levels that can be as large as the Hilbert space truncation level in our analysis. In experiment, however, usually only a handful number of levels can be simultaneously controlled. To this end, we introduce \textit{binary} control parameters $\bm{m} = (m_1,\ldots,m_{\mathcal{N}})$ that are designed to represent which of the $\mathcal{N}$ states can be controlled in an experiment. For example, universal control over the Fock states would entail $m_n = 1~\forall~n$, so that $\sum_n m_n = \mathcal{N}$. More practically, we can assume that only a smaller subset $\mathcal{N}_G < \mathcal{N}$ of the states can be controlled, in which case $\sum_n m_n = \mathcal{N}_G$ (namely $\mathcal{N}-\mathcal{N}_G$ values of the binary parameters $m_n$ are set to zero). We determine which of the states are controllable (i.e. which of the $m_n$'s are non-vanishing) during training using an appropriate loss function that tries to maximum performance while also respecting the constraint $\sum_n m_n = \mathcal{N}_G$. Note that we assume $\bm{m}$ carries no layer index $i$; this represents the experimental situation where the same set of bosonic Fock states can be controlled from one layer to the next. The trainable phase vector $\vec{\bm{\zeta}}$ can (and generally does) vary with layer index $i$.

\begin{figure}
    \centering
    \includegraphics[width=\linewidth]{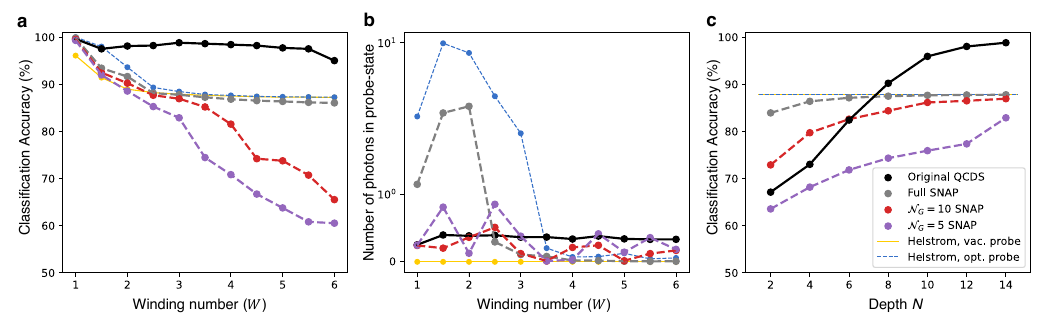}
    \caption{\textbf{Quantum computational displacement sensing with a single bosonic mode sensor using SNAP gates}. \textbf{a.} Classification accuracy vs. spirals winding number $W$ for SNAP gate protocol and using the bosonic QSP protocol (at depth $N=14$). Also plotted is the maximum accuracy achievable as defined by the Helstrom bound when using a vacuum probe state, and when using an arbitrary optimized probe state, but still with operations on a single bosonic mode. \textbf{b.} Probe state photon number for the various protocols in \textbf{a.} as a function of spirals winding number $W$. \textbf{c.} Classification accuracy as a function of the depth of the protocol, for $W = 3$.}
    \label{appfig:snap}
\end{figure}

We use the protocol defined by Eq.~(\ref{appeq:qcdssnap}) to perform binary classification for the spirals task as a function of winding number $W$, as considered in the main text. We train the protocol parameters $\vec{\bm{\zeta}},\vec{\bm{\beta}}$ for $N=14$ layers. The results are shown in Fig.~\ref{appfig:snap}a, alongside the results of our QCDS protocol; note that the latter uses a qubit-cavity sensor and implements bosonic QSP (echoed conditional displacements and single-qubit rotations). We find that the SNAP gate protocol performance appears bounded at high winding numbers and substantially underperforms the bosonic QSP protocol that uses an entangled qubit. Restricting to control over fewer states $\mathcal{N}_{G}=5,10$ restricts performance even further, with $\mathcal{N}_G=10$ outperforming the $\mathcal{N}_{G}=5$ protocol, as expected from the increased number of trainable parameters.

To explore this result further, in Fig.~\ref{appfig:snap}a we also show the maximum accuracy that can be achieved for this series of tasks when performed using a single bosonic mode only; this is given by the Helstrom bound~\cite{helstrom1969quantum}. We consider two cases: first, the simple case where the bosonic mode is in a vacuum probe state prior to the sensing operation $D(\bm{\alpha})$. The Helstrom bound then determines the maximum classification accuracy for this task assuming an arbitrary measurement performed following sensing. For a second bound, we allow the initial probe state to be optimized using an arbitrary unitary operation over the bosonic mode Hilbert space. This latter Helstrom bound allows an optimized probe state and therefore should exceed the vacuum state Helstrom bound; this is indeed observed in Fig.~\ref{appfig:snap}a, at low $W$. In Fig.~\ref{appfig:snap}c, we plot the classification accuracy as a function of the depth of the protocol, which shows that the SNAP based protocol does not exceed the Helstrom bounds. However, the key observation is that both these bounds are below the bosonic QSP performance. This implies that when using a single bosonic mode only for the entire QCDS protocol, the maximum achievable performance for the spirals task is constrained. We believe this is due to the fact that the tasks we consider here require simultaneously extracting information contained in both quadratures of the bosonic mode, which requires the use of an ancilla mode (as is done for our bosonic QSP protocol, or for other schemes like dense-coding using a TMS interferometer). 

The probe state photon numbers generated by the various SNAP gate protocols are also shown as a function of the spiral winding number in Fig.~\ref{appfig:snap}b). The probe state photon numbers are generally on the order of or larger than those for our bosonic QSP protocol, especially at lower $W$. At higher $W$, we see that the full SNAP protocol, and more importantly the arbitrary unitary protocol (which is not photon number constrained) both end up learning protocols with very small probe state photon numbers (approaching a vacuum probe state), with the resulting performance approaching the vacuum state Helstrom bound. These results suggest that the underperformance is not simply a matter of probe state photon number resources. 

We next consider what may be possible if a SNAP-inspired gate could be used that could also allow entangling the bosonic receiver with a qubit, as we do for the bosonic QSP protocol.

\subsubsection{QCDS using conditional SNAP gate interferometry}

We now consider a generalization of SNAP gates that allow for the application of phases on the bosonic mode conditioned on the state of a control qubit. We refer to this as the conditional SNAP or CSNAP gate,
\begin{align}
    {\rm CSNAP}(\bm{\zeta},\bm{\xi}) = \sum_{n=0}^{\mathcal{N}} e^{i m_n\zeta_{n}} \ket{n,g}\bra{n,g} + \sum_{n=0}^{\mathcal{N}} e^{i m'_n\xi_{n}} \ket{n,e}\bra{n,e},
\end{align}
To the best of our knowledge, such a CSNAP gate has not been realized in the cQED literature thus far. Note that if $\bm{\zeta}=\bm{\xi}$, ${\rm CSNAP}$ reduces to the standard unconditional SNAP gate acting on the bosonic mode. In the more general case, the CSNAP gate is able to entangle a qubit with the bosonic mode, similar to the ECD gate. We have introduced binary control parameters $\bm{m},\bm{m}'$ that determine the states that can be controlled by the CSNAP gate; by allowing $\bm{m}$ and $\bm{m}'$ to be distinct, we allow for different states to be controllable for the two different qubit states.

The QCDS protocol using CSNAP gates then takes the form:
\begin{align}
    \ket{\psi(\bm{\alpha})} = R(\theta',\phi'){\rm CSNAP}({\bm{\zeta}}')U^{\dagger}(\vec{\bm{\zeta}},\vec{\bm{\beta}})D(\bm{\alpha})U(\vec{\bm{\zeta}},\vec{\bm{\beta}})\ket{0,g}
    \label{appeq:qcdscsnap}
\end{align}
where the trainable unitary $U$ now  takes the form $U(\vec{\theta},\vec{\phi},\vec{\bm{\zeta}},\vec{\bm{\beta}}) = \prod_{i=0}^{N-1} R(\theta_i,\phi_i){\rm CSNAP}({\bm{\zeta}}_i)D(\bm{\beta}_i)$, including trainable single qubit rotations $R(\theta_i,\phi_i)$ similar to our bosonic QSP protocol. A final CSNAP gate and qubit rotation is followed by measurement of the qubit, which is trained to output the predicted class label, again exactly as our bosonic QSP protocol.

\begin{figure}
    \centering
    \includegraphics[width=\linewidth]{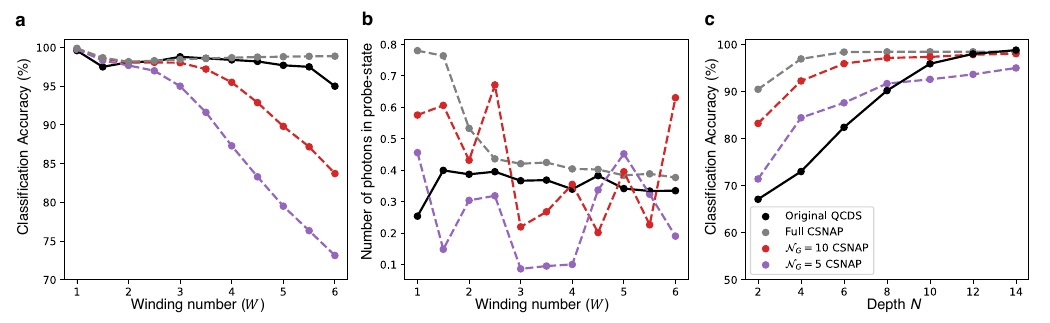}
    \caption{\textbf{Quantum computational displacement sensing with a qubit-bosonic mode sensor using CSNAP gates}. \textbf{a.} Classification accuracy vs. spirals winding number $W$ for CSNAP gate protocol and using the bosonic QSP protocol. \textbf{b.} Probe state photon number for various protocols in \textbf{a.} as a function of spirals winding number $W$. \textbf{c.} Classification accuracy as a function of the depth of the protocol, for $W = 3$.}
    \label{appfig:csnap}
\end{figure}

In Fig.~\ref{appfig:csnap}a, we plot the results of the spirals classification task as a function of winding number using the QCDS protocol with trainable CSNAP gates. We see now that the CSNAP protocol with full control over the bosonic Hilbert space is able to reach a similar performance to the bosonic QSP protocol. However, if we restrict control to $\mathcal{N}_G=5,10$ states, the performance is substantially lower, with the $\mathcal{N}_G=10$ protocol outperforming the $\mathcal{N}_G=5$ protocol once more. In Fig.~\ref{appfig:csnap}b, we plot the probe state photon number for the CSNAP protocols and the bosonic QSP protocol as a function of winding number $W$. We note that the photon numbers are all relatively close, since we have chosen the same displacement \textit{magnitude} for the CSNAP protocol as was used in our bosonic QSP protocol. The fact that the full CSNAP performance is very similar to the bosonic QSP performance hints that there may not be a significant advantage either way in terms of probe state photon number requirements. In Fig.~\ref{appfig:csnap}c, we present the classification accuracy as a function of the depth $N$ for the various protocols, at $W=3$. Here we see that CSNAP gate, if realized, could potentially provide more expressivity at lower depths than the bosonic QSP protocol, although this dependence would need to be studied across tasks to make a definitive conclusion.

These results suggest that the CSNAP generalization of SNAP gates, if realized, could provide a powerful scheme for QCDS. However, for the spirals task considered here, it appears that control over several bosonic mode states is required to reach similar performance to the bosonic QSP protocol.

\subsection{Distributed quantum computational displacement sensors}
\label{app:future:multi_mode_multi_qubit}

\begin{figure}[h!]
    \centering
    \includegraphics[width=0.6\textwidth]{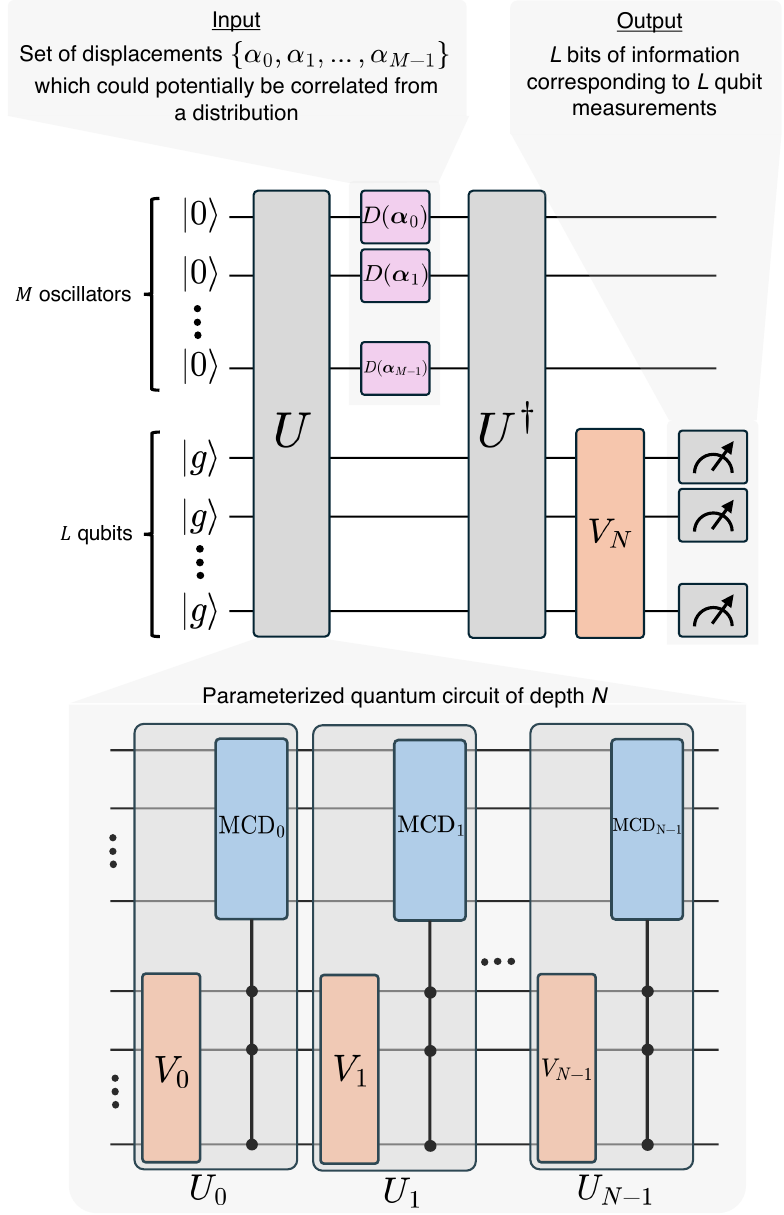}
    \caption{\textbf{Potential circuit diagram for a multi-mode multi-qubit quantum computational displacement sensor}. In the general setting, the $M$ bosonic modes each sense a displacement, which could originate from a classical probability distribution. The $L$ qubits, by entangling with the modes, can run a quantum computing algorithm, here characterized by a unitary $U$. The protocol ends with the qubits measured in their computational basis. Therefore the output of the sensor is a bitstring of length $L$.}
    \label{fig:potential}
\end{figure}

In this Appendix, we outline architectures for potential future implementations of quantum computational displacement sensors which can perform more sophisticated sensing tasks. The general architecture we envision is illustrated in Fig.~\ref{fig:potential}, which comprises of $M$ sensing bosonic modes, along with $L$ ancilla qubits. Such a system can in general compute functions of the sensed displacements $\{\alpha_0, \alpha_1, \dots \alpha_{M-1} \}$. The output is the bitstring generated by the measurements of the qubits at the end.

The unitary $U$ of the protocol is decomposed by a sequence of gates. Motivated by our protocol, each unit consists of an entangling gate, and a gate only in the qubit Hilbert subspace. The entangling gate, which we label as the ``multi-mode conditional displacement gate":
\begin{equation}
    \mathrm{MCD}(\boldsymbol{\beta}_{0i}, \boldsymbol{\beta}_{1i} \dots \boldsymbol{\beta}_{(M-1)i}) = \sum\limits_{i=0}^{2^L-1}D(\boldsymbol{\beta}_{0i})D(\boldsymbol{\beta}_{1i})\dots D(\boldsymbol{\beta}_{(M-1)i}) \ket{i}\!\bra{i},
\end{equation}
where $D(\bm{\beta}_{mi})$ is a displacement on the $m$th mode,
\begin{align}
    D(\bm{\beta}_{mi}) \equiv \exp \{ (\beta_{x,mi} + i\beta_{p,mi})\hat{a}_m^{\dagger} - (\beta_{x,mi} - i\beta_{p,mi})\hat{a}_m \},
\end{align}
and where we label the qubit states $\ket{b_0, b_1, \dots b_{L-1}}$ by integer $\ket{i}$ with corresponding binary expansion. Therefore the unitary imparts a displacement, conditioned on state $\ket{i}$, on each mode. We can consider the type of functions of displacements which can be represented by considering the action of this gate on the vacuum state of the bosonic modes. Then, similar to our protocol, this state picks up a geometrical phase proportional to the sum $\sum\limits_{m=0}^{M-1} (\alpha_{j} \beta_{j}^* - \alpha_{j}^* \beta_{j})$, where $\alpha_j$ and $\beta_j$ are complex numbers, with $\alpha_j$ the sensed displacement. This is therefore a weighted linear combinations of the sensed displacements. By appropriately choosing the conditional displacements for each ket $\ket{i}$, each state picks up a corresponding weighted sum of displacements.

Interspersed with these entangling gates, we envision qubit-only gates $V_i$, which is the generalization of the single qubit rotation gates of our protocol. For a given platform, these gates can be further broken down into simpler gates (such as a cascade of single-qubit and two-qubit gates). In total, this sequence of gates can represent a complicated interferometry protocol, capable of computing relevant features of the sensed displacement. The output is a sequence of qubit measurements. Since there are $2^L$ possible outcomes, such a protocol could in principle perform a $2^L$-class classification task. However, such a construction might be highly sensitive to any imperfections of the protocol, such qubit readout error, decoherence, or imperfect gates. It therefore might be more practical to use $L$ qubits to classify among $\sim L$ tasks. In this situation, all binary outcomes can be binned into the various class labels. This sets up redundancy, since multiple outcomes map to the same class label. Such a mapping can potentially be chosen to be robust to the imperfections of the system, which might result in different mappings under different noise scenarios. This mapping can also be done implicitly -- for instance -- via a classical neural network which maps the bitstring to class labels (which can be trained end-to-end along with parameters of the protocol)~\cite{khan2025quantum_novel}. We emphasize that such an ansatz might not be optimal for all sensing scenarios or quantum platforms. Other protocols, based on theoretical results, might be better suitable. For instance, Ref.~\cite{singh2025towards} introduces and numerically evaluates several protocols for bosonic state preparation and entangling interactions between bosonic modes. Such protocols might naturally be extended for quantum computational sensing tasks.

On the other hand, a conventional quantum sensor will reconstruct the vector of displacements $\{\alpha_0, \alpha_1, \dots \alpha_{M-1} \}$, which is then postprocessed on a classical computer to perform the corresponding task. Depending on the implementation, the estimate can be of poor signal-to-noise ratio. For instance, a phase-preserving QLA on each mode will add noise independently on each mode. Such an approach can perform poorly when the task involves estimating a global linear function of the displacements~\cite{zhuang2019physical}.

Finally, an open question is how such protocols can be trained. Due to the hardness of simulating quantum system on classical computers, training these protocols can be expensive. A potential solution could involve choosing an appropriate ansatz which can be efficiently classically simulable. While such a circuit cannot (by design) achieve a purely quantum computational advantage, it can still achieve a quantum computational-sensing advantage (such as our experiment). This might not be a strong restriction since many quantum sensing protocols do not rely on the full Hilbert space to achieve a quantum sensing advantage. Furthermore, a hybrid quantum-classical approach can be used, where a classical optimizer updates the parameters of the protocol based on measurement outcomes.

\end{document}